\documentclass[10pt,a4paper]{article}
\usepackage{geometry}
\usepackage{graphicx} 
\graphicspath{{./images/}}
\usepackage{titlepic}
\usepackage{float}
\usepackage{wrapfig}
\usepackage{tabularx}
\usepackage{multirow}
\usepackage{multicol}
\usepackage{comment}
\usepackage[utf8]{inputenc}
\usepackage{lineno}  
\usepackage{graphicx} 
\usepackage{amsmath} 
\usepackage{booktabs}    
\usepackage{threeparttable}  
\usepackage{caption}     
\usepackage{multirow}  
\usepackage{url} 
\usepackage{hyperref}
\hypersetup{
    colorlinks=true,
    linkcolor=blue,
    filecolor=blue,      
    urlcolor=blue,
    pdftitle={AION-10: Technical Design Report}, 
    }

\makeatletter
\renewcommand{\maketitle}{%
  \begin{center}
    {\LARGE\bfseries\@title\par}
    \vspace{1.0cm}
    
    {\large
      K. Bongs$^{a}$, A. Brzakalik, U. Chauhan, S. Dey, O. Ennis, S. Hedges$^{b}$, 
      T. Hird, M. Holynski, S.~Lellouch, M. Langlois$^{c}$, B. Stray$^{c}$;~$^1$
      
      \vspace{0.4cm}
      
      B. Bostwick$^{d}$, J. Chen, Z. Eyler, V. Gibson, T. L. Harte, C. C. Hsu, 
      M. Karzazi, C. Lu, B.~Millward, J. Mitchell, N. Mouelle, 
      B. Panchumarthi$^{e}$, J. Scheper, U. Schneider$^{*}$, X. Su$^{f}$, Y.~Tang, 
      K. Tkal{\v c}ec$^{*}$, M. Zeuner$^{g}$, S. Zhang, Y. Zhi$^{h}$;~$^2$
      
      \vspace{0.4cm}
      
      K. Clarke, A. Vick$^{*}$;~$^3$
      
      \vspace{0.4cm}
      
      C. F. A. Baynham, O. Buchm{\"u}ller$^{*}$, D. Evans, L. Hawkins, R. Hobson, 
      L.~Iannizzotto-Venezze, A. Josset, D. Lee, E. Pasatembou, 
      B. E. Sauer, M. R. Tarbutt, T.~Walker;~$^4$
      
      \vspace{0.4cm}
      
      L. Badurina$^{i}$, A. Beniwal$^{j}$, D. Blas$^{k}$, J. Carlton, J. Ellis$^{*}$, C. McCabe, 
      G. Parish, D.~Pathak~Govardhan, V. Vaskonen$^{l}$;~$^5$
      
      \vspace{0.4cm}
      
      T. Bowcock, K. Bridges$^{*}$, A. Carroll, J. Coleman, G. Elertas$^{*}$, S. Hindley, 
      C. Metelko, H.~Throssell$^{*}$, J. N. Tinsley$^{*}$;~$^6$
      
      \vspace{0.4cm}
      
      E. Bentine, M. Booth, D. Bortoletto, C. Foot$^{*}$, N. Callaghan$^{*}$, 
      C. Gomez-Monedero, K.~Hughes, A. James, T. Leese, A. Lowe$^{*}$, 
      J. March-Russell, J. Sander$^{*}$, J. Schelfhout, I.~Shipsey$^{\dagger}$, D. Weatherill, D. Wood;~$^7$
      
      \vspace{0.4cm}
      
      S.N. Balashov$^{*}$, M.G. Bason, K. Hussain$^{m}$, H. Labiad, P. Majewski, 
      A.L. Marchant, D.~Newbold, Z. Pan$^{*}$, Z. Tam$^{*}$, T.C. Thornton, 
      T. Valenzuela, M.G.D. van der Grinten$^{*}$, I.~Wilmut~$^8$
    }
  \end{center}
  
  \vspace{0.8cm}
  \noindent\rule{\textwidth}{0.4pt}
  \vspace{0.4cm}
  
  \begin{flushleft}
    \small
    $^1$Physics and Astronomy, University of Birmingham, Edgbaston, Birmingham, B15 2TT, UK\\
    $^2$Cavendish Laboratory, J J Thomson Avenue, University of Cambridge, Cambridge, CB3 0HE, UK\\
    $^3$ASTeC, STFC Daresbury Laboratory, Warrington, WA4 4AD, UK\\
    $^4$Department of Physics, Blackett Laboratory, Imperial College London, Prince Consort Road, London, SW7 2AZ, UK\\
    $^5$Physics Department, King's College London, Strand, London, WC2R 2LS, UK\\
    $^6$Department of Physics, University of Liverpool, Merseyside, L69 7ZE, UK\\
    $^7$Department of Physics, University of Oxford, Parks Road, Oxford, OX1 3PU, UK\\
    $^8$Rutherford Appleton Laboratory, UKRI-STFC, Harwell Campus, Didcot, OX11 OQX, UK
  \end{flushleft}
  
  \vspace{0.4cm}
  \noindent\rule{\textwidth}{0.4pt}
  \vspace{0.4cm}
  
  \begin{flushleft}
    \footnotesize
    $^a$Present address: Institute of Quantum Technologies, German Aerospace Center (DLR), Wilhelm-Runge-Stra{\ss}e 10, 89081 Ulm, Germany\\
    $^b$Present address: Nomad Atomics, 33 Elizabeth Street, Richmond, Victoria 3121, Australia\\
    $^c$Present address: Jet Propulsion Laboratory, California Institute of Technology, Pasadena, California 91109, USA\\
    $^d$Present address: Kirchhoff-Institut f{\" u}r Physik, Universität Heidelberg, Im Neuenheimer Feld 227, 69120 Heidelberg, Germany\\
    $^e$Present address: Department of Physics \& Astronomy, Northwestern University, 2145 Sheridan Road, Evanston, Illinois 60208-3112, USA\\
    $^f$Present address: Physikalisches Institut, Universit{\" a}t Tübingen, Auf der Morgenstelle 14, 72076 T{\" u}bingen, Germany\\
    $^g$Present address: Ludwig-Maximilians-Universit{\" a}t München, Geschwister-Scholl-Platz 1, 80539 München, Germany\\
    $^h$Present address: Department of Physics, University of Virginia, 382 McCormick Road, Charlottesville, VA 22904, USA\\
    $^i$Present address: Walter Burke Institute for Theoretical Physics, California Institute of Technology, Pasadena, CA 91125, USA\\
    $^j$Present address: Intersect Australia, Sydney, Australia\\
    $^k$Present address: Institut de F\'{i}sica d'Altes Energies (IFAE), The Barcelona Institute of Science and Technology, Campus UAB, 08193 Bellaterra (Barcelona), Spain and Instituci\'{o} Catalana de Recerca i Estudis Avan\c{c}ats (ICREA), Passeig Llu\'{i}s Companys 23, 08010 Barcelona, Spain\\
    $^l$Present address: Keemilise ja Bioloogilise F\"{u}\"{u}sika Instituut, R\"{a}vala pst. 10, 10143 Tallinn, Estonia\\
    $^m$Also at the University of Liverpool\\
    $^*$Leading contributors to this work\\
    $^\dagger$Deceased
  \end{flushleft}
  
  \vspace{0.4cm}
  
}
\makeatother

\begin{document}

\title{
\includegraphics[width=0.5\textwidth]{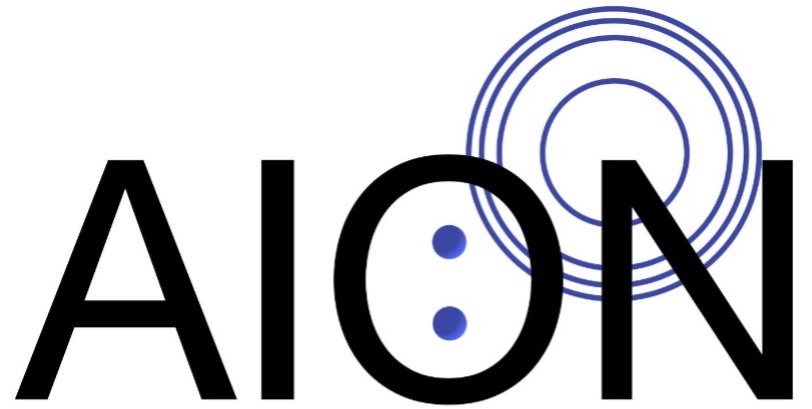}
\newline
\newline
AION-10: Technical Design Report for a 10m Atom Interferometer in Oxford}

\date{\today}
\maketitle
\begin{abstract}
This Technical Design Report presents AION-10, a 10-meter atom interferometer to be located at Oxford University using ultracold strontium atoms to make precision measurements of fundamental physics. AION-10 serves as both a prototype for future larger-scale experiments and a versatile scientific instrument capable of conducting its own diverse physics programme.

The design features a 10-meter vertical tower housing two atom interferometer sources in an ultra-high vacuum environment. Key engineering challenges include achieving nanometer-level vibrational stability and precise magnetic field control. Solutions include active vibration isolation, specialized magnetic shielding, and a modular assembly approach using professional lifting equipment.

Detailed analysis confirms the design meets all performance requirements, with critical optical components remaining within our specifications 97\% of the time under realistic operating conditions. Vacuum and vibration measurements in the host building validate that the instrument will achieve the precision needed for quantum sensing applications.

This work establishes the technical foundation for scaling atom interferometry to longer baselines while creating a cutting-edge facility for precision measurements that could advance our understanding of fundamental physics.\\
~~\\
AION-REPORT/2025-04
\end{abstract}

\newpage
\tableofcontents

\newpage

\section{Introduction}
\label{sec:introduction}
The discoveries by the LIGO and Virgo long-baseline laser interferometer experiments of gravitational waves (GWs) emitted during the mergers of stellar-mass black holes~\cite{LIGOScientific:2016aoc} and neutron stars~\cite{LIGOScientific:2017vwq} motivate experiments searching for GWs in other frequency ranges. Examples include space-borne laser interferometer experiments such as LISA~\cite{LISA:2017pwj} and pulsar timing arrays (PTAs)~\cite{InternationalPulsarTimingArray:2023mzf} that probe lower frequency ranges. Atom interferometers~\cite{Buchmueller:2023nll} such as AION~\cite{Badurina:2019hst} have the potential to fill a gap in the frequency coverage between terrestrial and space-borne laser interferometers~\cite{Dimopoulos:2007cj,Graham:2012sy,Graham:2016plp}, opening a window onto mergers of intermediate mass black holes~\cite{Ellis:2023iyb}. These mergers could be stepping stones in the formation of the supermassive black holes at the centres of most galaxies, so searching for GWs in complementary frequency ranges has the potential to provide a deeper understanding of the most massive compact objects in the universe.

In addition to searching for GWs, atom interferometers have many other applications in fundamental physics, including searches for the interactions of waves of ultra-light dark matter with atoms~\cite{Graham:2015ifn,Arvanitaki:2016fyj,Badurina:2021lwr}, violations of the gravitational equivalence principle~\cite{Biedermann:2014jya, Rosi:2017ieh,Asenbaum:2020era}, probing the properties of dark energy~\cite{Burrage:2014oza,Hamilton:2015zga,Sabulsky:2018jma} and searching for possible ``fifth forces”~\cite{Abdalla:2024sst}. For these reasons, increasing effort around the world is going into developing long-baseline atom interferometers~\cite{TVLBAIMOU}. We note that atom interferometry is also an important driver of research in quantum sensor technologies with applications to navigation and remote sensing~\cite{Stray:2022grf}, and has much in common with technologies for quantum computing. 

The AION Collaboration is a team with members from multiple UK institutions planning to develop and construct a family of atom interferometers using \textsuperscript{87}Sr, working in partnership with the MAGIS experiment~\cite{MAGIS-100:2021etm} in the US and co-operating with other international collaborators. The next phase of the AION project is to build a $10\,\textrm{m}$ baseline instrument capable of taking useful science measurements, and to prove out the technology required to move to longer-baseline experiments ($100\,\textrm{m}$ and $1\,\textrm{km}$ on Earth, and the AEDGE mission in space~\cite{AEDGE:2019nxb}). As such, the AION programme aims to play a pioneering role in the rapidly advancing field of atom interferometry, working at the forefront of quantum sensing and fundamental physics exploration.

The AION Collaboration plans to use \textsuperscript{87}Sr atoms~\cite{Badurina:2019hst}, because their narrow $698\,\textrm{nm}$ optical transition facilitates exceptional precision performance, making them ideal for high-sensitivity measurements.
Delivering AION's ambitious scientific goals necessitates overcoming key engineering challenges: achieving ultra-high vacuum, maintaining nanoradian-level vibrational stability, and controlling magnetic environments with high precision. These challenges form the focus of this Technical Design Report (TDR), which summarises the engineering progress of the project on the $10\,\textrm{m}$ instrument and sets out the roadmap for commissioning the experiment's tower and vacuum system. In parallel with the AION $10\,\textrm{m}$ design work presented here, R\&D work is also being carried out on developing atom sources~\cite{AION:2023fpx} and interferometry techniques~\cite{AION:2025igp} for the project in several laboratories around the UK. This complementary work is beyond the scope of this TDR, and is not covered in detail here.

The structure of this TDR is as follows. In Section~\ref{sec:specs} we outline the specifications and requirements, in Section~\ref{sec:design} we describe the engineering designs of the components of the AION $10\,\textrm{m}$ detector, and in Section~\ref{sec:assembly} we describe the planned assembly and build procedure. Section~\ref{sec:stability} presents a vibrational stability analysis, Section~\ref{sec:magnetic} analyses the magnetic system, and Section~\ref{sec:vacuum} describes the vacuum system. The TDR is summarised in Section~\ref{sec:conclusions}.

\section{Specification Requirements for AION-10}
\label{sec:specs}

\subsection{Overview of Requirements}

AION-10 is a gradiometer based on differential measurements between two atom interferometers utilising the ultra-narrow clock transition in fermionic \textsuperscript{87}Sr \cite{Badurina:2019hst, AION:2025igp}. The instrument is intended both as a scientific platform and as a development stage towards larger-scale interferometers \cite{AION:2023fpx}. In addition to the direct practical requirements, for instance the required viewports, in-vacuum optics, and the connection nodes to further equipment (cameras, atom sources), there are many requirements stemming from potential systematic errors. 

In the ideal case without other technical limitations or systematic errors, the relevant sensitivity of an individual atom interferometer is typically inversely proportional to the number of large-momentum transfers ($N_{\rm LMT}$) and the square root of the number of atoms ($N_{\rm a}$):

\begin{equation}
d^{\text{best}} \propto \frac{1}{N_{\rm LMT}}\frac{1}{\sqrt{N_{\rm a}}},
\end{equation}
where $d^{\text{best}}$ represents the smallest detectable signal amplitude. 
Both $N_{LMT}$ and $N_a$ will improve over time through upgrades to exchangeable components (lasers and atom sources) and optimized interferometer sequences that will undergo continuous development. The permanent infrastructure, namely the $10\,\textrm{m}$ tower structure, vacuum vessel, magnetic system, and in-vacuum optics, must in contrast constrain systematic errors to levels below the final targeted resolution \cite{AION:2023fpx}.
The design presented in this report is driven by rigorous specifications that enable optimal performance for gravitational wave and dark matter detection. 


\subsection{Mechanical stability (Sections \ref{sec:design}, \ref{sec:stability})}

Mechanical vibrations of optical components in the path of the interferometer beam (steering mirrors, telescope lenses, retro mirror) lead to changes in the optical path length between the atom clouds and hence to additional phase noise in the differential measurements --- for details see the relevant Sections.

We note also that the envisioned phase-shear readout maps the interferometer phase onto the spatial fringe pattern of the atomic density~\cite{Sugarbaker:2013}. For a typical spacing of 1000~$\mu$m, a resolution goal of $\pi/10000$ translates to a fringe displacement of approximately $100\,\textrm{nm}$. Therefore, the camera assemblies must either remain stable within $100\,\textrm{nm}$ or allow real-time monitoring of their relative motion to this precision. This influences the vibration isolation platform design (Section~\ref{sec:isolation}) and camera alignment systems (Section~\ref{sec:cameras}), with performance validated in Section~\ref{sec:stability}.

The minimum positional accuracy for optical components will guide the decision whether to use adjustable mounts. Since in-vacuo adjustability adds complexity and failure risk, its use will be limited as far as possible.

In addition to the short-term vibrational stability, an instrument of this size ($10\,\textrm{m}$) will unavoidably be subject to thermal drifts over timescales of several hours. The instrument must sustain operational conditions continuously over such periods and allow for auxiliary dimensional monitoring to be able to compensate for slow drifts in the data processing.

\subsection{Magnetic Field Requirements (Section \ref{sec:magnetic})}
The fermionic \textsuperscript{87}Sr atom has a Zeeman structure due to its nuclear angular momentum $I = 9/2$. This creates multiple possible clock transitions, which are all sensitive to magnetic fields. In particular, the M = 9/2 to 9/2 clock transition exhibits a frequency shift of 489~Hz/G~\cite{boyd_nuclear_2007}.
In order to avoid parasitic excitations along unwanted transitions, such as the M = 9/2 to 7/2 transition, two requirements need to be fulfilled:\\
 (1) A well-defined constant quantisation axis provided by a horizontal magnetic bias field of up to 10~G is needed along the relevant parts of the interferometer beam pipe to resolve neighbouring Zeeman sublevels and optimise state preparation and interrogation.\\
(2) The interferometer laser must be linearly polarised with its polarisation angle less than $10\,\textrm{mrad}$  from the applied magnetic field, so that the M = 9/2 to 9/2 transition is driven while the parasitic M = 9/2 to 7/2 ($\sigma^-$) transition is strongly suppressed.
 A $10\,\textrm{mrad}$ misalignment between magnetic field and polarisation direction would already correspond to a $10^{-4}$ ratio of Rabi frequencies on the undesired vs the desired Zeeman transitions, corresponding to a $10^{-4}$  pulse infidelity.
 
 The light polarisation will be set by the in-vacuo optics, which will be difficult to align to within the required tolerance to a single set of bias coils.
Hence a tunable horizontal bias field  with homogeneity better than $5\,\textrm{mG}$ and $5\,\textrm{mrad}$ (angular) is required. It must be applicable in both horizontal axes to be able to align its direction  with the linear light polarisation.

In addition, time-dependent magnetic field noise must remain below 1~$\mu$G/$\sqrt{\text{Hz}}$ in order to suppress phase noise in the interferometric signal. 

These requirements are met by the magnetic shielding and field design described in Sections \ref{sec:magshielding} and \ref{sec:magguide}.

\subsection{Vacuum Requirements (Section \ref{sec:vacuum})}
To avoid atom loss and decoherence from collisions with residual gas,  the pressure in the interferometer pipe should be in the extreme high vacuum (XHV) region (ideally close to  $p\approx 10^{-11}\,$mbar). This requirement is central to the vacuum system architecture described in Section \ref{sec:vacuum}. The vacuum requirements in the parts needed for the preparation of the interferometer laser beams (input beam pipe, beam conditioning pipe (BCP), beam transfer pipe) are less stringent high vacuum (HV) requirements ($p<10^{-5}\,$mbar).
Maintaining a clean assembly environment is essential to minimize the risk of deposition of debris on optical surfaces during final pump-down, as diffraction off, e.g., dust particles would give rise to phase-front errors in the interferometer beam. Practical cleanliness procedures are detailed in Section~\ref{sec:vacconnect}.

\subsection{Design Implications and System Integrity}

These requirements underpin the engineering designs presented throughout this document. As discussed in Section \ref{sec:schematic}, AION-10 comprises two vertically aligned $5\,\textrm{m}$ interferometers housed in a $10\,\textrm{m}$ XHV system. 
Detailed analyses in the subsequent sections demonstrate how the system meets each requirement, ensuring that AION-10 fulfils its role as a high-precision scientific instrument and as a prototype for future instruments such as AION-100 and AEDGE \cite{Badurina:2019hst}.

\section{AION-10 Design Elements}
\label{sec:design}
\subsection{Instrument Schematic}
\label{sec:schematic}
\begin{figure}
    \centering
    \includegraphics[width=0.825\textwidth]{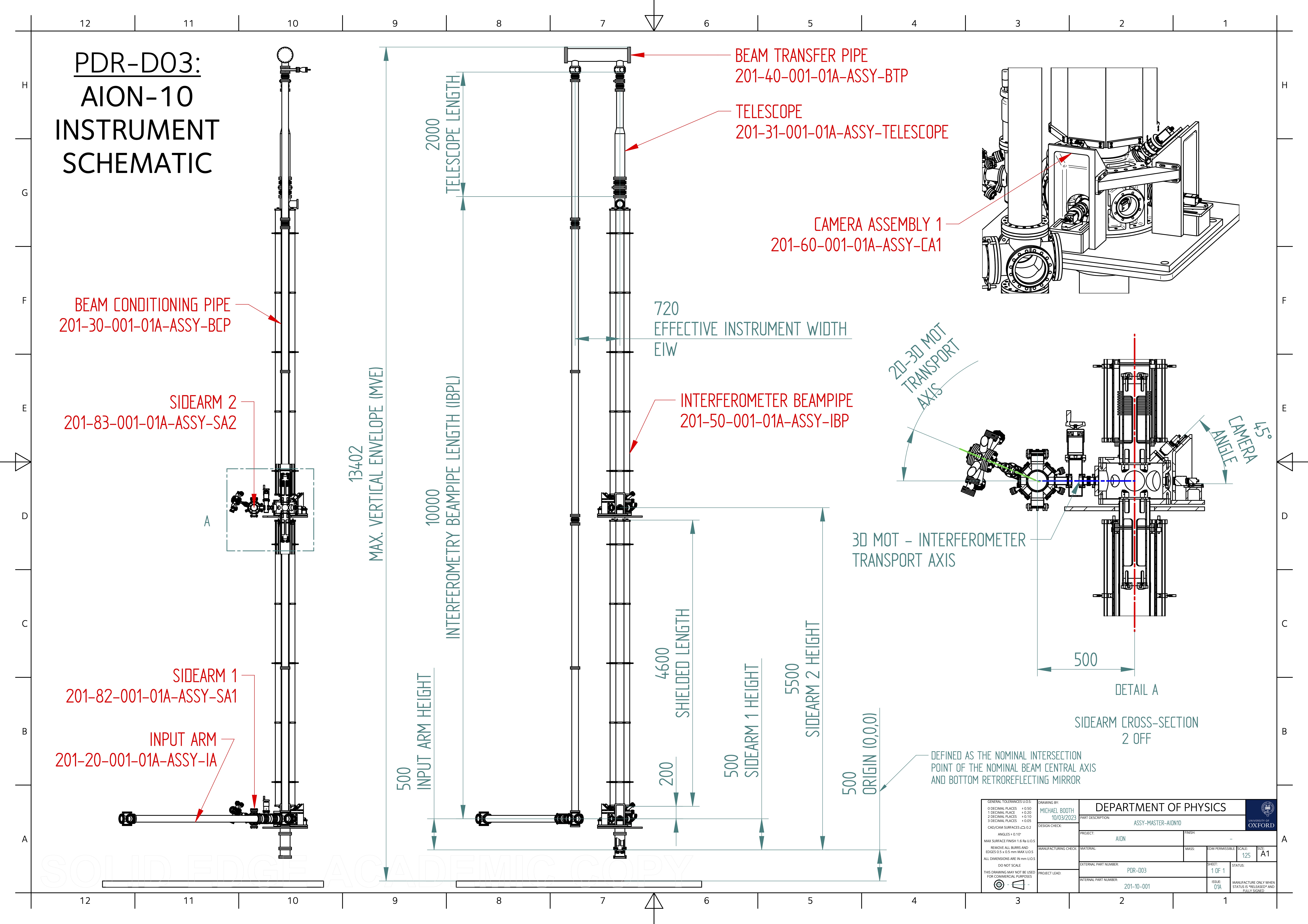}
    \caption{Schematic engineering drawing of AION-10.}
    \label{fig:schematic1}
\end{figure}

The AION-10 project envisages an aligned pair of $5\,\textrm{m}$ baseline atom interferometers in a $10\,\textrm{m}$ vertical vacuum tube. Fig.~\ref{fig:schematic1}, produced for the project's Provisional Design Review, shows the layout of the science instrument (not including its support structure). The instrument uses a narrow-linewidth interferometer laser at $698\,\textrm{nm}$ to perform interferometry with clouds of \textsuperscript{87}Sr atoms launched into a pair of vertical $5\,\textrm{m}$ beam pipes, with camera assemblies at the base of each pipe to measure the final phase shifts of the falling atoms. The interferometer laser beam enters the instrument through the input arm and is then directed up through the beam conditioning pipe, where it is reflected by a pair of mirrors in the beam transfer pipe back down through the interferometry beam pipe. It is then reflected again off the bottom retroreflecting mirror (BRM), which is mounted on the phase-shear detection platform at the base of the instrument, to point back up into the interferometry beam pipes. 

The atom clouds are prepared in two dedicated atom sources, called sidearms, that contain strontium ovens, 2D and 3D  magneto-optical traps, and additional traps  to produce the ultra-cold atom clouds that are then transported through XHV connections into the interconnect chambers in the interferometer beam pipe. 
As discussed in Section \ref{sec:specs}, these sidearms are crucial for the final resolution of the interferometers. As they are treated as exchangeable components and will undergo continued development, they are not included in this design review.

Two of the most important requirements from the project engineering perspective are the vacuum levels and the vibrational stability requirements. The vacuum level in the main beam pipe must be ${\cal O}(10^{-11})$~mbar, and the vibrational stability during interferometry sequences must keep some key optical components in the range of tens of nanoradians/nanometres from their design orientations/positions. The vacuum requirement is less stringent between the input arm and the upper lens of the telescope, and there will be an interface between HV and XHV at the upper telescope lens.
The rest of Section \ref{sec:design} details the various sub-assemblies that form the design of AION-10, reviewing the current levels of maturity, decisions taken and assumptions made for each. The order of presentation largely follows the interferometer laser beam as it travels through the vacuum system.

\subsection{Input Arm}
\label{sec:inputarm}
\begin{figure}
    \centering
    \includegraphics[width=0.6\textwidth]{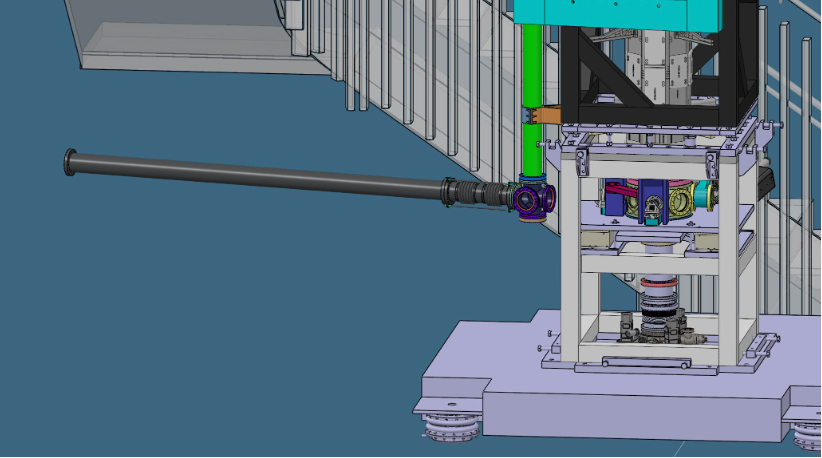}
    \caption{Engineering drawing of the laser input arm to the beam conditioning pipe.}
    \label{fig:inputarm}
\end{figure}

The input arm (shown in Fig.~\ref{fig:inputarm}) transfers the laser beam from the laser laboratory to the interferometer. It is in the early stages of design, as the decision on free space or fibre-optic laser delivery to the beam conditioning pipe has not yet been made. For now we assume that we will use in-vacuo free-space delivery, as it is the more difficult option from an engineering perspective. We may simplify later the specification to fibre-optic delivery.  

The input arm is expected to consist of a straight vacuum pipe with a viewport at the end to insert the laser beam and potentially to contain two convex lenses of identical focal lengths $f$ to deliver the beam via a $4f$ telescope to the beam conditioning pipe. It will be connected to the vertical beam conditioning pipe via a six-way cross that will house the steering mirror and corresponding adjustment mechanisms and provide additional vacuum ports for beam monitoring optics and pump apertures.



\subsection{Beam Conditioning Pipe}
\label{sec:bcp}
\begin{figure}
    \centering
    \includegraphics[width=0.15\textwidth]{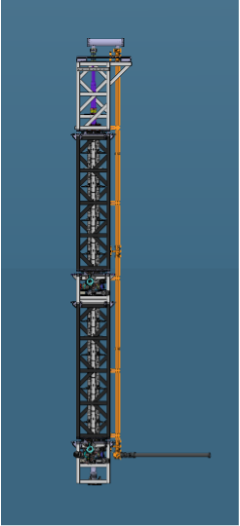}
    \hspace{5mm}
        \includegraphics[width=0.45\textwidth]{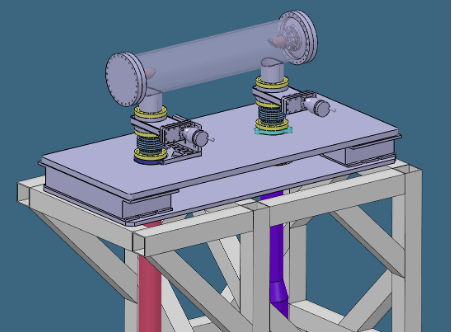}
        \hspace{5mm}
            \includegraphics[width=0.13\textwidth]{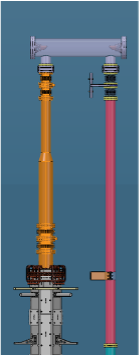}
    \caption{Left panel: Beam conditioning pipe on the full tower. Centre panel: Beam transfer pipe. Right panel: Highlight of the telescope section.}
    \label{fig:BCP}
\end{figure}

The beam conditioning pipe (BCP) (shown in the left panel of Fig.~\ref{fig:BCP}) is one of the simpler sections of the vacuum system. It consists of a tube under less stringent HV levels than the interferometer beam pipe, on the order of $10^{-5}\,$~mbar. Its primary purpose is to provide a vacuum environment to transmit the laser without picking up wavefront errors due to air turbulence or dust particles and to house dedicated beam-cleaning optics to diffract away existing wavefront perturbations to prepare the input beam for the main telescope (see below). We have designed lens holders with adjustment mechanisms and bellows on either side. These can be placed at any point along the BCP, and the final positions will be determined by a trade-off between stability simulations and beam cleaning performance, taking into account the final free space or fibre-optic beam delivery discussed above.
We have placed lenses at the bottom, middle and top of the BCP in CAD to investigate this, and will finalize vibration stability simulations on each location once the beam delivery matures.

\subsection{Beam Transfer Pipe}
\label{sec:btp}
The function of the beam transfer pipe (shown in the centre panel of Fig.~\ref{fig:BCP}) is to reflect the interferometry laser beam around 180 degrees via two mirrors into the main telescope at the top of the interferometer beam pipe. 
At least one of the two mirrors will be dynamically adjusted during each interferometer sequence in order to enable, together with the bottom retroreflecting mirror, a tunable effective pivot point for the interferometer laser to compensate for Coriolis effects~\cite{glick2024coriolis}.
The second mirror is likely to use  a piezo-actuated hexapod system or comparable setup  to compensate slow drifts of the tower geometry due to, e.g.,\ temperature drifts. 
Analysis of the exact requirements for adjustability and precision is still in progress, but they are expected to be far less stringent than for the main telescope (see below). 


\subsection{Telescope}
\label{sec:telescope}
The main telescope (right panel of Fig.~\ref{fig:BCP}) is a Keplerian telescope consisting of two plano-convex lenses with $f_1 = 55\,$mm and $f_2 = 1947\,$mm that expands the interferometer laser beam to a waist of 1cm throughout the interferometer beam pipe. The first lens could feature an aspheric first surface to minimize optical path differences across the final beam. The telescope is placed at the top of the pipe, with two lenses positioned approximately $2\,\textrm{m}$ apart within the assembly. The module will have bellows on either end to allow adjustment of the lens holders, and to aid assembly of the tower. The lower lens holder may also act as part of a differential pumping aperture that separates the XHV area below and the HV area above~\cite{WARBURTON1990350}. As the vacuum pumping strategy develops, a decision will be taken on whether to seal the lens holder or keep the differential pumping aperture.


Limiting the pointing fluctuations on the main interferometer beam to $<35\,$nrad, such that the resulting phase gradients remain much smaller than the effect of the phase-shear readout, translates into stringent \textit{stability} requirements for the telescope lenses, namely maximal transversal shifts of $<40\,$nm  ($<70\,$nm) for the first (second) lens. Furthermore, the maximum tilts of the lenses are $<1.8\, \mu$rad ($<70\,$nrad).
Consequently, extensive simulation work is crucial to predict the vibrational response of the lenses to external inputs from the building, ensuring that the design adheres to the intended specifications. This simulation work is documented in Section \ref{sec:stability}.



\subsection{Interferometer Beam Pipes}
\label{sec:ibp}
The pair of $5\,\textrm{m}$ interferometry beam pipes (shown in the centre of Fig.~\ref{fig:schematic1}) are the workhorses of the instrument. They define the length of the baseline of the experiment and must provide as `clean' an environment as possible for the atoms during interferometry. As such they must facilitate the XHV levels, and also provide a stable and controlled magnetic field inside the pipe. In order to meet these targets the beam pipes consist of several layers that are, from innermost to outermost:
\begin{enumerate}
    \item The vacuum pipe assembly;
    \item Heater tape and insulation;
    \item The magnetic coil assembly;
    \item Magnetic shielding.
\end{enumerate}

The vacuum pipe assembly is a standard vacuum system with CF flanges for connections, designed to maintain the vacuum levels required for the experiment. Directly attached to these are heater tapes and insulation. Reaching a pressure of $10^{-11}$~mbar will require an extensive bakeout of the vacuum system, and the beam pipe is the most significant portion of its volume. The heater tape will provide the heat for bake-out, and the insulation will reduce the heat load transferred to the rest of the structure. Outside this there will be magnetic field coils, designed to provide a known and consistent magnetic field across the beam pipes, see Section \ref{sec:magguide}. Finally, the magnetic shielding will shield the interior magnetic field from external fields. As with the telescope, there will be bellows placed at either end of the vacuum pipes, to aid assembly. The design and assembly of this structure is explained in more detail in Section \ref{sec:magshielding}.

\subsection{
Phase-Shear Detection Platform}
\label{sec:phaseshear}

\begin{figure}
    \centering
    \includegraphics[width=0.75\textwidth]{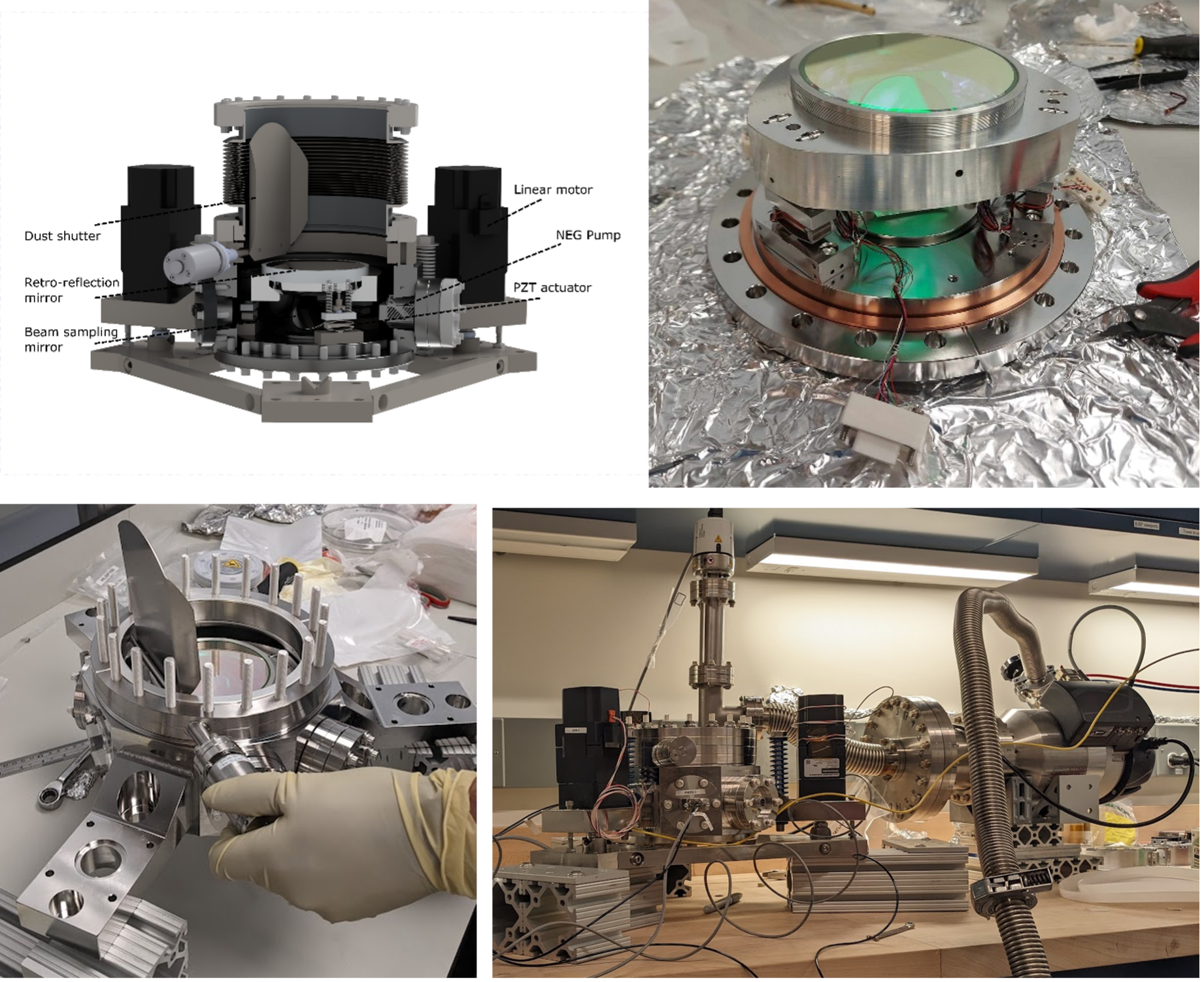}
    \caption{CAD model ({\it upper left panel}) and assembly phases ({\it upper right, lower left, lower right panels}) of the bottom retro-reflecting mirror system for the 10m MAGIS prototype at Stanford, which serves as a basis for the design of the corresponding system for AION-10. {\it Upper right panel}: piezo-actuated assembly with mirror mounted, {\it lower left panel}: completed assembly with mirror cover, {\it lower right panel} vacuum testing of complete assembly.}
    \label{fig:BRM1}
\end{figure}

This Section details the work involved in assembly and pre-commissioning of the phase-shear detection platform for the $10\,\textrm{m}$ MAGIS prototype at Stanford, see Fig.~\ref{fig:BRM1}. This platform houses and controls the BRM. The system is ready to be assembled for a $100\,\textrm{m}$ MAGIS experiment at Fermilab. The design, development, and assembly were accomplished in collaboration with Stanford and Northwestern universities and Fermilab.
The AION experiment has an opportunity to leverage the knowledge, experience and design of the MAGIS phase-shear detection platform and adapt it for planned interferometers in the UK.

The phase-shear detection platform is an XHV chamber located at the bottom of the interferometers at MAGIS-10, MAGIS-100 and AION, see Fig.~\ref{fig:fulltower}, blue section. The platform’s purpose is to house and control the BRM. This mirror’s precise and fast angular control is achieved with three in-vacuum piezoelectric transducer (PZT) actuators. Precision and speed are necessary for Coriolis force compensation~\cite{glick2024coriolis}, which is a leading systematic error, and for implementing the phase-shear detection method~\cite{Sugarbaker:2013}. The mirror angle can be directly measured in two ways: strain gauge sensors attached to PZTs that measure their extension,  and by using light reflected off the back surface of the mirror into a position-sensitive detector. Both these methods can be employed in a closed-loop control of PZTs. A cutaway view of the system’s CAD model is shown in Fig.~\ref{fig:BRM1}, top left. For coarse alignment of the interferometry beam, the inclination of the vacuum chamber can be adjusted using three linear actuators mounted to the outside of the chamber~\cite{MAGIS-100:2021etm}. 

The assembled platform at the MAGIS prototype has a measured pressure of 5x10\textsuperscript{-8}mbar at the pre-commissioning phase, with the measurement limited by the leak detector’s resolution. The goal of 10\textsuperscript{-11}mbar is expected to be achieved after baking for 2 weeks or more at a relatively low temperature of $<$85°C (limited by PZT wires) and following activation of the installed NEG pump. 

Preliminary performance data were taken and analysed using an electronic feedback loop that employs strain gauge sensors attached to the PZT actuators to monitor their extension. The measured settling time for the PZT actuators was $65\,\textrm{ms}$ for the $100\,\textrm{nm}$ position error threshold, below the specified settling time of $< 100\,\textrm{ms}$. The established closed-loop control demonstrated stable performance and eliminated the PZT hysteresis. The goal of angular stability for the PZT actuation was $< 80\,\textrm{nrad}$ (rms) for monitoring based on the strain gauges and, ultimately, $< 50\,\textrm{nrad}$ when using an optical lever to monitor the mirror angle. The data taken with the electronic feedback loop enabled surpassed the stability specification of $< 50\,\textrm{nrad}$, achieving an rms stability of $33\,\textrm{nrad}$ ($\sim 10$s). A prototype optical lever was also built, and its optimisation is ongoing. Upon request, the data and analysis can be provided. 

In conclusion, the constructed phase-shear detection platform has achieved its noise and precision specifications, and work is ongoing to improve these parameters further by employing an optical feedback loop. 

The phase-shear detection system is a vital interferometer component. The described platform can be adapted for the $10\,\textrm{m}$ AION interferometer with minor modifications. The successful assembly and testing of the platform means one of the main deliverables of the MAGIS-UK work package is ready for installation. It also demonstrates effective collaboration between the two sister projects, leveraging the gained knowledge, experience and know-how to mitigate risks and prevent duplicating efforts.

\subsection{Interconnect Chambers (ICC)}
\label{sec:icc}
\begin{figure}
    \centering
    \includegraphics[width=0.42\textwidth]{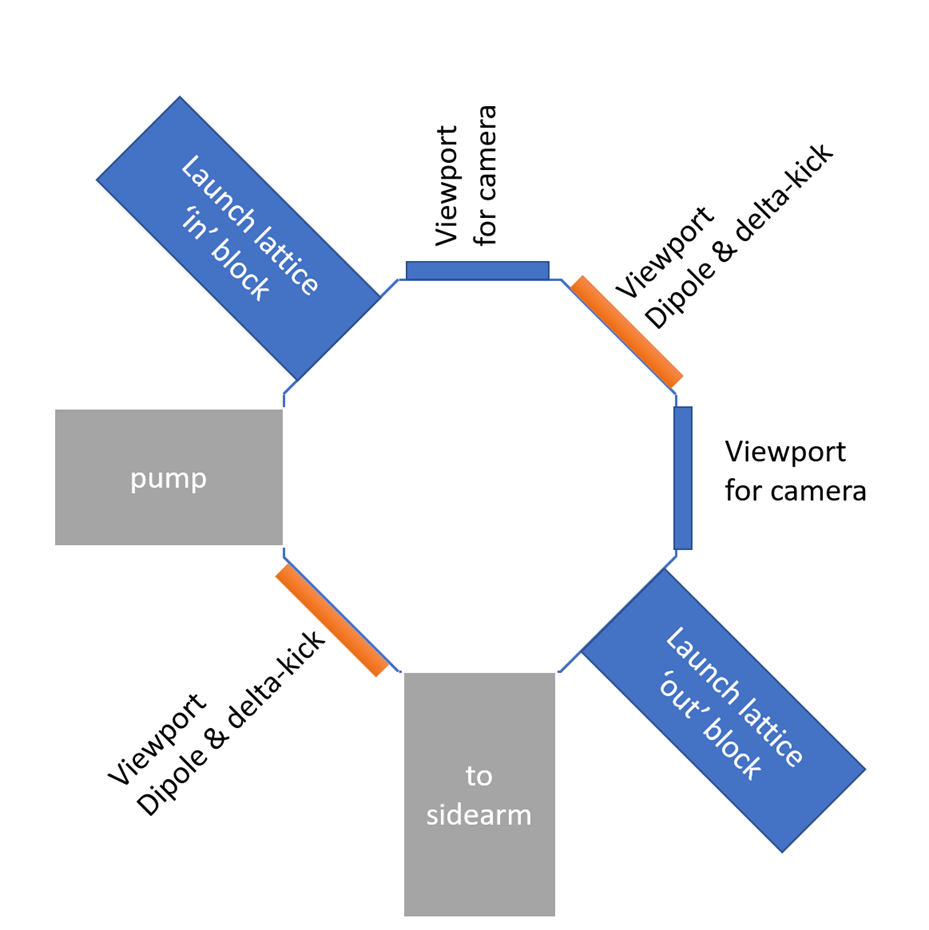}   \includegraphics[width=0.53\textwidth]{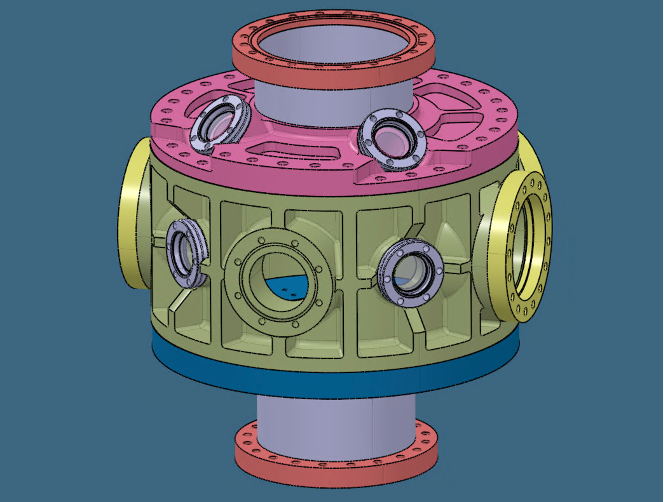}
\caption{{\it Left panel}: Schematic diagram of an interconnect chamber, showing the various connections in the horizontal plane. {\it Right panel}: CAD design of an interconnect chamber. The smallest grey viewports correspond to the camera viewports in the left panel.}
    \label{fig:ICCDia}
\end{figure}

At the bottom of each interferometer beam pipe, there will be connection nodes called interconnect chambers (ICC). As illustrated in the left panel of Fig.~\ref{fig:ICCDia}, these are multipurpose vacuum chambers that provide interfaces to various components. As well as the connection to the sidearms for transporting atom clouds to the beam pipe (detailed below), they also provide viewports for the external laser optics for the launch lattice and dipole trap and delta-kick cooling systems, cameras, and a port for vacuum pumps. In addition to the camera viewports shown schematically in the left panel of Fig.~\ref{fig:ICCDia}, two more viewports are positioned at 45° to the vertical above the first two, as seen in the CAD design in the right panel. The interior of the chamber minimises surface area for vacuum performance, and the exterior is designed to minimise total material used to reduce the magnetic effect of the chamber as much as possible.

\subsection{Launch Lattice}
\label{sec:launchlattice}
\begin{figure}
    \centering
    \includegraphics[width=0.5\linewidth]{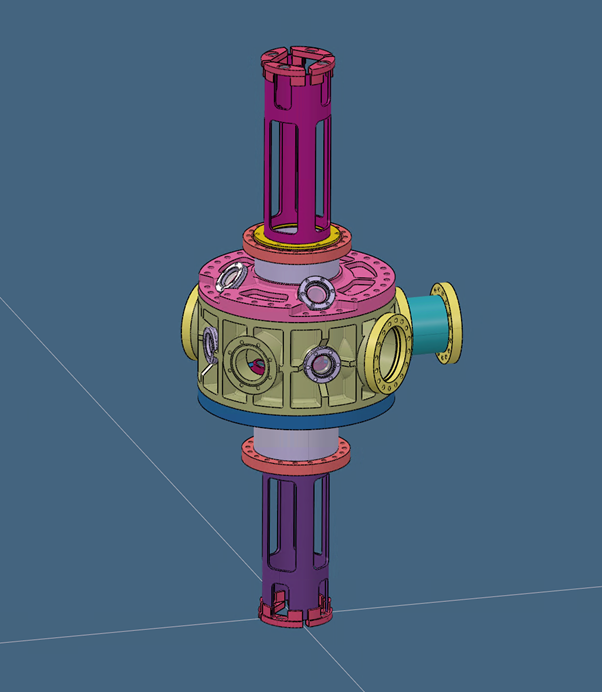}
    \caption{Launch lattice ``scaffolding" protruding from an interconnect chamber.}
    \label{fig:LLFull}
\end{figure}

Key optical elements in the instrument are the two launch lattices, where two interfering laser beams form a moving optical lattice that launches the atoms upwards into the interferometer beam pipes. The laser beams are prepared using a combination of adjustable external and fixed in-vacuum optics. The in-vacuum part consists of a series of mirrors inside each interconnect chamber and on a `scaffold' that extends approximately $0.5\,\textrm{m}$ above and below each interconnect chamber. The conceptual design of the scaffold is shown in Fig.~\ref{fig:LLFull}. The design is not final, and will be informed by the requirements of the optics as they become available.

\begin{figure}
    \centering
        \includegraphics[width=0.5\textwidth]{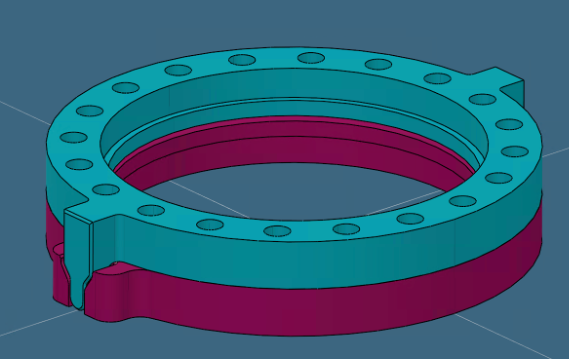}\\
    \includegraphics[width=0.25\textwidth]{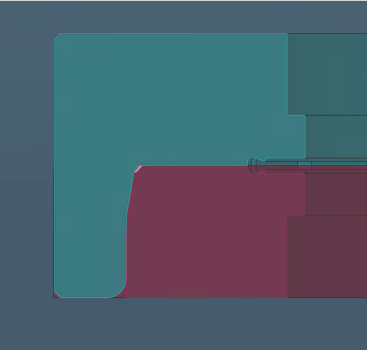}
    \includegraphics[width=0.26\textwidth]{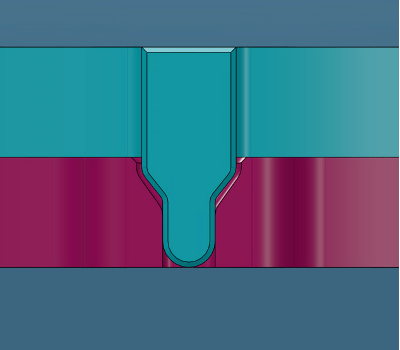}
    \caption{{\it Upper panel}: Conceptual design of the custom vacuum flanges for the launch lattice. {\it Lower panels}: Details of the flanges.}
    \label{fig:LLDet}
\end{figure}

We want to increase the effective aperture of the beam pipes as much as possible, which requires leaving a small gap between the scaffold and the vacuum pipe. Our chosen vacuum system design consists of several discrete modules to make the craning operations easier for assembly, see Section \ref{sec:stairwell}. The launch lattice needs to have a connection to each interconnect chamber, while the scaffold extends into each beam pipe. To avoid a tricky lifting operation to slot the scaffold into the beam pipes, we have chosen to split the design and attach the scaffolding to the beam pipe modules. This adds a layer of complexity to the design, and includes a new requirement, namely that it be repeatable when being put back together after being taken apart. It will be necessary to set up the launch lattice in laboratory conditions to verify that all the optical elements are in the correct position and, in the case of a split lattice, it will need to be disassembled, added to different sub-assemblies and reassembled when the vacuum connections are made, once the tower has been built. As seen in Fig.~\ref{fig:LLDet}, we plan to use custom vacuum flanges with features to position and constrain the two parts of the launch lattice relative to each other. The design includes slots and tabs on opposing CF flanges with tight machining tolerances to provide repeatable positional accuracy during reassembly. It will also include knock-out holes so the flanges can be disassembled without damaging the seal faces.

\subsection{Sidearm Assemblies}
\label{sec:sidearm}

\begin{figure}
    \centering
    \includegraphics[width=0.5\textwidth]{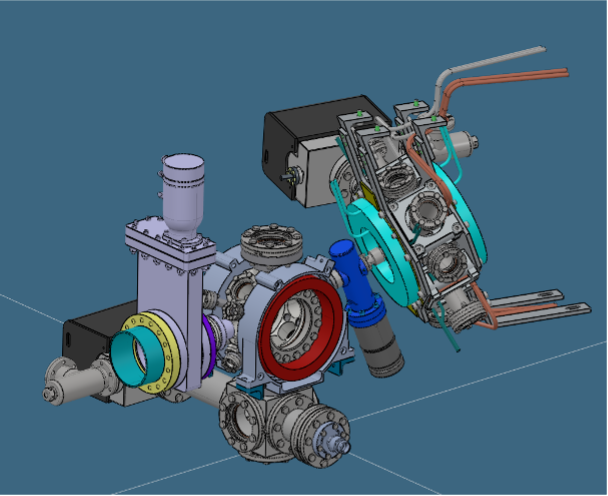}
    \caption{CAD diagram of a sidearm assembly: see~\cite{AION:2023fpx} for details.}
    \label{fig:Sidearm}
\end{figure}

Fig.~\ref{fig:Sidearm} shows a CAD diagram of one of the first-generation sidearms, the assemblies responsible for cooling strontium and delivering it to the interferometer beam pipe. Five of these sidearm assemblies have been built and are being used for R\&D in institutions contributing to AION~\cite{AION:2023fpx}. It is expected that the AION-10 tower would use an upgraded version of the existing sidearms, but the design of the tower is such that an existing unit could be installed without modification.

\subsection{Camera Assemblies}
\label{sec:cameras}

\begin{figure}
    \centering
    \includegraphics[width=0.5\textwidth]{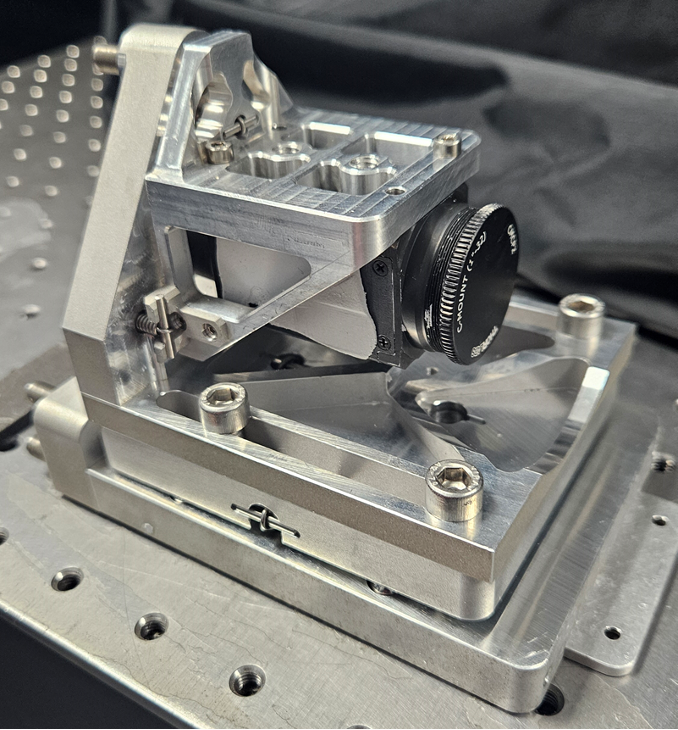}
    \caption{Completed camera adjuster assembly.}
    \label{fig:Cam3}
\end{figure}

\begin{figure}
    \centering
    \includegraphics[width=0.5\textwidth]{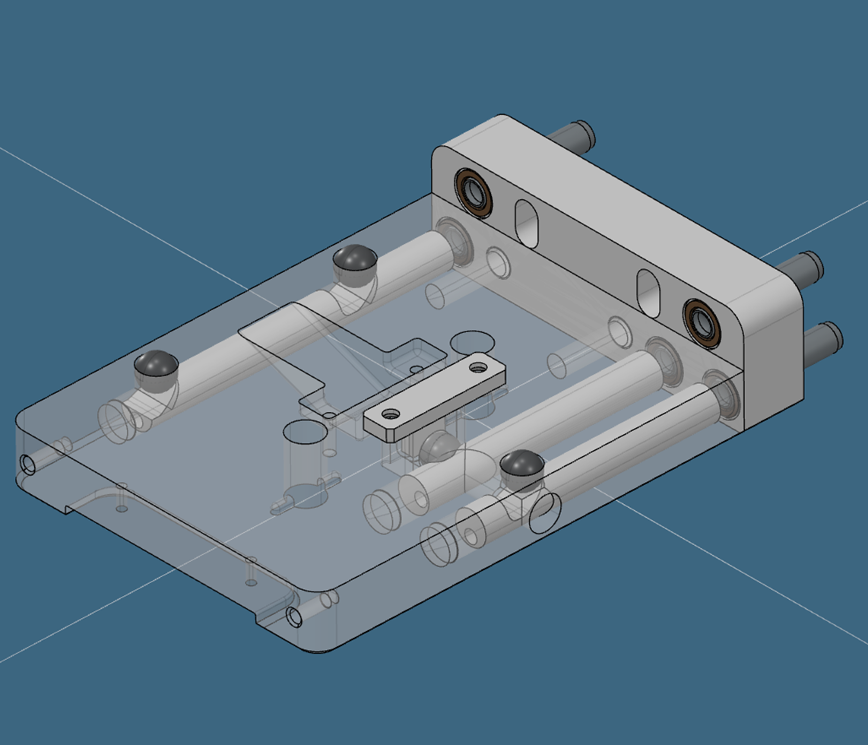}
    \includegraphics[width=0.5\textwidth]{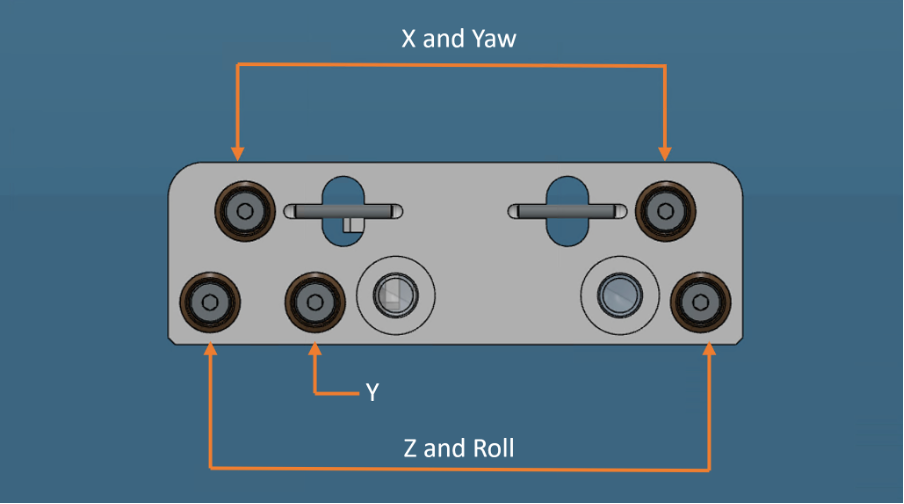}
    \caption{Upper panel: CAD drawing of the camera adjuster mechanism. Lower panel: Schematic diagram of the camera adjuster mechanism.}
    \label{fig:Cam1}
\end{figure}

The camera systems are used to image the final atom clouds at the end of the interferometer sequence and are therefore among the more critical parts of the design. The requirements of the system are that the camera units be extremely stable relative to one another ($100\,\textrm{nm}$), see Section \ref{sec:specs}, and that they also be  finely adjustable to ensure they can be precisely focussed at the atom cloud. They are close to a final level of design maturity, with full 6 degrees of freedom fine adjustment capability. A completed camera adjuster assembly is shown in Fig.~\ref{fig:Cam3}. The adjustment mechanism uses a series of fine-threaded ball-ended grub screws to push on pins inside the main assembly block. This then pushes ball bearings in a perpendicular direction on ramps. (The grub screws and matching bushes are provided by Thorlabs~\cite{THOR}.) A CAD drawing of the camera adjustment mechanism is shown in the upper panel of Fig.~\ref{fig:Cam1} and a schematic diagram in the lower panel.

The whole assembly is pre-loaded with springs to maintain tension against the screws. The precision of the system has been confirmed to be sufficient for use on AION-10, this has been verified by adjusting the camera mount platform to be parallel to the base at all points on the surface to a tolerance of $1\,\textrm{$\mu$m}$, from a randomly adjusted starting point. It was possible to make this adjustment within 2 minutes. Under thermal testing it was found that heat was not transferred effectively from the camera out to the bulk of the assembly. This was not a huge surprise as the 3 ball bearings mounting the upper assembly effectively provide point contacts. Nevertheless, this needs attention, and the plan is to add thermal straps to connect the upper and lower parts of the assembly to provide a path for heat transfer. The final improvement to make is for access to the camera’s cables at the back. A solution has already been drawn for this, to expand the cut-out and include a large chamfer. The stability requirement has been addressed using the  isolation strategy  detailed in the next section.

This work was carried out in conjunction with the MAGIS project, as part of the AION deliverables to that project. 


\subsection{Isolation Strategy}
\label{sec:isolation}
Following the simulations detailed below it became clear that to keep the camera assemblies for each interconnect chamber within tolerance with respect to each other, some active vibration isolation of the camera platform is required. However, due to limitations on total force tolerated by active vibration isolation  units, we have devised a setup  enabling us to build the tower without the active isolation units and to add them at a later stage of assembly. This will avoid the risk of overloading and damaging the units during instrument assembly. We have specified Herzan AVI 200M/400M isolators to be sufficient~\cite{HERZAN}. These work as a pair to support a mounting plate that holds the camera assemblies for each interconnect node. The following subsection details the installation procedure and design for mounting the isolators.

\subsubsection*{Isolator Mount Design and Integration}

\begin{figure}
    \centering
    \includegraphics[width=0.45\textwidth]{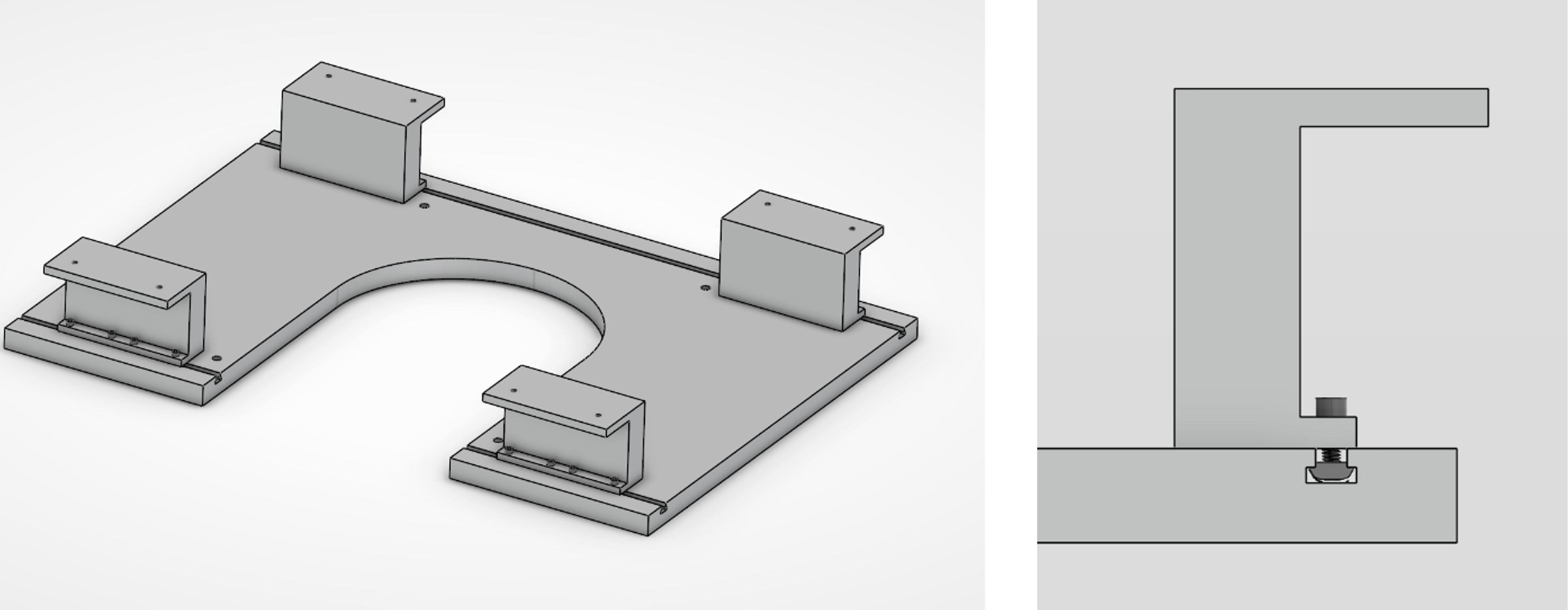}
    \includegraphics[width=0.45\textwidth]{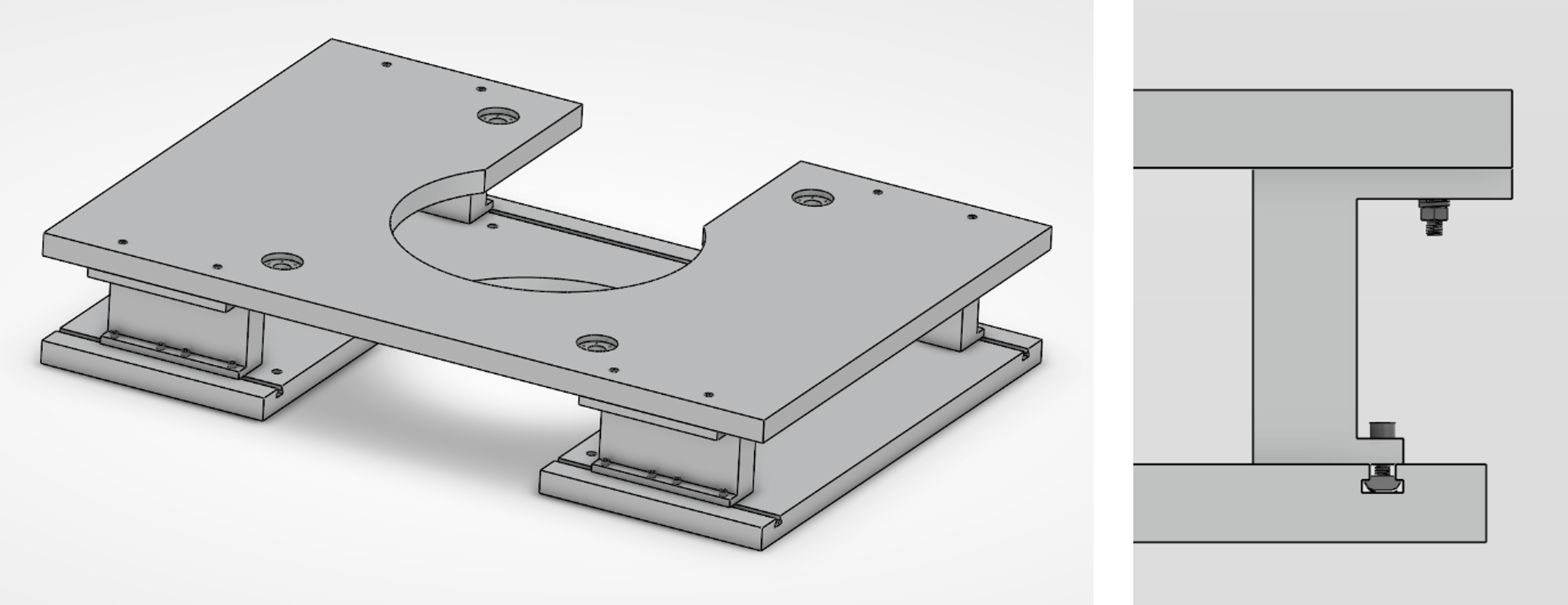} \\
    \includegraphics[width=0.45\textwidth]{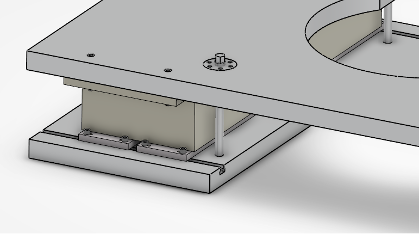}
    \includegraphics[width=0.45\textwidth]{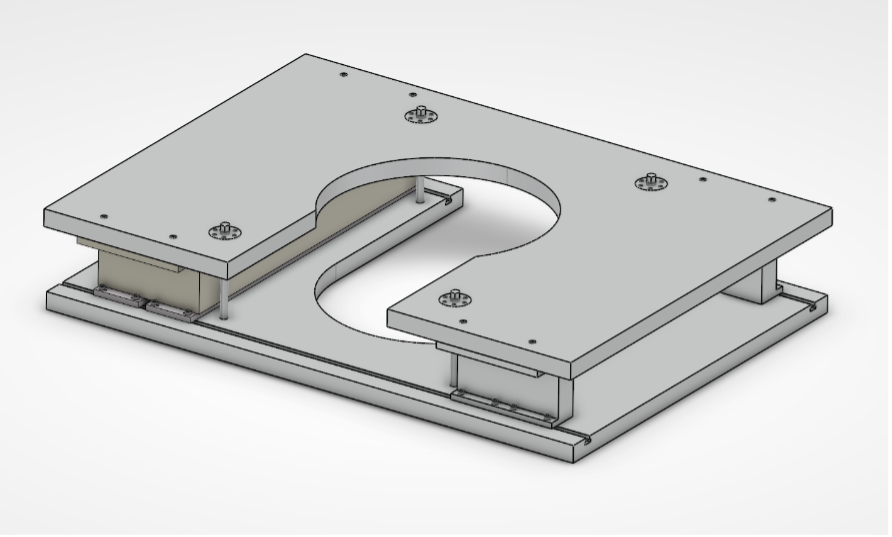} \\
    \includegraphics[width=0.65\textwidth]{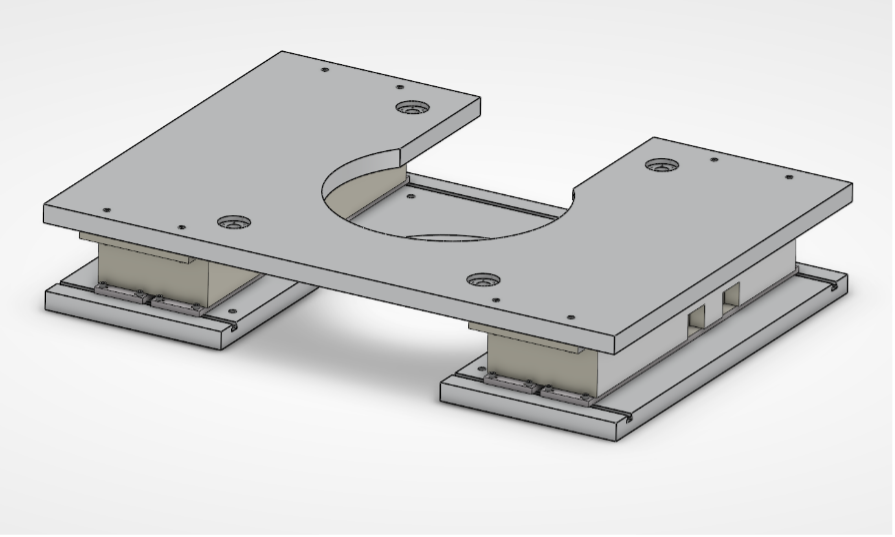}
    \caption{Stages in the assembly of the isolator mount. {\it Top left panels}: Isolator lower assembly with spacer blocks and tee-slot. {\it Top right panels}: Full assembly and upper mounting. {\it Middle left panel}: Rod detail. {\it Middle right panel}: Partial assembly with one isolator unit installed. {\it Bottom panel}: Finalised assembly.}
    \label{fig:isomount1}
\end{figure}

In the absence of isolation units for the tower build, spacer blocks will be installed between the upper and lower plates, as shown in the top left panels of Fig.~\ref{fig:isomount1}. They will be attached to the lower plate via a t-slot feature that allows the blocks to be translated horizontally into position with pre-installed Bosch Rexroth T-Slot stones. 
The upper plate can then be mounted directly on top, as shown in the top right panels of Fig.~\ref{fig:isomount1}. 

The spacer blocks can be exchanged for two isolation units with the following actions. Four bushes will be mounted to the upper plate, a threaded rod will be installed through each bush and locate into the lower plate (middle left panel of Fig.~\ref{fig:isomount1}). Once located, the bolts attaching the upper plate to the spacer blocks can be removed. This upper bolted connection can be seen in the middle right panel of Fig.~\ref{fig:isomount1}. 


Rotating the threaded rod will raise the upper plate. The bellows on the interferometry beam pipes can take up the slack as the platforms are raised. The bolted connection of the spacer blocks to the lower plate can be loosened, and the spacer blocks can be removed from each side. In this raised position, the isolation units can be installed. They can also have pre-installed T-slot stones and can be mounted in the same way as the blocks.  The finalised assembly is shown in the bottom panel of Fig.~\ref{fig:isomount1}.


All of the models in the Herzan AVI series measure the same height. Where they differ is in their length and width. The CAD images show the AVI 400M version which is the mid-sized series. If a change in series occurs, the position of the t-slot and mounting holes will be updated. This version weighs approximately $45\,\textrm{kg}$, so manual handling of the units will require two people.   
When positioned the lower bolted connections can be secured. The upper plate can then be lowered by backing off the threaded rods. This will allow the upper plate to return to position but now on top of the isolation blocks. The upper bolted connection between the upper plate and isolation block can be secured. The threaded rods can be removed from the upper plate and stored for the next occurrence.  



\subsubsection*{Isolator Platform Design}

\begin{figure}
    \centering
    \includegraphics[width=0.45\textwidth]{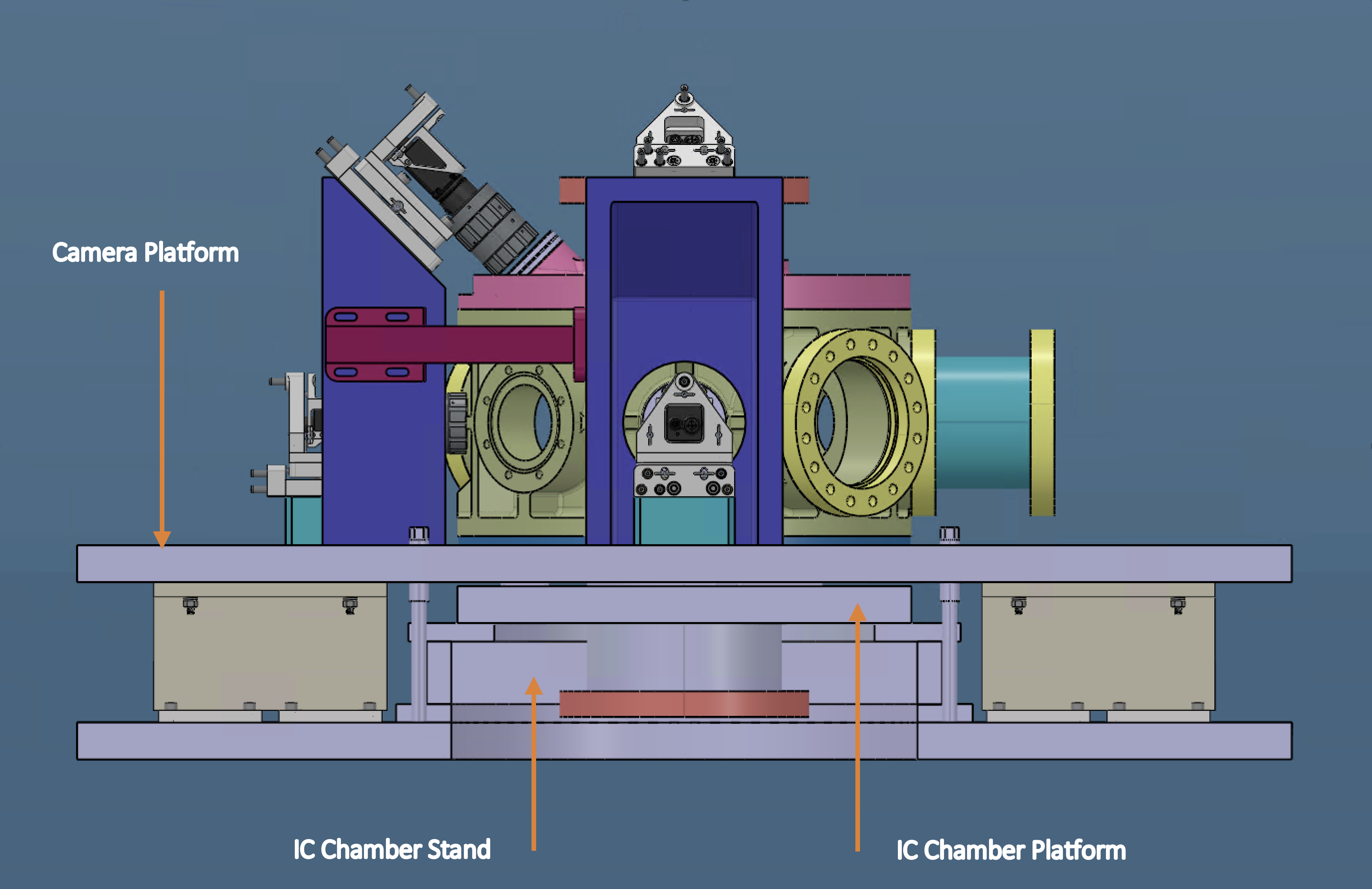}
    \includegraphics[width=0.45\textwidth]{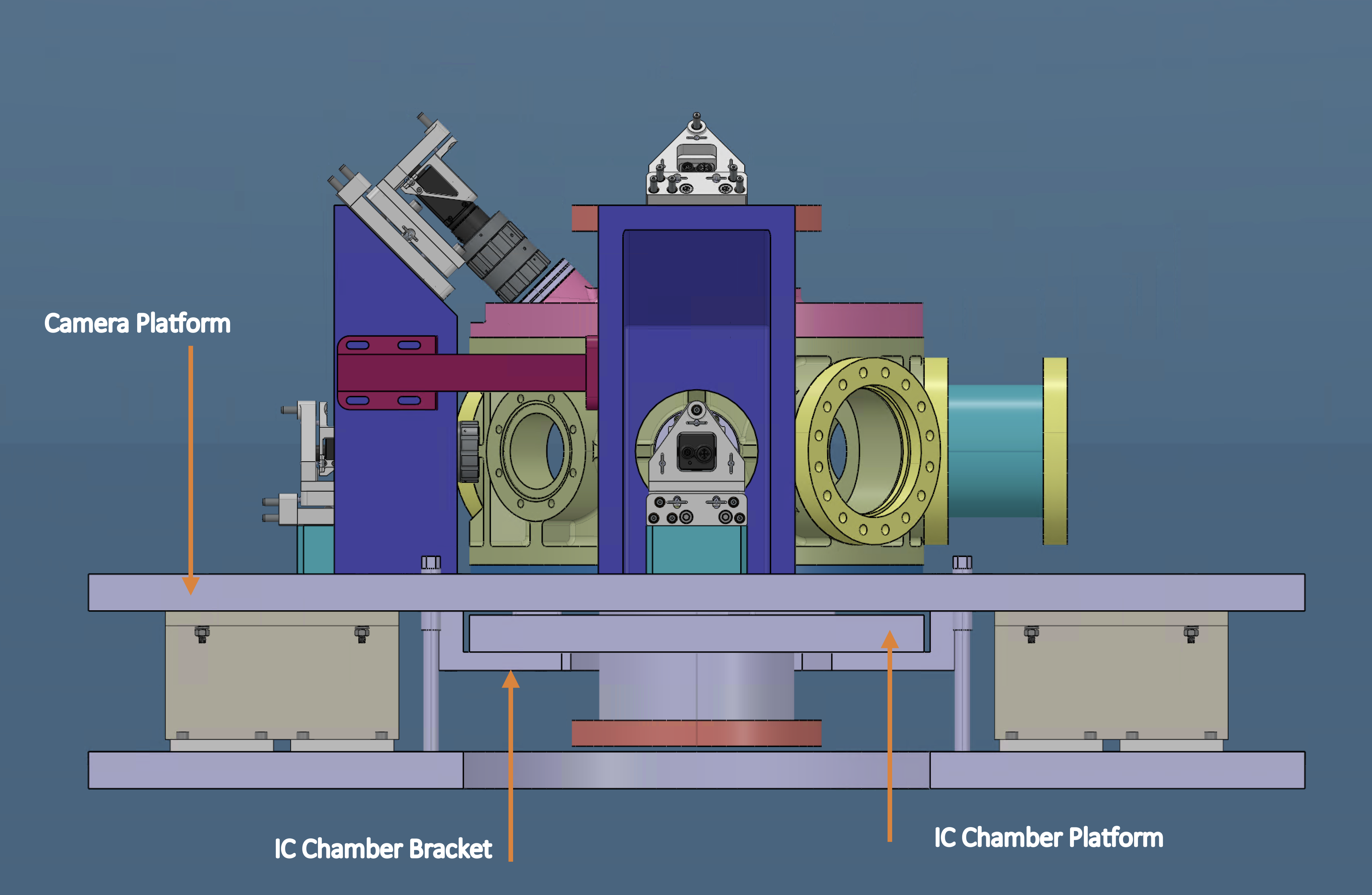}
    \caption{Possible configurations for the isolator platform. {\it Left panel}: The camera assemblies are mounted on the isolation plate only and the interconnect chamber is mounted on the lower un-isolated plate with a stand. {\it Right panel}: The upper isolated plate is unchanged, but the interconnect chamber is now mounted on this plate using bracketry.}
    \label{fig:isoplatform}
\end{figure}

At the current design stage it is not known whether we will be required to isolate vibrationally only the camera assemblies or also the in-vacuum launch lattice optics. The conclusion to this question will be informed by simulations and an analysis of the stability requirements of the launch lattice. Our designs for the platform can handle both, however, mitigating the risk that the specification might change late in the design process. This is achieved by having two inter-compatible platform designs available; it will also be possible to switch between them after installation.
In the first configuration, shown in the left panel of Fig.~\ref{fig:isoplatform}, only the camera assemblies are mounted to the isolation plate and the interconnect chamber is mounted to the lower un-isolated plate with a stand. In the second configuration, shown in the right panel of Fig.~\ref{fig:isoplatform}, the upper isolated plate is unchanged, but the interconnect chamber is now mounted on this plate using bracketry. 

\subsection{External Optical Interferometer Network}
\label{sec:externinet}

\begin{figure}
    \centering
    \includegraphics[width=0.6\textwidth]{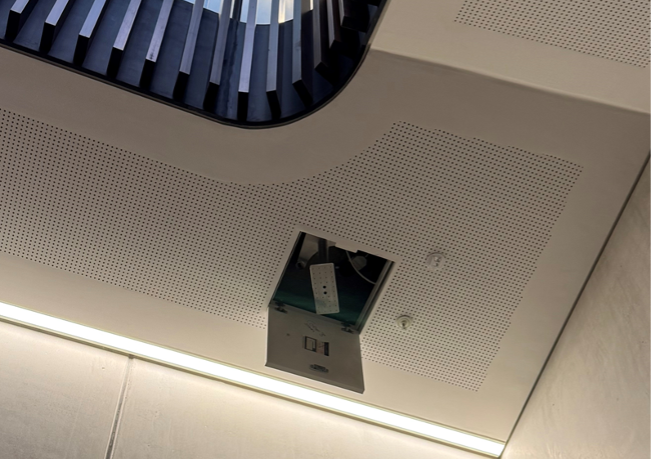}
    \hspace{5mm}
    \includegraphics[width=0.117\textwidth]{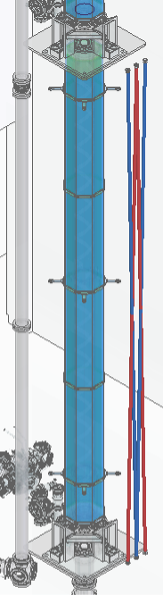}
    \caption{Left panel: Optical breadboard installed in the ceiling of the B2 level of the Beecroft stairwell. Right panel: Illustration of a potential external interferometry network between two camera mount assemblies in the Beecroft stairwell.}
    \label{fig:optint}
\end{figure}

The external interferometer network is planned to use commercial optical interferometer systems using phase differences in laser signals to measure accurately sub-micron displacements of parts of the instrument caused by, for instance, thermal drifts. We have started to develop the ability to test the system in the Beecroft stairwell (see below). We have installed concrete anchors in the ceiling of the B2 level, and have installed an assembly holding a breadboard to mount retroreflectors on, as seen in the left panel of Fig.~\ref{fig:optint}.  With this installed it will be possible to start to take real world measurements around the stairwell and understand the system's accuracy and precision.
The right panel of Fig.~\ref{fig:optint} illustrates a potential external interferometry network between two camera mount assemblies in the Beecroft stairwell.

\pagebreak
\section{Build and Assembly Procedure}
\label{sec:assembly}
\subsection{Beecroft Stairwell}
\label{sec:stairwell}

\begin{figure}[h]
    \centering
    \includegraphics[width=0.5\textwidth]{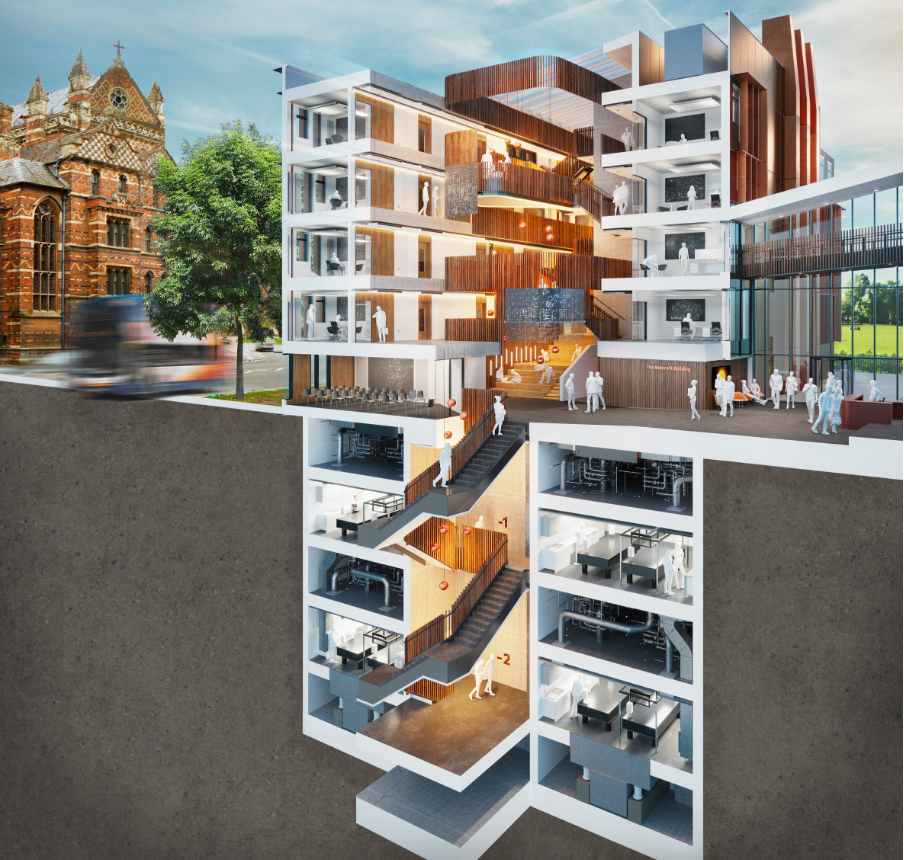}
    \caption{Illustration of the Oxford Beecroft building and its basement floors. (Image credit: Hawkins/Brown Architects)}
    \label{fig:beecroft}
\end{figure}

The AION-10 instrument is to be installed in the Beecroft building in Oxford, which is illustrated in Fig.~\ref{fig:beecroft}. This is a relatively new part of the Oxford Physics Department that is designed to be vibrationally isolated from the outside environment. The location within the Beecroft building is the stairwell descending from the ground floor atrium down to the underground laboratories. For the rest of this report, we refer to the floors as follows (in descending order): G for ground level, B1 for the first basement level, and B2 for the second and lowest basement level. The base of AION-10 will sit on the floor of the B2 level.

\begin{figure}[h!]
    \centering
    \includegraphics[width=0.3\textwidth]{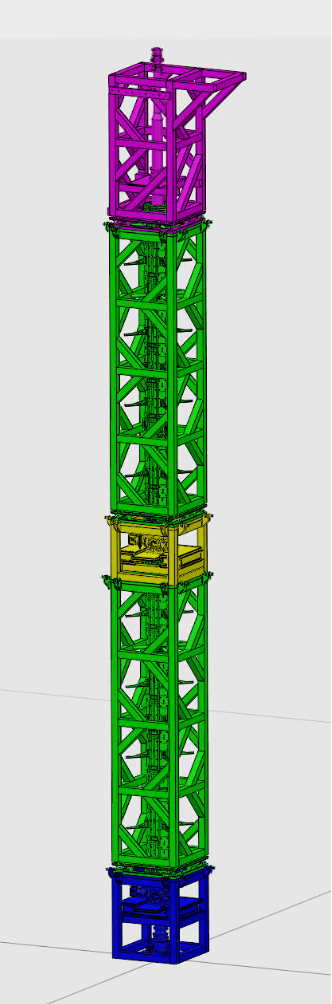}        \includegraphics[width=0.217\textwidth]{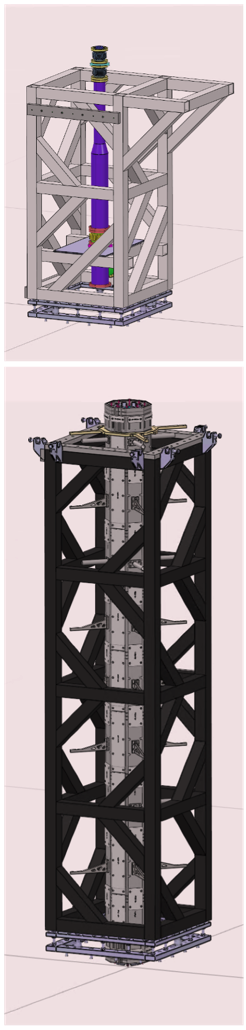} \\
        \includegraphics[width=0.27\textwidth]{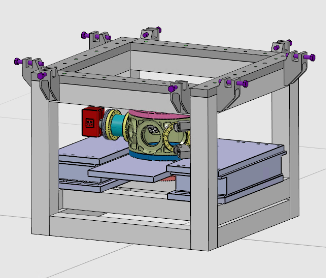}
    \includegraphics[width=0.25\textwidth]{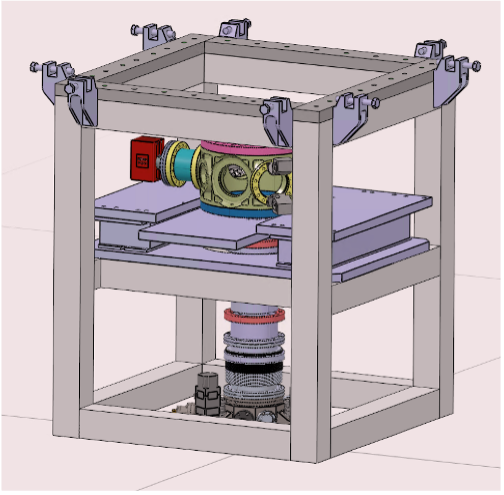}
    \caption{{\it Upper left panel}: The full tower assembly, with the modules highlighted. {\it Top right panel}: The telescope module. {\it Middle right panel}: A beam pipe module. {\it Lower left panel}: The upper interconnect module. {\it Bottom right panel}: The lower interconnect module.}
    \label{fig:fulltower}
\end{figure}

The plan is to build the interferometer beam pipes and their supporting tower structures in five “modules” that can be transported and craned into the Beecroft stairwell and assembled by a lifting company~\cite{CYCLONE}. As seen in Fig.~\ref{fig:fulltower}, they will include two identical beam pipe modules, a lower ICC/phase shear detection platform module (that will not need to be craned), an upper ICC module and a telescope module. Together they form the structural tower of the instrument and a large part of the instrument itself. The design will be such that the structure can be bolted together with the vacuum system components separate and all vacuum connections capped to avoid contamination of the cleaned vacuum surfaces. It is not planned that they will arrive under vacuum with gate valves, due to a lack of space underneath the magnetic shielding. This will enable the lifting company to be on site for the minimum possible time and reduce the associated costs. The vacuum system can then be connected and tested, and the rest of the instrument can be assembled with the structure.

\subsection{Module Assembly}
We now describe the conceptual assembly procedures of the five tower modules that make up the instrument. Details of the supports and connections between the tower frame and the instrument are not yet complete. The procedure assumes the instrument layer structure shown in Fig.~\ref{fig:bplayer}.

\subsubsection*{Beam Pipe Modules}
\begin{figure}[h!]
      \centering
      \includegraphics[width=\linewidth]{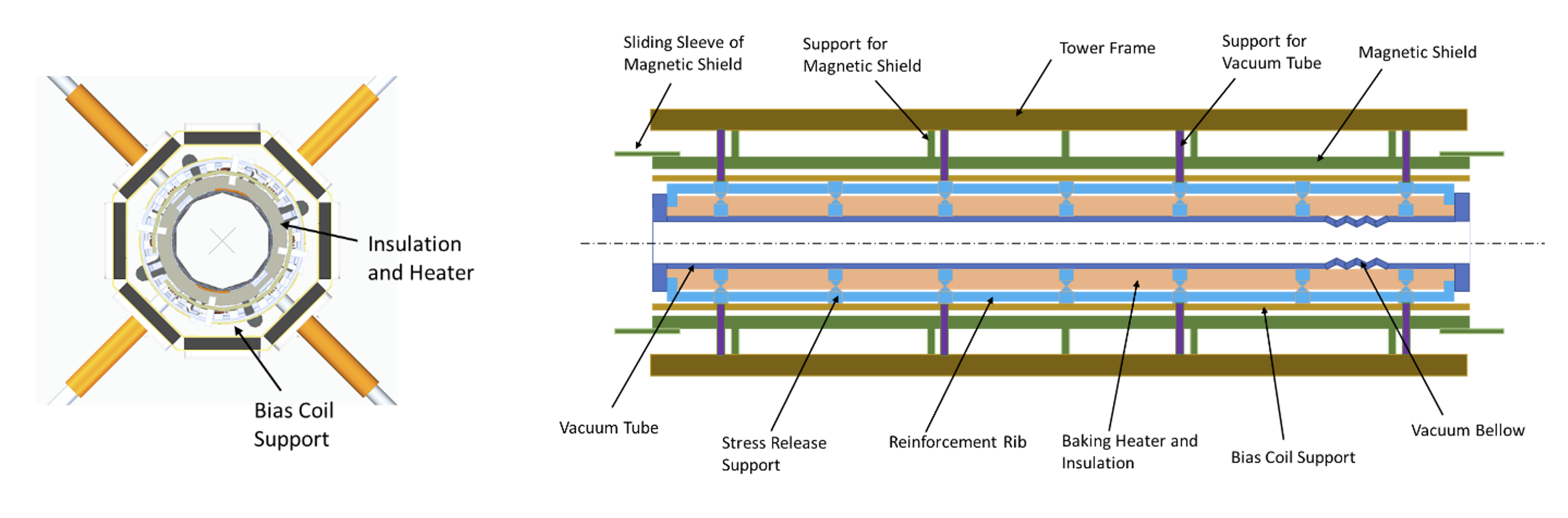}
      \caption{{\it Left panel}: The cross-section of the beam pipe. {\it Right panel}: The layer structure of the instrument.}
      \label{fig:bplayer}
  \end{figure}

\begin{figure}[h!]
    \centering
    \includegraphics[width=0.75\linewidth]{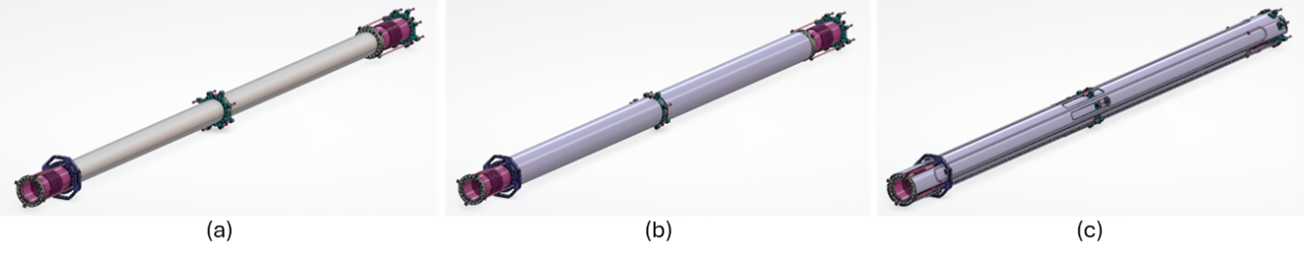}
    \caption{(a): A vacuum pipe sub-assembly. (b): Pipe wrapped with heating elements and insulation. (c): Pipe with coil formers and coils installed.}
    \label{fig:vacsubassy}
\end{figure}

The beam pipe modules will be the largest of the tower elements.  They will arrive at the Beecroft pre-assembled, and include the vacuum tube, bakeout tape and thermal insulation, field coils and magnetic shielding as shown in Fig.~\ref{fig:vacsubassy}, and the beam pipe will be surrounded by the tower structure, see Fig.~\ref{fig:bplayer}. The tower structure will be formed of aluminium box sections welded into a truss structure. This assembly will be built offsite (at a location to be specified) and transported to the Beecroft staging area (see below). 
The lower and upper main modules are to be assembled horizontally. The vacuum pipe sub-assembly should already be capped and sealed at both ends to maintain internal cleanliness before being transported to the assembly venue. The compressed bellows are held in place with threaded tie rods that also act to position the flanges relative to each other, see more details below. The vacuum pipe is first wrapped with heating elements and insulation, potentially with the help of a jig that rotates about the axis of the vacuum pipe, before the coil formers and field coils are installed. 

\begin{figure}
     \centering
     \includegraphics[width=0.9\linewidth]{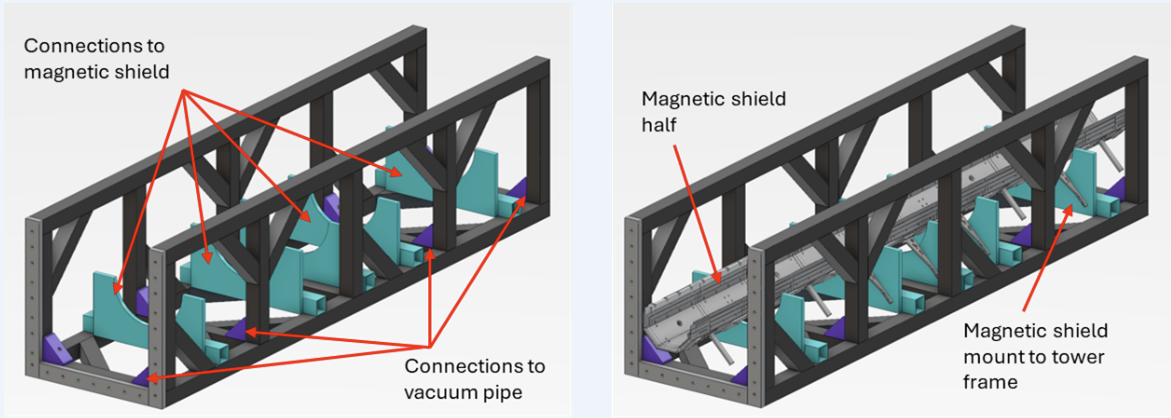}\\
         \includegraphics[width=0.44\linewidth]{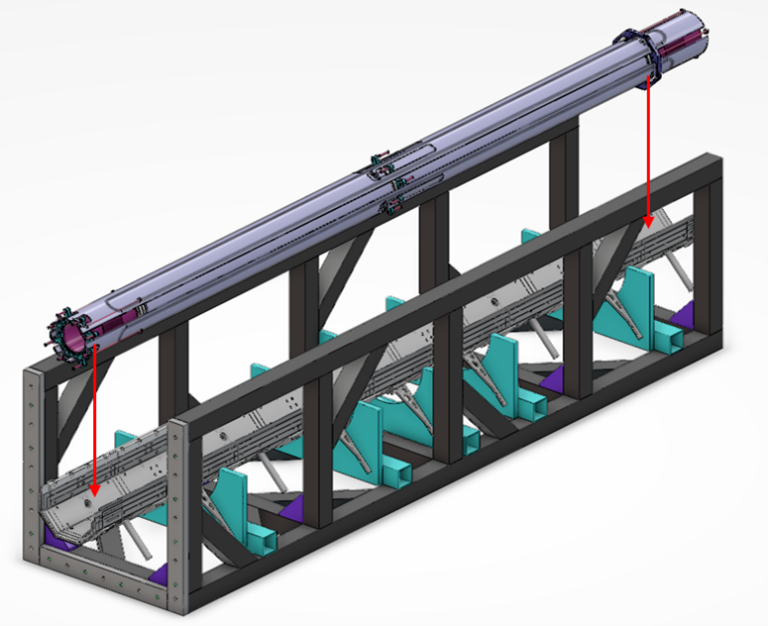}
     \includegraphics[width=0.46\linewidth]{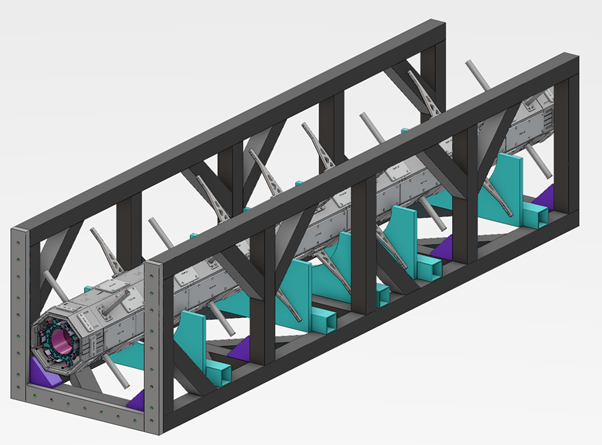}\\    \includegraphics[width=0.9\linewidth]{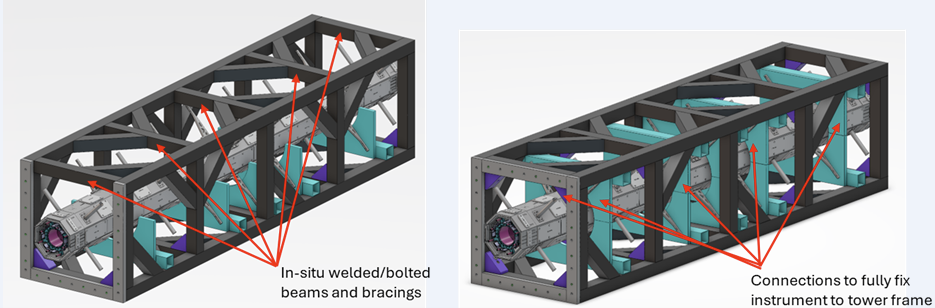}

     \caption{{\it Top left panel}: U-shaped welded support structure. {\it Top right panel}: U-shaped structure with half of the magnetic shield attached. {\it Middle left panel}: Vacuum pipe sub-assembly lowered into the magnetic shield and fixed to the frame. {\it Middle right panel}: The second half of the magnetic shield is installed. {\it Bottom left panel}: The square profile of the module is completed by welding the remaining beams. {\it Bottom right panel}: Final connections are made to fix the instrument to the tower frame.}
     \label{fig:BP1}
 \end{figure}

As the modules are around $5\,\textrm{m}$ long, a significant amount of space would be required to thread the instrument through a closed profile welded frame structure for assembly. Therefore, the following assembly sequence will be followed, as illustrated in Fig.~\ref{fig:BP1}. First the tower frame will be welded as an open U-shaped structure, so that half of the magnetic shield can be fixed in the frame, and the vacuum pipe can then be lowered into the magnetic shield and fixed to the frame. The second half of the magnetic shield is attached to complete the magnetic shield assembly with the vacuum pipe inside. Beams are then installed onto the frame either by welding in-situ  or bolted to complete the closed square profile of the tower module. The connections between the tower frame and the magnetic shield and vacuum pipes are then reinforced. The space required for assembling the module can be greatly reduced with this method. 
 
It is possible that the U-shaped open-profile welded structure may distort more than a closed square profile during welding. Details on welding fixtures and adjustment mechanisms for the connections between the frame and the instruments will continue to be developed. 

\subsubsection*{Interconnect Chamber Modules}
\begin{figure}
    \centering
    \includegraphics[width=0.75\linewidth]{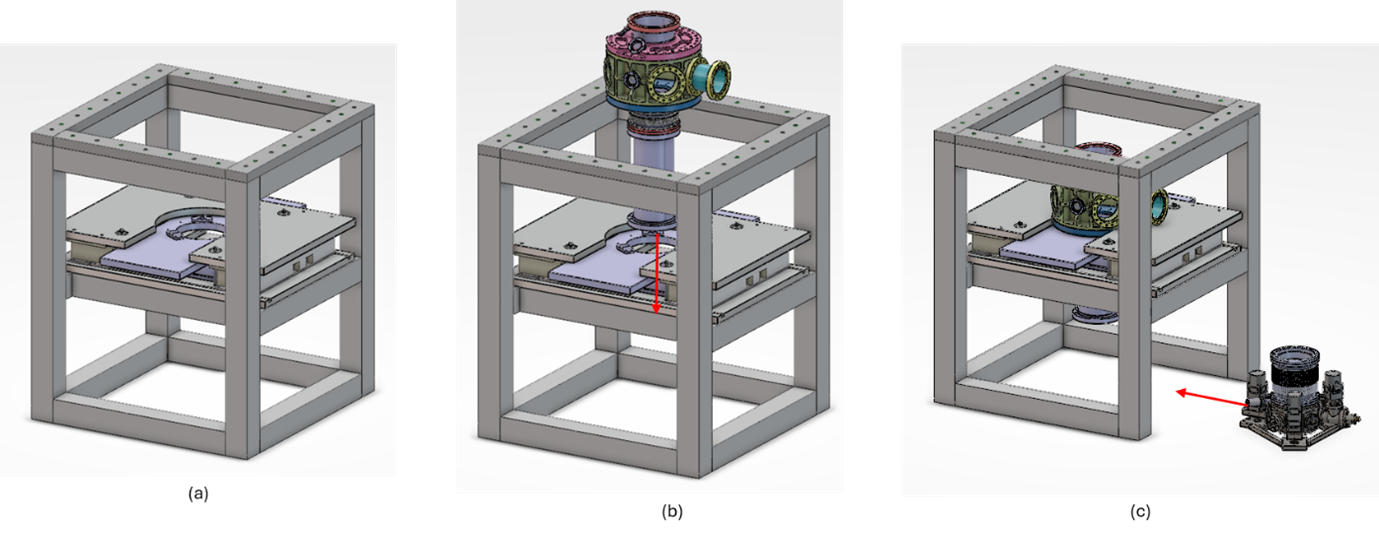}
    \caption{(a): The welded square profile frame of the lower interconnect module. (b): Lowering the lower interconnect chamber through the opening of the frame. (c): Sliding the BRM chamber from the side for initial installation and servicing.}
    \label{fig:ic1}
\end{figure}

The lower interconnect module will include the phase shear detection platform (housing the BRM) and the lower interconnect chamber. As this module will be at the base of the instrument, and the height is under $1.3\,\textrm{m}$, it will be possible to transport the module to the B2 level in the freight lift. It is therefore expected that the module will be in position when the lifting company arrives to set up their equipment, to reduce their time on site and the number of craning operations required. On the other hand, the upper interconnect chamber module will be craned in with the other modules, and will contain the upper interconnect chamber only. As with the lower module it will have platforms to hold active isolation units for the camera assemblies, which will be blocked off for assembly. It is anticipated that the interconnect chamber will be mounted in the tower structure for craning, but as the design and the scaffolding strategy evolves, it may be decided to install these later.
These modules can be assembled vertically as illustrated in Fig.~\ref{fig:ic1}. The frames are first welded as square profiles, and the interconnect chambers in both modules can be lifted into the frame through the opening at the end of the modules. We plan to leave one side of each module as a bolted connection, so as to have the ability to replace the interconnect chambers if an upgrade becomes necessary. The BRM chamber can also be slid into the base of the tower separately after the lower module is in place. This can be done by having a removable beam at the foot of the module, as with the ICCs. The sliding mechanism can also be used during servicing of the retroreflecting mirror chamber. 

\subsubsection*{Telescope Module}
\begin{figure}
    \centering
    \includegraphics[width=0.45\linewidth]{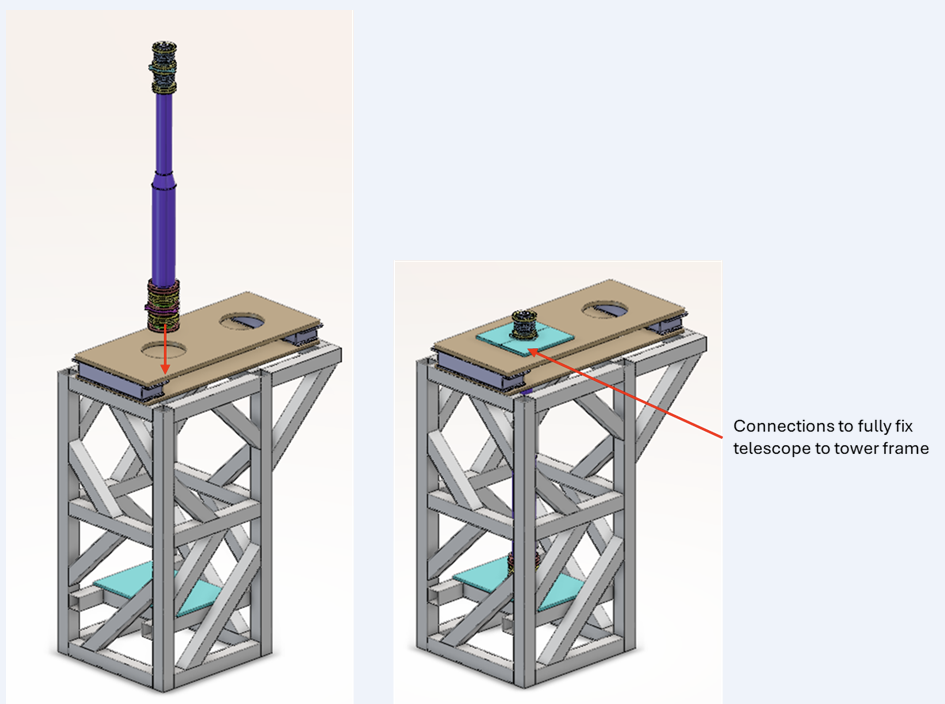}
         \includegraphics[width=0.45\linewidth]{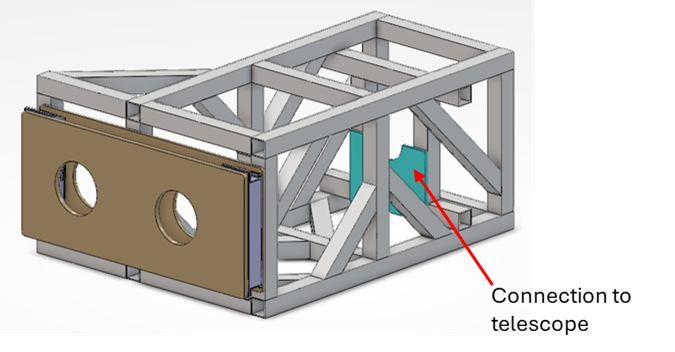}
    \caption{{\it Left and middle panels}: Vertical assembly of the telescope module. {\it Right panel}: Horizontal assembly of the telescope module.}
    \label{fig:tscpe1}
\end{figure}

The uppermost of the main tower modules will include the telescope vacuum system and optics, so will be the final module to be lifted in by crane. Once this module is installed the structural tower is complete, the rest of the instrument will be attached to the tower, the vacuum system will be connected and pump down can begin.
As the top module has a length of around $2\,\textrm{m}$, there is sufficient space to thread the telescope through the module frame for assembly, as illustrated in the left and middle panels of Fig.~\ref{fig:tscpe1}. The tower frame can therefore be welded as a closed square profile, which reduces the possible distortion in the structure during welding. 

\begin{figure}
    \centering
    \includegraphics[width=0.9\linewidth]{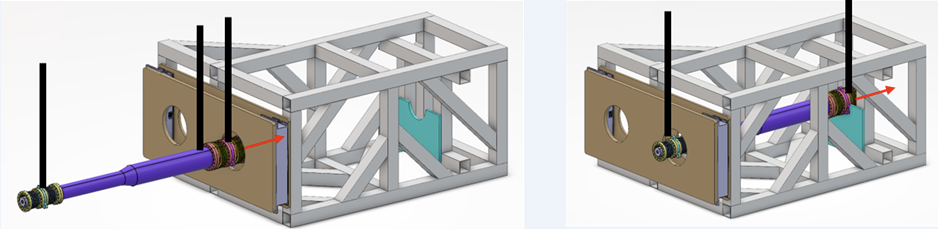}\\
     \includegraphics[width=0.45\linewidth]{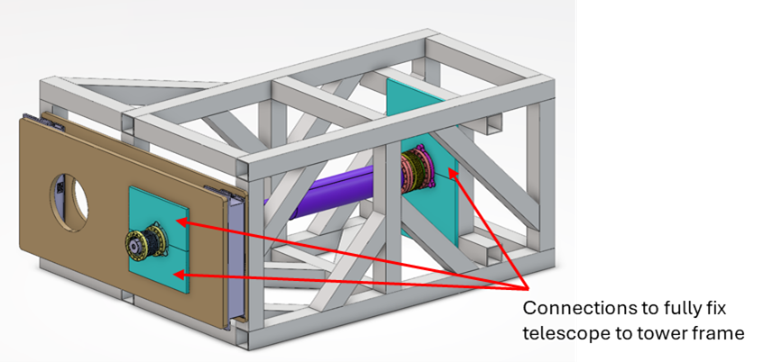}
          \includegraphics[width=0.45\linewidth]{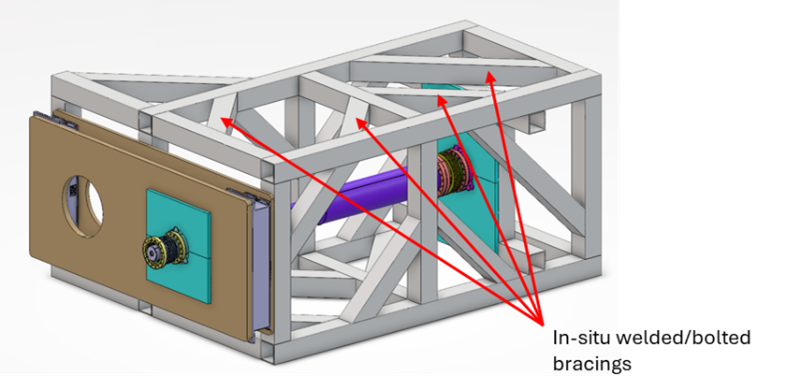}
    \caption{{\it Upper panels}: Lifting operation for horizontal assembly. {\it Lower left panel}: Connections installed to fix telescope. The indicated lifting points are representative, they will not be on the flexible end of bellows. {\it Lower right panel}: Bracings attached to complete the frame.}
    \label{fig:tscpe3}
\end{figure}

The telescope module can be assembled either horizontally or vertically. For {vertical assembly}, a ceiling height of at least $5\,\textrm{m}$ will be needed to accommodate the length of the module and the telescope, as well as space for an overhead crane. Lifting points are needed on the telescope module to facilitate the assembly process. The telescope can then be lifted through the frame’s opening at the top and fully fixed to the tower frame. 
In the case of {horizontal assembly}, no specific ceiling height is required. One side of the diagonal bracing on the frame structure can be left out for easy access of lifting accessories to put the telescope in place, as illustrated in Fig.~\ref{fig:tscpe3}. Extra handling and lifting points may need to be added to the telescope to facilitate the assembly process. 
Extra connections between the telescope and the tower frame are fitted in to fully fix the telescope to the frame after it is moved into place. Diagonal bracings that reinforce the module frame structure are finally welded in-situ or bolted onto the side of the frame where the bracings were left out for lifting equipment access. 

The connections shown in the figures are conceptual and their designs will be further developed to tailor-fit the chosen assembly process. 
A vertical assembly sequence is preferred as the assembly setup is simpler with fewer assembly steps. Fewer lifting points are also required on the telescope. There are potential sites within the Department of Physics of Oxford University with sufficient ceiling height and suitable lifting equipment for this process. The telescope section will be capped off during all assembly processes to preserve cleanliness. These will be made using clear viewports, so that once the vacuum system is mounted into the tower structure the lenses can be aligned before assembly to the tower.

\subsubsection*{Staging Area}
\begin{figure}[h!]
    \centering
    \includegraphics[width=0.6\linewidth]{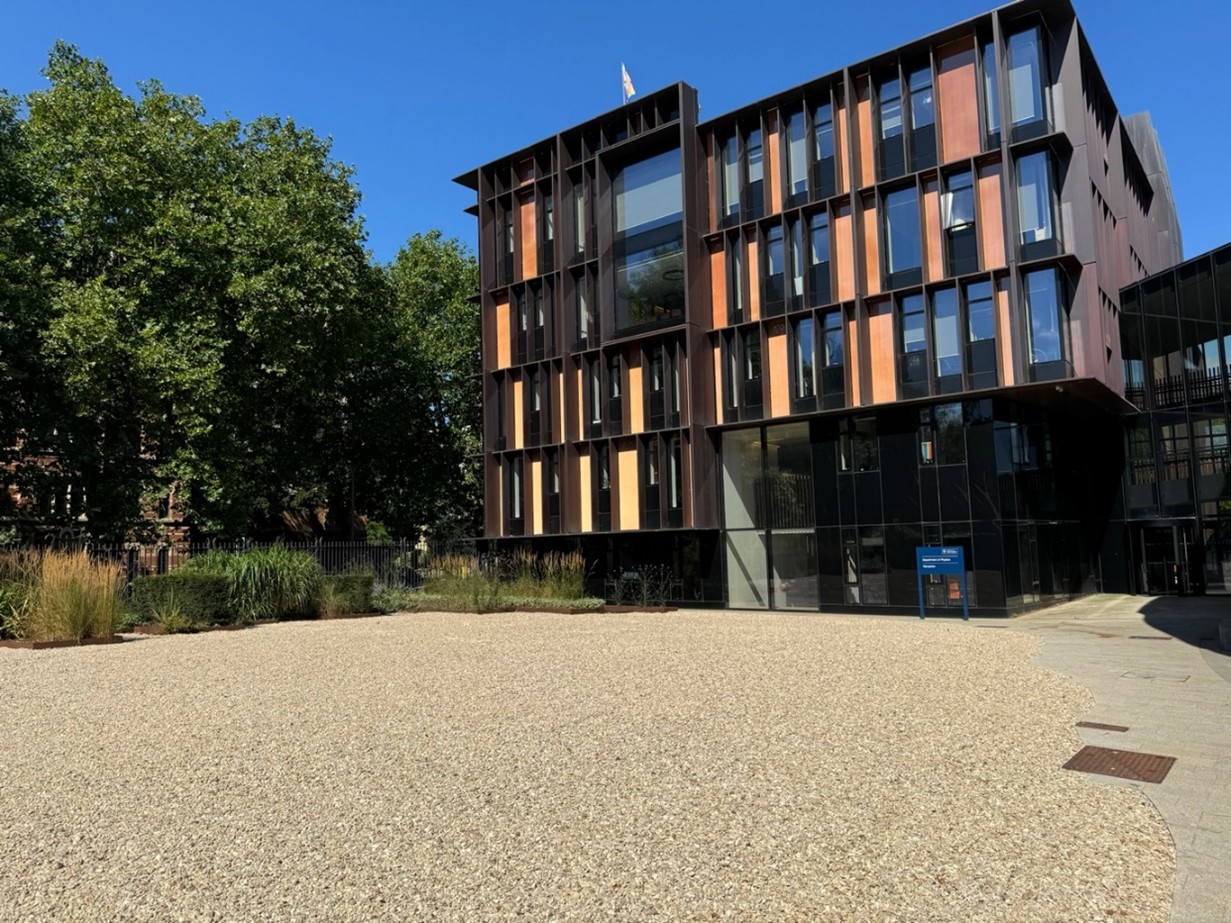}
    \caption{The exterior of the Beecroft building, showing the planned AION-10 staging area.}
    \label{fig:beecroftext}
\end{figure}

Once the modules arrive, they will be placed in a large gravelled staging area just outside the Beecroft, shown in Fig.~\ref{fig:beecroftext}, close to the window through which they will be craned. There are plans to carry out some landscaping in the area, but we will work with site services at Oxford either to delay this work until after AION-10 commissioning, or to leave us sufficient space for the staging area. It is expected that the final preparations for lifting will take place in a tent that will have positive pressure inside to reduce dust and contaminants as much as possible.

\subsubsection*{Lifting}

\begin{figure}
    \centering
    \includegraphics[width=1\linewidth]{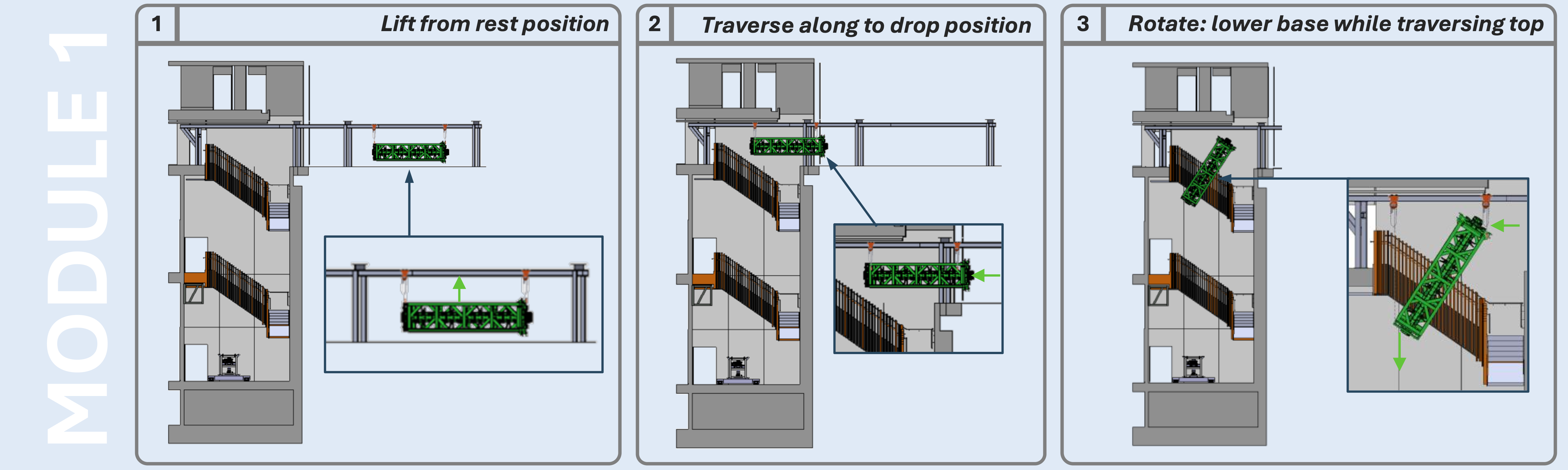}
    \includegraphics[width=1\linewidth]{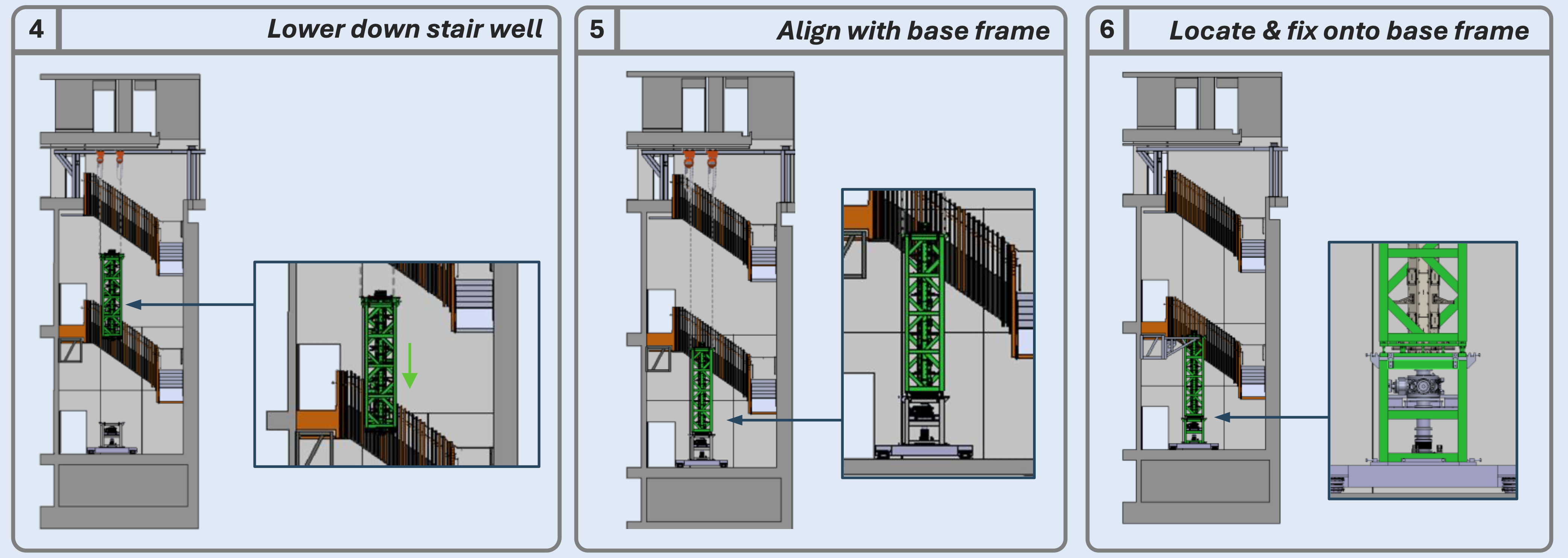}
    \includegraphics[width=0.5\linewidth]{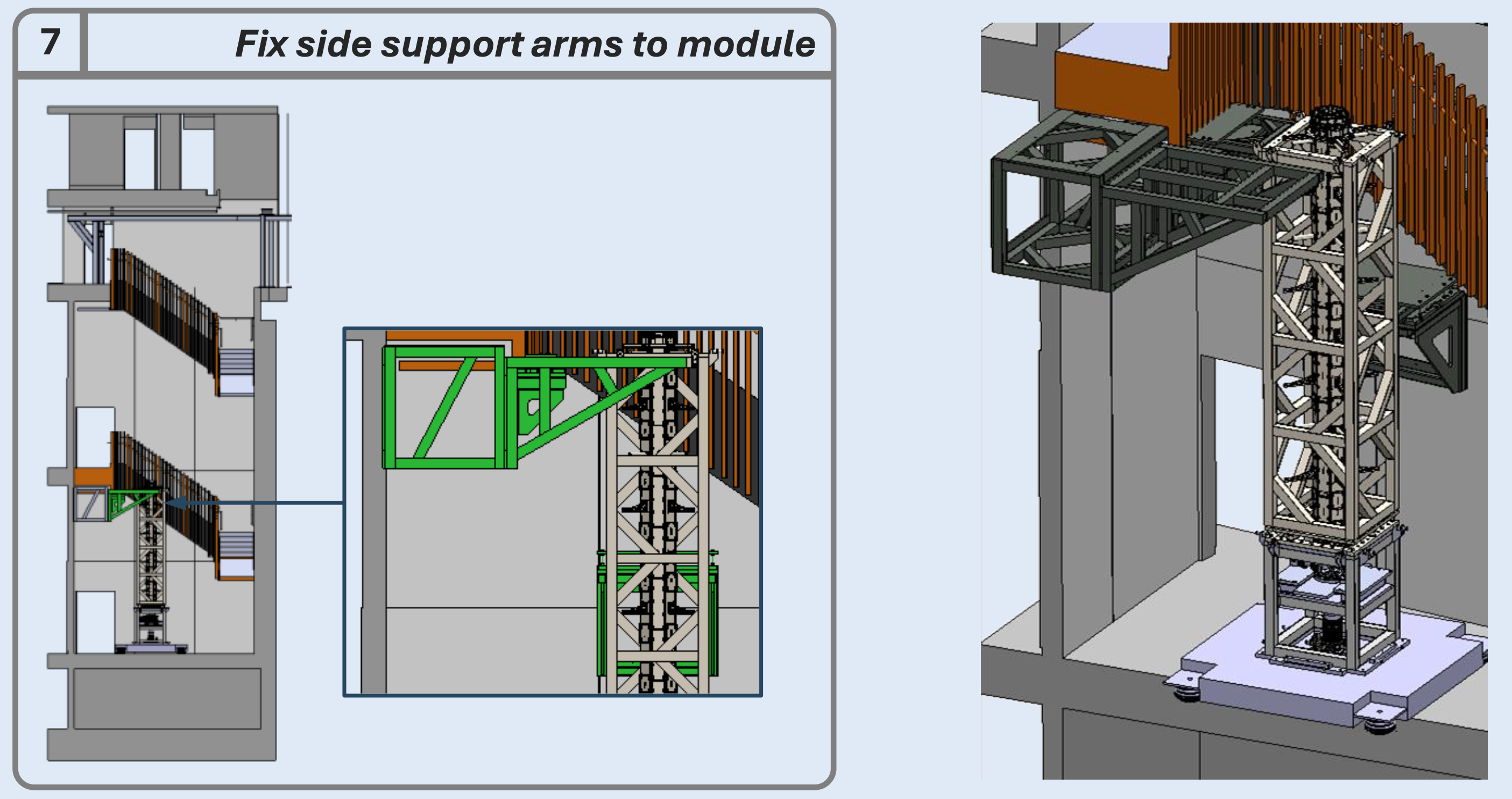}
    \caption{Example of the procedure for lifting one of the modules into the Beecroft stairwell~\cite{CYCLONE}. }
    \label{fig:mod1lift}
\end{figure}

\begin{figure}[h!]
    \centering
    \includegraphics[width=0.8\linewidth]{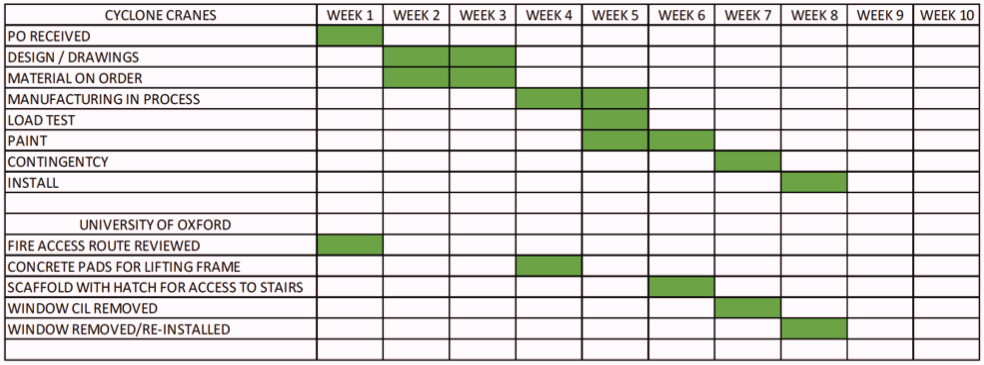}
    \caption{Proposed Gantt chart for AION-10 lifting operations~\cite{CYCLONE}.}
    \label{fig:Gantt}
\end{figure}

A professional lifting and crane company will be required to undertake the installation and craning of the modules through the Beecroft window into the stairwell. While there are trained lifting specialists at Oxford University, the complex and difficult nature of the installation necessitates paying for specialised operators. The project has approached a company (Cyclone Cranes~\cite{CYCLONE}) to do the lifting work. They have  provided a quote below £40k 
for installing temporary cranes in the Beecroft building, performing all the required lifts, and handling all health and safety issues, load testing, documentation and regulations adherence. The lifting sequence involves lowering each module (aside from the base module) into the stairwell, through an exterior window. Once the module has traversed horizontally along a gantry to the ``drop position", it will be rotated to the vertical and lowered. The module will then be bolted to the previously lowered module and the sequence will continue until the telescope module is added. Each module has a levelling base to align them as they are lowered and secured. See Fig.~\ref{fig:mod1lift} for an example of a module lift plan from Cyclone Cranes, and Fig.~\ref{fig:Gantt} for their Gantt chart for the work. Their expected time frame has them on site for one week, preceded by a week to remove the window and sill, and a week to install the scaffolding. To finalise the plan with the company we need to liaise with the structural engineers for the Beecroft to ensure the availability of mounting points able to handle the expected loads. Additionally, we will need to approach a scaffolding company, and work closely together to find a solution that facilitates the smoothest installation possible. It is expected that we will need multiple configurations of scaffolding for different parts of the installation process: for instance, we need access to different parts of the tower for the lift than we do for the commissioning of the sidearms.

\subsubsection*{Connection to Wall Supports}
\begin{figure}
    \centering
    \includegraphics[width=0.6\linewidth]{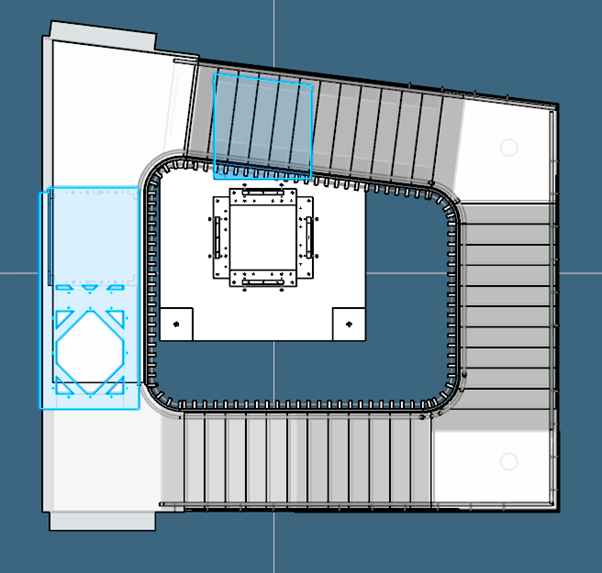}
    \caption{Overhead view of the Beecroft stairwell before tower installation.}
    \label{fig:stairwell}
\end{figure}

\begin{figure}
    \centering
    \includegraphics[width=0.45\linewidth]{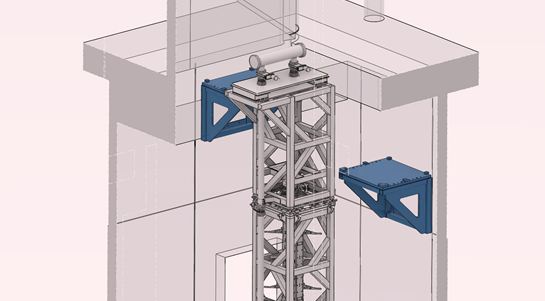}
    \includegraphics[width=0.45\linewidth]{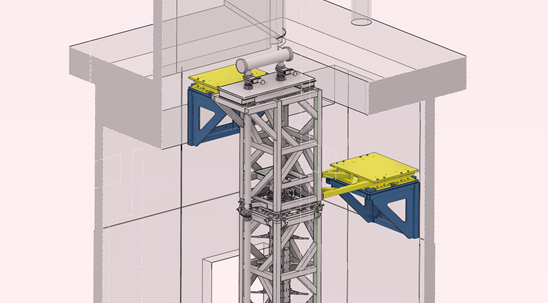}
    \caption{Left panel: Primary support structure (blue) with tower in place. Right panel: Secondary Support Structure (yellow) in place.}
    \label{fig:wallsupport}
\end{figure}

Fig.~\ref{fig:stairwell} shows an overhead view of the Beecroft stairwell before tower installation, and the planned wall support elements (highlighted in blue and yellow) are shown in Fig.~\ref{fig:wallsupport}. The fasteners between the stairwell walls and the supports to the tower are expected to be concrete anchors such as the Hilti HUS3-A 6~\cite{HILTI}. These are threaded and screwed directly into the wall, and are a common way to fix to concrete. This is another area where close collaboration with the structural engineer will be required. It is expected that the concrete anchors and the main support structures will be installed before the modules are craned in. Once the modules are craned in and fixed to each other and the base, the secondary mounts will be installed to attach the structure to the walls. This primary/secondary support design allows the primary supports to be in place before the tower is craned in, without getting in the way of the lifting operations.

\subsection{Vacuum Connection Strategy}
\label{sec:vacconnect}
Once the tower has been craned into place, assembled and secured to the Beecroft stairwell, the capped vacuum system modules will be connected to complete the ultrahigh vacuum (UHV) part of the instrument manifold. The following section details the steps involved.

\subsubsection*{Clean Areas Around Interconnects}
\begin{figure}
    \centering
    \includegraphics[width=0.6\linewidth]{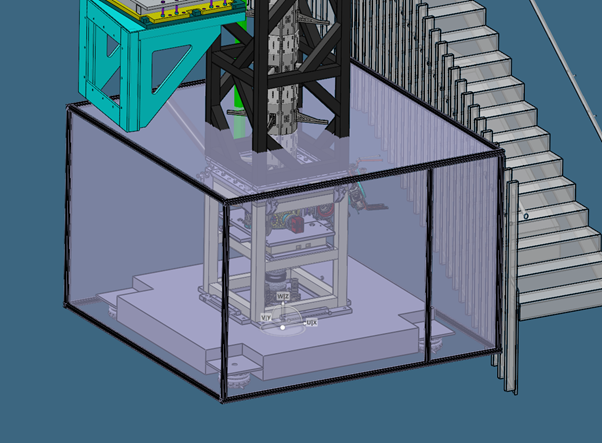}
    \caption{Clean box surrounding the lower interconnect chamber.}
    \label{fig:cleanbox}
\end{figure}

The Beecroft stairwell is not a clean area and it is not feasible to make it so for the vacuum connection process. Therefore, we propose to construct clean boxes around the interconnect chambers and key vacuum connections to reduce the chances of contamination, as illustrated in Fig.~\ref{fig:cleanbox}. We plan to have a pair of nested boxes around each vacuum connection: a smaller one with holes for hand access, and a larger one for personnel to enter. These boxes will have positive pressure applied to them so contaminants will tend to exit, and we expect to be able to provide a nitrogen purge system to provide more positive pressure in the vacuum system itself, and avoid moisture ingress, while connecting the flanges. 

\subsubsection*{Bellows}
\begin{figure}
    \centering
    \includegraphics[width=0.5\linewidth]{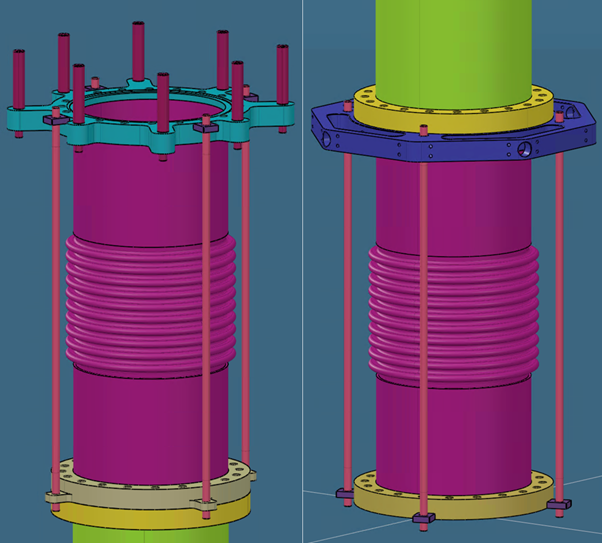}
    \caption{Upper and lower instrument beam pipe bellows.}
    \label{fig:bellows}
\end{figure}

As we plan to build the tower with a disconnected vacuum system, we need to be able to retract some of the flanges for assembly before returning them to their design positions for connection and commissioning. The beam pipes have been designed to accommodate this. We will install retractable bellows at both ends of the interferometer beam pipe assemblies, see Fig.~\ref{fig:bellows}, with threaded tie rods for positioning and to hold the assembly in its compressed position. In pre-assembly these tie rods will be tightened, so that the capped-off modules will arrive at the Beecroft ready to assemble. 

\subsubsection*{Caps and Connections}
Once the tower is installed and the clean boxes are in position, the vacuum system will be connected. At this stage the caps on the vacuum connection being made will be removed, and the tie rods on the bellows loosened to expand the bellows to the design position. The tie rods should be fully disconnected before tightening the flanges in order to avoid any load transfer to the vacuum system. The copper seals will then be installed and the bolts on the flange tightened to finish the connection. 

\subsubsection*{Optics Protection}
\begin{figure}
    \centering
    \includegraphics[width=0.5\linewidth]{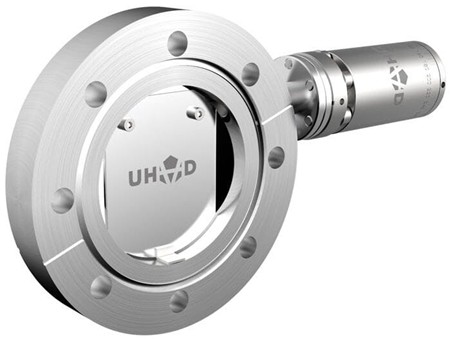}
    \caption{Example of a viewport shutter incorporated in the ultrahigh vacuum design~\cite{UHVshutter}.}
    \label{fig:viewportshutter}
\end{figure}

The cold atoms team have indicated to the engineering team that particles of dust larger than $10\,\textrm{$\mu$m}$ on the in-vacuum optics could cause wave-front aberrations that would degrade the results of the experiment.
Eradicating this risk entirely would entail cleaning and assembling the instrument in an ISO3 clean room. This is unfeasible for assembly of the $5\,\textrm{m}$ modules in the Beecroft stairwell. As well as the mitigation strategies detailed above, we will also provide an extra layer of protection for the telescope optics and the BRM in the form of viewport shutters, see the example in Fig.~\ref{fig:viewportshutter}~\cite{UHVshutter}. These are in-vacuum actuated shutters that create a protective cover for each critical lens/mirror in the instrument, from the vertical installation of the modules to just before turning the lasers on. These shutters should protect the optics from dust deposition during the pump down and bakeout, if any dust has stayed inside the system or entered during the connection process.  It is expected that moving the shutters to their open positions would be one of the final steps in the procedure of commissioning.

\subsection{Remaining Assembly}
\subsubsection*{Telescope}
Once the main tower has been assembled and the vacuum connections made to the beam pipes, the rest of the vacuum system can be assembled. The telescope connection to the main beam pipe will be made next, with the differential pumping aperture forming the end of the UHV part of the system.

\subsubsection*{Beam Transfer Pipe}
The beam transfer pipe on top of the instrument will be added next. It is anticipated that this will arrive as a pre-built assembly, with the optics and actuation tested in lab conditions before being added to the tower. The assembly has gate valves on each side but will not be pumped down for assembly, as it has a relatively small volume compared with the rest of the HV system, and the effort is not necessary since the rest of the pipe will be pumped down after installation.

\subsubsection*{Beam Conditioning Pipe}
The beam conditioning pipe will be installed next, completing the vacuum system manifold. The beam conditioning pipe will connect the beam transfer pipe at the top of the instrument with the input arm, which should already be in place before the tower installation. It contains optics that will condition the beam from the input arm so it is ready to enter the telescope and interferometer. This pipe does not need to be made from $5\,\textrm{m}$ sections, therefore it does not need to be craned in with the modules and it can be assembled into the tower later.

\subsubsection*{Interconnect Chamber Connections}
The various parts connecting to the interconnect chamber will be assembled. This includes the sidearms, the camera systems, the out-of-vacuum launch lattice parts and the field coils around the interconnect chamber.

\subsubsection*{Commissioning}
\begin{figure}
    \centering
    \includegraphics[width=0.3\linewidth]{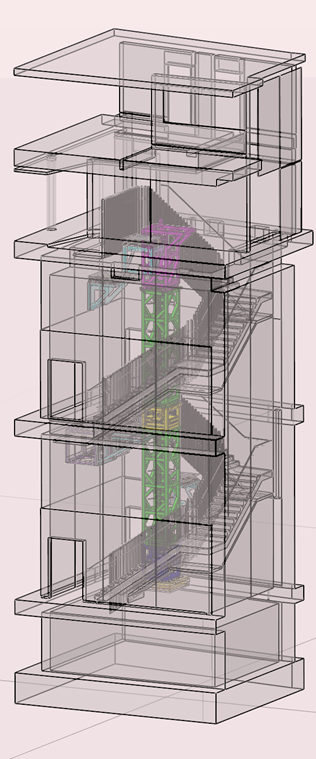}
    \includegraphics[width=0.29\linewidth]{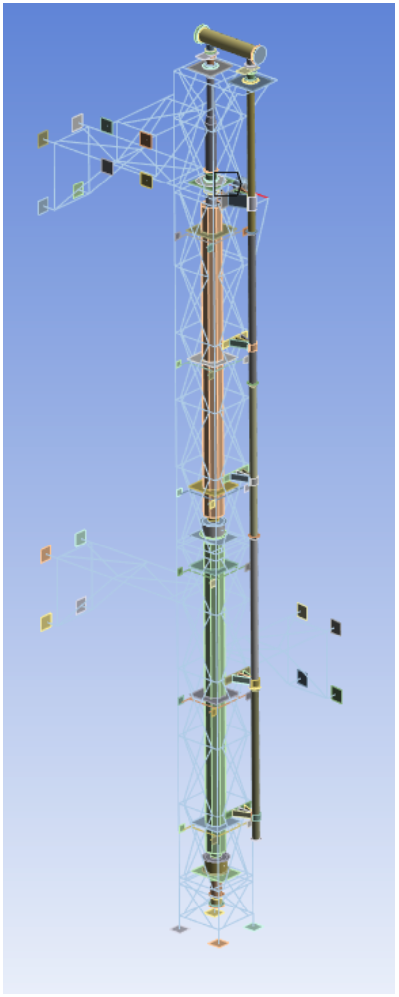}
    \caption{Left panel: The complete AION-10 tower following installation in the Beecroft stairwell. Right panel: 1D and 2D element hybrid analysis model.}
    \label{fig:towerwbeecroft}
\end{figure}

Once the instrument is fully assembled and installed in the Beecroft stairwell, as illustrated in the left panel of Fig.~\ref{fig:towerwbeecroft}, commissioning will begin. The vacuum will be pumped down, leak checked and any leaks fixed. This process could take weeks. Afterwards the various laser systems will be installed, tested and calibrated. The launch lattice systems will need to be calibrated and checked to see if the assembly and bakeout has moved any of the in-vacuum optics, and the camera systems will need to be installed and adjusted. The retroreflecting mirror assembly will be tested using an optical lever, as will the upper steering mirrors.

\section{Stability Analysis}
\label{sec:stability}
\subsection{Analysis of the Support Tower}
\subsubsection*{Beam Profile}
Modal analyses have been conducted to investigate the performance of support towers with different beam member profiles. The profiles compared are a box section of $100 \times 100\,\textrm{mm}$ with $10\,\textrm{mm}$ wall thickness, a box section of $100 \times 100\,\textrm{mm}$ with 5mm wall thickness, and $90 \times 90\,\textrm{mm}$ profile beams that can be assembled easily with bolts. Cross bracing patterns are used in the support tower models.

\begin{table}[htbp]
\centering
\caption{Masses and first 10 modal frequencies of support towers with different beam member profiles.}
\label{tab:modal_frequencies}
\begin{tabular}{lccc}
\toprule
\textbf{Mode} & \multicolumn{3}{c}{\textbf{Frequency (Hz)}} \\
\cmidrule(lr){2-4}
 & \textbf{Box section} & \textbf{Box section} & \textbf{Profile} \\
 & \textbf{100 × 100 × 10~mm} & \textbf{100 × 100 × 5~mm} & \textbf{90 × 90~mm KJN990500} \\
 & 4146~kg & 2715~kg & 4500~kg \\
\midrule
1st mode & 22.4 & 20.3 & 22.4 \\
2nd mode & 34.2 & 31.7 & 34.1 \\
3rd mode & 41.1 & 37.7 & 40.4 \\
4th mode & 45.9 & 40.8 & 45.6 \\
5th mode & 48.1 & 41.4 & 48.1 \\
6th mode & 54.8 & 47.9 & 53.9 \\
7th mode & 56.8 & 49.8 & 55.7 \\
8th mode & 60.5 & 55.7 & 60.4 \\
9th mode & 60.9 & 59.1 & 61.4 \\
10th mode & 65.4 & 64.4 & 65.4 \\
\bottomrule
\end{tabular}
\label{tab:10modefreq}
\end{table}

A simplified 3D model of the tower was developed, using a hybrid of 1D and 2D elements, where 1D elements model the beam members and 2D elements the pipes and the plates, see the right panel of Fig. \ref{fig:towerwbeecroft}. This reduced the required computing time. The beam conditioning pipe and the side support structures are modelled in these analyses. The first 10 modes calculated are listed in Table~\ref{tab:10modefreq}.
 
A support tower constructed using  $90 \times 90\,\textrm{mm}$ profile beams performs similarly to $100 \times 100 \times 10\,\textrm{mm}$ box sections, but profile beams have approximately 8\% more mass. $100 \times 100 \times 5\,\textrm{mm}$ box sections are significantly lighter than the other two options, and its modal frequencies are generally around 10\% less than the box section $100 \times 100 \times 10\,\textrm{mm}$.
The costs of box sections and profile beams are listed and compared in Table~\ref{tab:beam_costs}. The box sections that are most readily available commercially and of similar size are $101.6 \times 101.6 \times 6.35\,\textrm{mm}$ and $101.6 \times 101.6 \times 3.3\,\textrm{mm}$. Box sections are generally cheaper than profile struts per unit length, while the modal performance is similar. Box sections of aluminium grade 6082 also have good weldability and are suitable for building the support tower structure.\\

\begin{table}[htbp]
\centering
\begin{threeparttable}
\caption{Cost comparison of beams with different cross-sections.}
\label{tab:beam_costs}
\begin{tabular}{lccc}
\toprule
\textbf{Cross section} & \textbf{Dimension} & \textbf{£/m (excl. VAT)} & \textbf{Alloy grade} \\
\midrule
Box section & 101.6 × 101.6 × 3.3~mm & 36.31 & 6082 T6\tnote{1} \\
Box section & 101.6 × 101.6 × 6.35~mm & 54.56 & 6082 T6\tnote{1} \\
Profile & 90 × 90~mm & 112.31 & Not specified\tnote{2} \\
\bottomrule
\end{tabular}
\begin{tablenotes}
\small
\item[1] Clickmetal (Aluminium Box Section, 2024)~\cite{Clickmetal}.
\item[2] Bosch Rexroth on RS Components (Bosch Rexroth Silver Aluminium Profile Strut, 90 × 90~mm, 10~mm Groove, 3000~mm Length, 2024)~\cite{BoschRexroth}.
\end{tablenotes}
\end{threeparttable}
\label{tab:crsseccost}
\end{table}

While the cost of box sections is lower than that of the profile beams, box sections are not as flexible in assembly as the profile beams, but it is still possible to mount brackets and plates relatively easily. Bolted profile beams are also more prone to loosening of bolted connections under vibration and stress, while welded box sections have more reliable joints under vibration.

\subsubsection*{Bracing Pattern}
\begin{figure}
    \centering
    \includegraphics[width=0.35\linewidth]{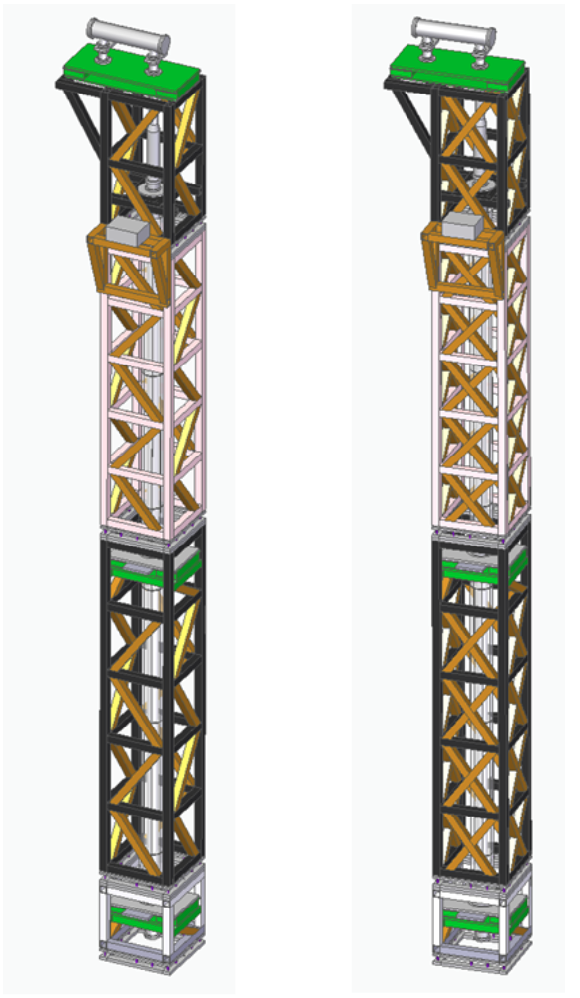}
    \caption{Left panel: Single bracing pattern. Right panel: Cross-bracing pattern.}
    \label{fig:towerwbeecroft2}
\end{figure}

Modal analyses have been conducted on the single and cross-bracing patterns illustrated in Fig.~\ref{fig:towerwbeecroft2} to compare the modal performances of the support tower. These all used the $100 \times 100 \times 10\,\textrm{mm}$ box section. 3D Models of free-standing support towers with nominal footprints of $0.9 \times 0.9\,\textrm{m}$ were analysed and compared, and the first 10 modes obtained are listed in Table~\ref{tab:bracingcomp}.\\

The mass of the cross-bracing model is around 11\% more than that of the single diagonal model, but the obtained modal frequencies are very similar. It can therefore be concluded that single diagonal bracings would have a better mass-to-stability-performance ratio than the cross-bracing pattern.

\begin{table}[htbp]
\centering
\begin{threeparttable}
\caption{Masses and first 10 modal frequencies of support towers with single diagonal and cross bracing.}
\label{tab:bracingcomp}
\begin{tabular}{lcc}
\toprule
\textbf{Parameter} & \textbf{Single diagonal} & \textbf{Cross bracing} \\
\midrule
Mass (kg) & 3415 & 3800 \\
\midrule
\multicolumn{3}{l}{~~~~~~~~~~~~~~\textbf{Modal frequencies (Hz)}} \\
1st mode & 2.30 & 2.35 \\
2nd mode & 2.34 & 2.39 \\
3rd mode & 12.7 & 13.4 \\
4th mode & 35.1 & 35.9 \\
7th mode & 37.0 & 36.9 \\
8th mode & 38.7 & 38.8 \\
9th mode & 39.4 & 40.3 \\
10th mode & 40.5 & 40.8 \\
\bottomrule
\end{tabular}

\end{threeparttable}
\end{table}

\subsubsection*{Stability of AION-10 Structure}
The purpose of the stability analysis of the AION-10 design was to evaluate the overall dynamic behaviour of the structure and compare the stability indicators to the design specifications. In the analysis carried out so far, three types of stability indicators are considered: 
\begin{itemize}
\item	Modal frequencies of the structure;
\item	Translational root mean square (RMS) or end-to-end values of response points;
\item	Angular RMS or end-to-end values of response points.
\end{itemize}

Modal frequencies of the structure were obtained by finite-element analysis (FEA) using methods of eigenvalue extraction. The modal frequencies are closely related to resonances of the structure and its overall stability. For designs under similar conditions, higher modal frequencies usually imply greater structure rigidity and better stability.
The translational and angular RMS/end-to-end values are indicators for a specific response point under given vibration input. They resemble amplitudes in deterministic signals but are calculated for random vibration inputs/responses. 
For a Gaussian signal/process, the RMS displacement from the mean position can be calculated as the standard deviation of the position data. A pair of imaginary boundaries can be drawn for the $\pm3\sigma$ region, within which there is 99.7\% probability that the centre of the camera will remain~\cite{lalanne2009mechanical,lutes2004random,lang2009}.

 
 

The distance between the boundaries in this case is defined as the end-to-end value of the camera vibration.

\subparagraph*{Modal Analysis}

\begin{figure}
    \centering
    \includegraphics[width=0.6\linewidth]{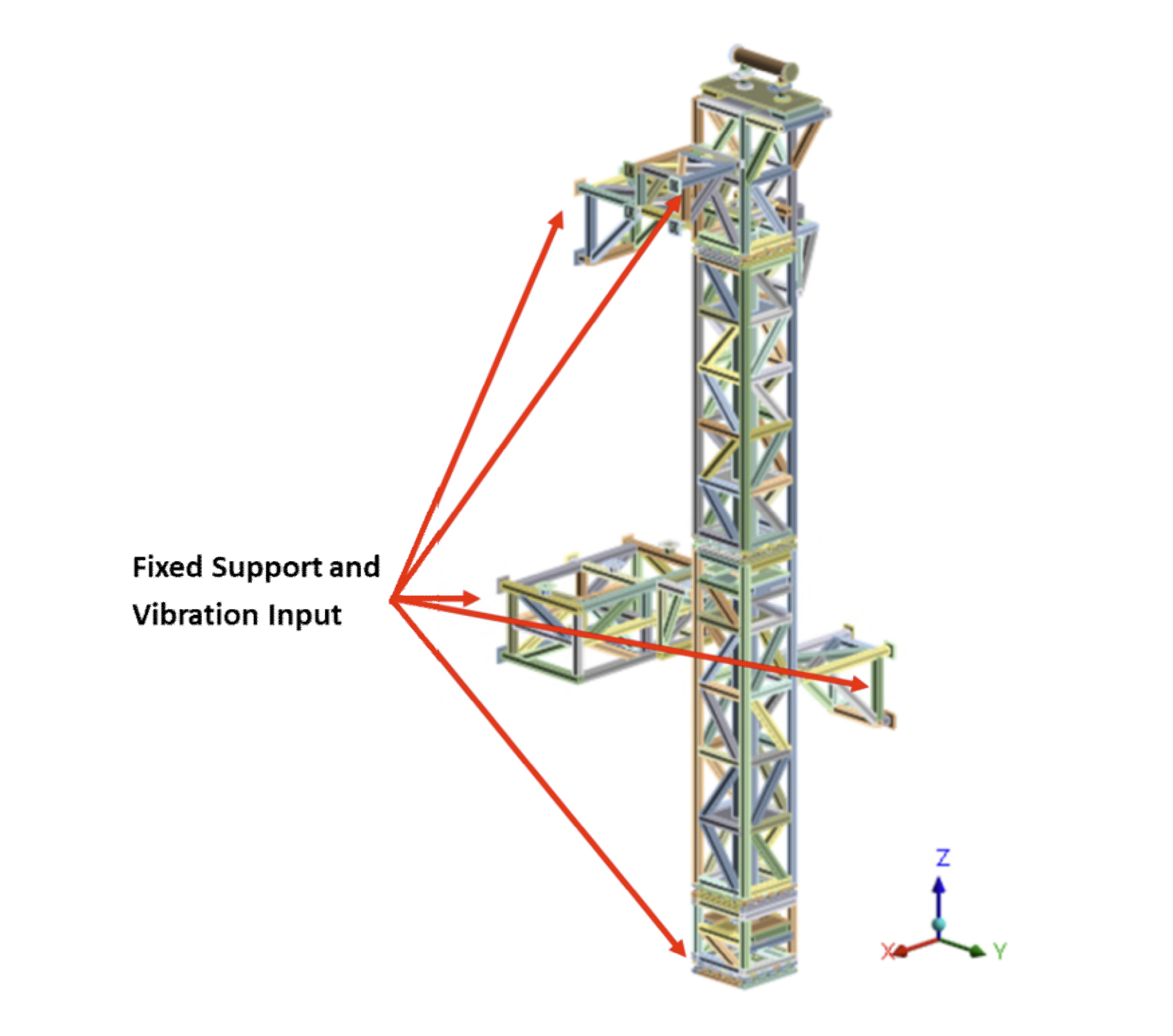}\\
    \caption{The tower model used for the modal analysis.}
    \label{fig:TowerModels}
\end{figure}

Modal analysis was carried out for the current AION-10 design to investigate the stability of the structure. 
We used a simplified version of the current AION-10 design shown in 
Fig.~\ref{fig:TowerModels}, that was meshed using 3D elements. The analysis modelled all major modules, including the main tower modules and all four wall supporting structures. To reduce the complexity of the calculation without compromising the outcome, it ignored insignificant features such as fixings, and simplified unimportant geometry. For boundary conditions, the model was fixed at the base and wall supports, which are the locations where the structure is attached to the building, as shown in Fig.~\ref{fig:TowerModels}. These contacts are also used as locations for inputting vibrations in the response calculation.

After the comparative studies above, a refined model was produced for proposed design. In the current model, $100 \times 100 \times 10\,\textrm{mm}$ box sections are used for the beam members with a single diagonal bracing pattern. The nominal footprint of the tower 
is $0.9 \times 0.9\,\textrm{m}$ in the current design to provide working space within the tower and to decrease the aspect ratio to make the stability of the support tower acceptable. 
The side support frame near the upper interconnect chamber is integrated with the side arm platform and therefore has a substantial mass as well as structural stiffness. 
The beam connection pipes are not modelled. Mesh refinements have been introduced for the connection interfaces between the main modules. The connections are modelled with cylinders to represent M20 screws. 
The results of the modal analysis for the current AION-10 design are listed in Table~\ref{tab:PDRcomparison}. The fundamental/first modal frequency of this design is $24.0\,\textrm{Hz}$. 
This frequency is considered to be acceptable taking into account the height and large geometry of the structure. 

\begin{table}[htbp]
\centering
\renewcommand{\arraystretch}{1.3}
\setlength{\tabcolsep}{9pt}
\caption{Modal Frequency Analysis of the current AION-10 design}

\begin{tabular}{|c|c||c|c|}
\hline
\textbf{\large Mode} & 
\textbf{\begin{tabular}[c]{@{}c@{}}Frequency\\[-0.2em] \small(Hz)\end{tabular}} & 
\textbf{\large Mode} & 
\textbf{\begin{tabular}[c]{@{}c@{}}Frequency\\[-0.2em] \small(Hz)\end{tabular}} \\
\hline\hline
1 & 24.0 & 11 & 65.1 \\
\hline
2 & 26.3 & 12 & 69.1 \\
\hline
3 & 34.1 & 13 & 74.3 \\
\hline
4 & 38.5 & 14 & 84.6 \\
\hline
5 & 42.3 & 15 & 87.6 \\
\hline
6 & 46.3 & 16 & 93.6 \\
\hline
7 & 47.4 & 17 & 100.1 \\
\hline
8 & 53.4 & 18 & 102.0 \\
\hline
9 & 54.8 & 19 & 102.8 \\
\hline
10 & 58.6 & 20 & 108.7 \\
\hline
\end{tabular}

\vspace{0.5em}
\begin{minipage}{0.85\textwidth}
\centering
\begin{minipage}{0.95\textwidth}
\vspace{0.5em}
\centering
\vspace{0.5em}
\end{minipage}
\end{minipage}
\label{tab:PDRcomparison}
\end{table}

\begin{figure}
    \centering
    \includegraphics[width=0.9\linewidth]{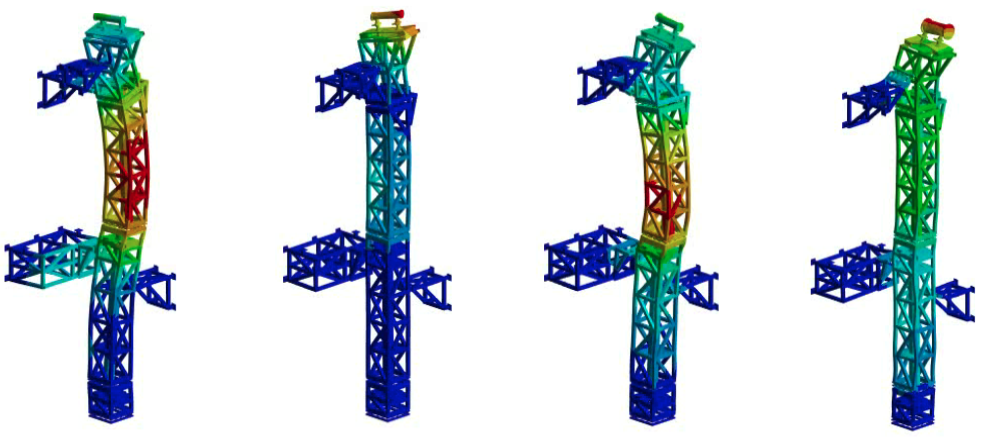}\\
    \caption{3D illustration of modes 1, 2, 4 and 5 (from left to right): $24.0\,\textrm{Hz}$ bending mode, $26.3\,\textrm{Hz}$ local mode, $38.5\,\textrm{Hz}$ bending mode, $42.3\,\textrm{Hz}$ elongation mode.}
    \label{fig:BendingModes}
\end{figure}

In addition to the modal frequencies, the analysis produced deformation shapes for each mode. Fig.~\ref{fig:BendingModes} shows the shapes of the 1st, 2nd, 4th and 5th vibration modes. The plots illustrate the high-deformation regions (red colour) at specific frequencies. By adding extra bracing to these regions, the stability of the structure may be further improved. However, any such modifications must take into account restrictions of geometry and will need to be balanced with cost and performance.

\subparagraph*{Telescope Stability Analysis}

\begin{figure}
    \centering
    \includegraphics[width=0.5\linewidth]{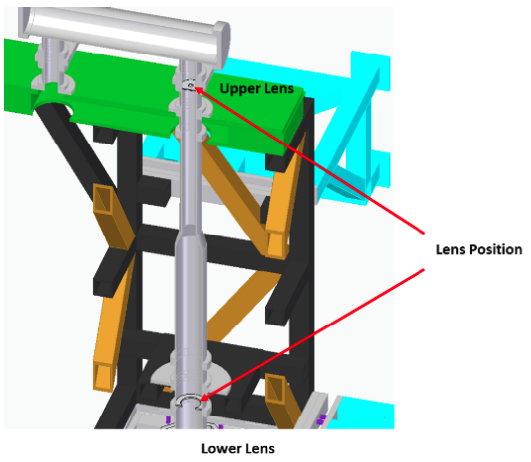}\\
    \caption{The telescope lenses.}
    \label{fig:TelescopeLenses}
\end{figure}

In order to evaluate further the stability of the AION-10 structure, the response-based stability indicators (RMS/end-to-end values) were calculated with a focus on the lenses of the telescope. We recall that the telescope section is an important module in the interferometer that will condition the laser beams for the main interferometer module. The main optics in the telescope consist of two lenses, as illustrated in Fig.~\ref{fig:TelescopeLenses}. To simplify the analysis, the lenses are modelled as rigidly connected parts with a direct connection to the outer case. 
These lenses are sensitive to vibration as the mechanical motion can cause the centre of the laser to fluctuate as well as affecting the focus of the beam. As stability indicators, the responses of these lenses and their relative motion were extracted for the centre of the optics, which can be later used for comparing to the design specifications.

\begin{figure}
    \centering
    \includegraphics[width=0.5\linewidth]{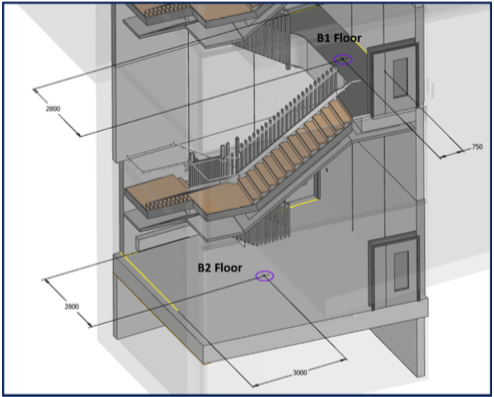}\\
    \caption{Vibration measurement locations indicated by circles.}
    \label{fig:VibrationMeasurement}
\end{figure}

\begin{figure}
    \centering
    \includegraphics[width=0.8\linewidth]{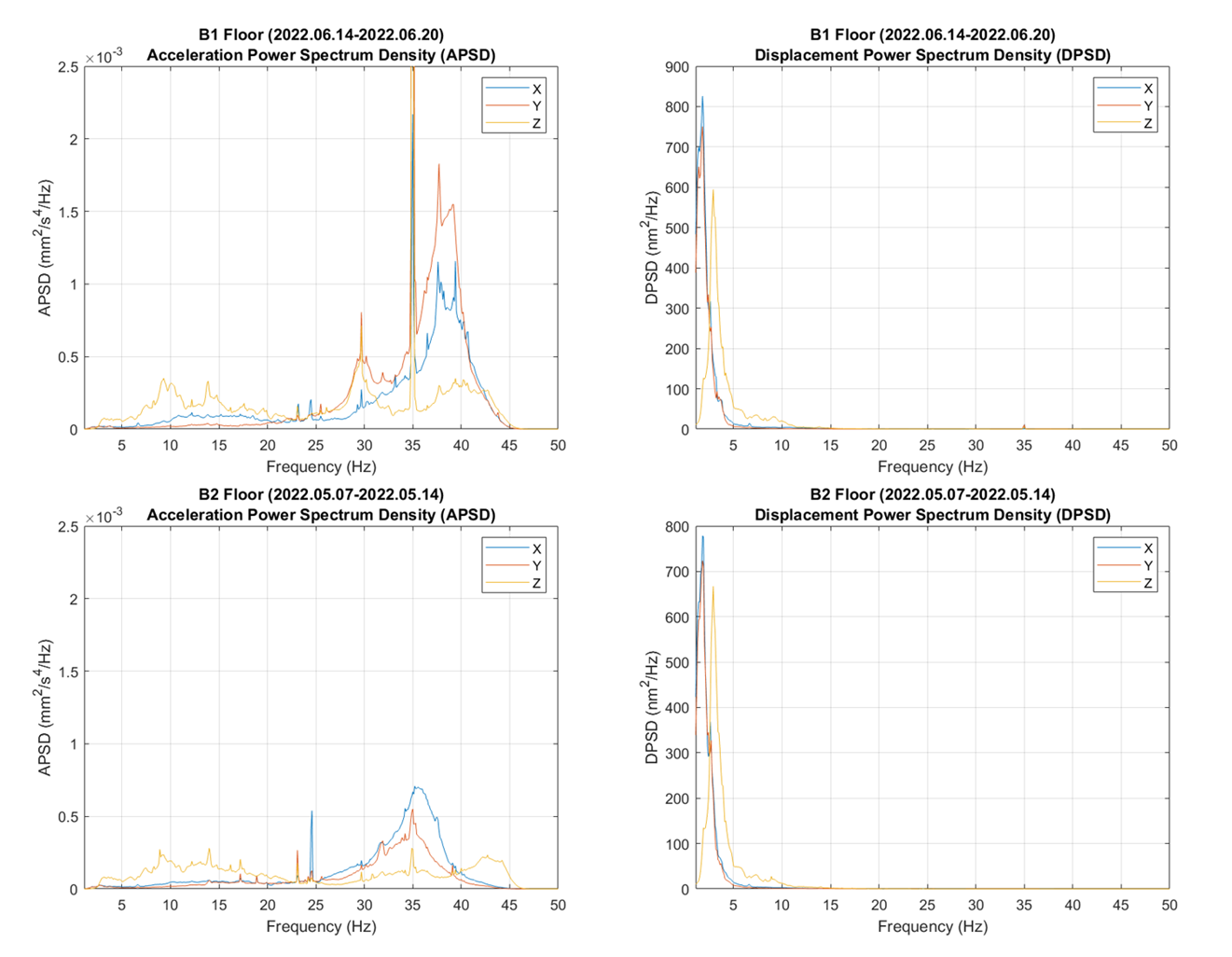}\\
    \caption{Acceleration and displacement power spectra of vibration data.}
    \label{fig:Acceleration+Displacement}
\end{figure}

The responses were calculated assuming a 3-axis single source vibration. The excitation was applied to the entire tower structure through all the fixed supports (four wall supports and one base support) illustrated in Fig.~\ref{fig:TowerModels}. Applying the same input to multiple support points is in practise equivalent to treating the stairwell as a rigid structure as all the fixed supports vibrate in the same way. The vibration input data used in this analysis were obtained from a test carried out in the Beecroft building during 14-20 June 2022. As shown in Fig.~\ref{fig:VibrationMeasurement}, the measurements were taken on both the B1 and B2 floors at the locations indicated by circles. As B1 is closer to the ground and entrance of the building, it shows a higher vibration level than B2. As a more conservative consideration, the B1 measurement data were used for this analysis as the vibration input. The acceleration and displacement power spectrum of the measured  vibration data are shown in Fig.~\ref{fig:Acceleration+Displacement}.

The results of the analysis are shown in Table~\ref{tab:telescoperesponse}. The responses at the upper and lower lenses are given as RMS and end-to-end values. In the case of the displacement RMS, the input vibration is $33.1-40.8\,\textrm{nm}$, whereas the response of the lenses is between $36.0-44.6\,\textrm{nm}$. A smaller RMS of $7.1-17.0\,\textrm{nm}$ was found in the relative motion of the lenses.\\

\begin{table}[htbp]
    \caption{Responses of telescope lenses}
    \centering
    \renewcommand{\arraystretch}{1.3}
    \begin{tabular}{l|cccc}
    \hline
    \multirow{2}{*}{\textbf{Parameter}} & \multicolumn{4}{c}{\textbf{RMS / End-To-End Values ($0.5-50\,\textrm{Hz}$, nm)}} \\
    \cline{2-5}
    & \textbf{Input} & \textbf{Upper lens} & \textbf{Lower lens} & \textbf{Relative} \\
    \hline
    X (nm) & 40.8 / 244.8 & 44.6 / 267.6 & 42.0 / 252.0 & 17.0 / 102.0 \\
    Y (nm) & 38.7 / 232.2 & 42.0 / 252.0 & 39.8 / 238.8 & 7.1 / 42.6 \\
    Z (nm) & 33.1 / 198.6 & 36.4 / 218.4 & 36.0 / 216.6 & 10.0 / 60.0 \\
    RX (nrad) & --- & 8.4 / 50.4 & 7.2 / 43.2 & 14.7 / 88.2 \\
    RY (nrad) & --- & 16.4 / 98.4 & 5.6 / 33.6 & 10.1 / 60.6 \\
    RZ (nrad) & --- & 10.4 / 62.4 & 5.5 / 33.0 & 11.8 / 70.8 \\
    \hline
    \end{tabular}
    \label{tab:telescoperesponse}
\end{table}

The displacement RMS for the upper lens is around the maximum transverse shift specified in Sec.~\ref{sec:telescope}, while all other values are well below the maximally allowed values.
Given that the targets from Sec.~\ref{sec:telescope} are worst case estimates, and that our amplification factor from input to output from the tower is extremely low, it is concluded that the structure shows satisfactory stability in this analysis. On the final tower we plan to include vibration sensors near key optics to confirm these simulations and react if the response is worse than predicted.\\

\subparagraph*{Comparison of Model without Magnetic Shield and Vacuum Tube}

During the structure design and analysis, it was noted that not all parts of the structure have the same stability specifications, with the cameras and optical components having  by far the most stringent stability requirements. These parts will need special vibration mitigation methods and reinforcement in place. On the other hand, components such as the magnetic shield have a much higher tolerance for vibration.

Taking into account the large difference in stability requirements, a natural question arose: will it bring any benefit from the engineering point of view if the vibration-insensitive modules are decoupled from the main tower and supported by a separate structure? These parts are also of considerable mass. By excluding them from the main frame, less mass will be structurally connected to the critical components, which may further improve the stability.

To address this question, a `lite model' was considered, in which both the magnetic shield and vacuum tube were disconnected from critical modules such as the interconnect chambers and telescope, that are most sensitive to vibrations. In the full structure, the total mass supported by the frame is around $2638\,\textrm{kg}$ whereas a mass of only $1745\,\textrm{kg}$ is supported in the lite model. Thus, by excluding the vibration-insensitive modules, a reduction of $893\,\textrm{kg}$ (33.9\%) in the supported mass is achieved in the lite version, which may result in improved stability performance. 
 


The same analysis settings and vibration inputs were used for the lite model as for the full model. 
As illustrated in Table~\ref{tab:fulllitecomp}, although the motions of the lenses are generally less in the lite model, slightly higher relative responses were found, which may be partly attributed to the absence of extra bracing due to the removal of magnetic shield and vacuum pipe.\\

\begin{table}[htbp]
 \caption{Comparison of responses for the full and lite models.}   \centering
    \small
    \renewcommand{\arraystretch}{1.2}
    \setlength{\tabcolsep}{4pt}
    \begin{tabular}{l|c|cc|cc|cc}
    \hline
    \multirow{3}{*}{\textbf{Parameter}} & \multicolumn{7}{c}{\textbf{RMS / End-To-End Values (0.5-50~Hz, nm)}}\\
    \cline{2-8}
    & \multirow{2}{*}{\textbf{Input}} & \multicolumn{2}{c|}{\textbf{Upper lens}} & \multicolumn{2}{c|}{\textbf{Lower lens}} & \multicolumn{2}{c}{\textbf{Relative}}\\
    \cline{3-8}
    & & \textbf{Full} & \textbf{Lite} & \textbf{Full} & \textbf{Lite} & \textbf{Full} & \textbf{Lite}\\
    \hline
    X (nm) & 40.8 / 244.8 & 44.6 / 267.6 & 45.0 / 270.0 & 42.0 / 252.0 & 41.7 / 250.2 & 17.0 / 102.0 & 18.2 / 109.2\\
    Y (nm) & 38.7 / 232.2 & 42.0 / 252.0 & 41.5 / 249.0 & 39.8 / 238.8 & 39.0 / 234.0 & 7.1 / 42.6 & 8.4 / 50.4\\
    Z (nm) & 33.1 / 198.6 & 36.4 / 218.4 & 36.1 / 216.6 & 36.0 / 216.6 & 36.0 / 216.0 & 10.0 / 60.0 & 16.1 / 96.6\\
    RX (nrad) & --- & 8.4 / 50.4 & 10.2 / 61.2 & 7.2 / 43.2 & 6.0 / 36.0 & 14.7 / 88.2 & 13.6 / 81.6\\
    RY (nrad) & --- & 16.4 / 98.4 & 16.2 / 97.2 & 5.6 / 33.6 & 5.3 / 31.8 & 10.1 / 60.6 & 3.7 / 22.2\\
    RZ (nrad) & --- & 10.4 / 62.4 & 10.3 / 61.8 & 5.5 / 33.0 & 5.5 / 33.0 & 11.8 / 70.8 & 15.0 / 90.0\\
    \hline
    \end{tabular}
    \label{tab:fulllitecomp}
\end{table}

The conclusion of this analysis was that, 
although excluding the magnetic shield and vacuum pipe may result in higher modal frequencies and some improvement in lens responses, the benefit is marginal compared to the extra engineering effort required to decouple mechanically the heavy vibration-insensitive modules. It has therefore been decided to stick to the full structure design for the AION-10 support tower.

\subsection{Multi-Input Vibration Model and Vibration Survey of Beecroft Building}
In the stability analyses presented above, the responses were calculated using a single-input algorithm. In this method, the same vibration input is applied at all support points, which simplifies the analysis significantly. This is in fact equivalent to fixing the support points to a rigid body. For AION-10, it means the stairwell is a rigid mega-structure and moves as a whole under the given excitation. 
However, as the stairwell has a depth of around $15\,\textrm{m}$, the deformation of the mega-structure under ground-borne and ambient vibrations can be notable. Various levels of vibration may also be found in different parts of the stairwell. To account for this in the stability analysis, it is necessary to consider multiple vibration inputs in the analysis model.
\subsubsection*{Vibration Survey of Beecroft Building}
In a multi-input vibration study the frequency response functions (FRFs) and auto/cross spectra of the inputs will be needed. The FRFs can be computed through FEA, whereas the auto/cross spectra of the inputs need to be obtained through site tests. It should be noted that as the cross spectra preserve relationships between different signals, it can only be calculated if the signals are obtained from a synchronized measurement. Instead, if vibrations at different locations are measured independently, it will be impossible to recover the cross spectrum without additional information. Synchronized measurements may also allow processes such as operational modal analysis to be performed, which can retrieve modal parameters and reconstruct mode shapes of the building.

The vibration data currently available for Beecroft building were taken in 2022 on the B1 and B2 floors without synchronization, and the spectra of the vibrations are illustrated in Fig.~\ref{fig:Acceleration+Displacement}. The data served well for the single-input analysis described above. To provide further insight into the vibrations of the stairwell in the Beecroft building, it is desirable to carry out additional vibration tests for the following reasons:
\begin{itemize}
\item 
Obtain vibration data for locations where the support structure of AION-10 will be attached;
\item Perform synchronized measurements, which will allow the data to be fed into the multi-input vibration model.
\end{itemize}

\begin{figure}
    \centering
    \includegraphics[width=0.8\linewidth]{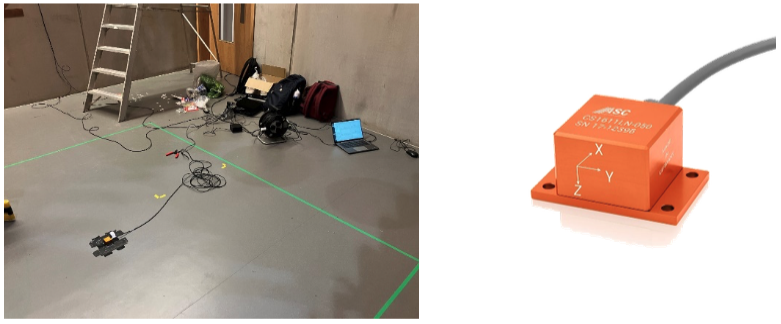}\\
    \caption{Vibration test set-up: data-acquisition system and vibration sensor.}
    \label{fig:VibrationTest}
\end{figure}

A vibration test was carried out for the Beecroft building in November 2024 with the aim of obtaining more data to inform the future design and analysis of the AION-10 tower. Fig.~\ref{fig:VibrationTest} shows the experimental set up for the vibration test, which employed three 3-axis voltage-output accelerometers. The sensor output was sampled by a 24bit 16-channel data-acquisition system at $5\,\textrm{kS/s}$. As the frequency range of interest is $0-200\,\textrm{Hz}$, the sample rate is adequate in this case.\\

\begin{figure}
    \centering
    \includegraphics[width=0.4\linewidth]{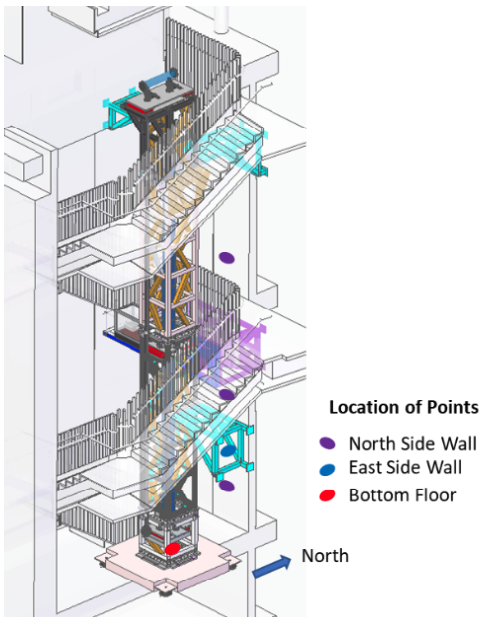}\\
    \includegraphics[width=0.8\linewidth]{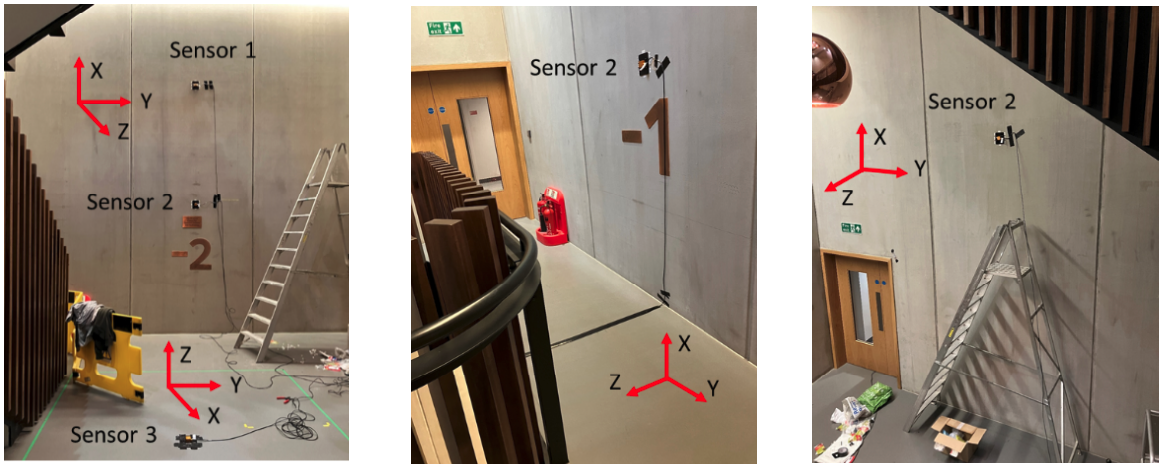}\\
    \caption{Points of vibration measurements and experimental set up.}
    \label{fig:VibrationTest2}
\end{figure}

Due to equipment restrictions, the measurements were taken for representative points that were accessible and close to the structure supports. A total of 3 tests were performed at 5 measurement points, as shown in Figure~\ref{fig:VibrationTest2}. In each test, the sensors were placed at three locations. Sensors 1 and 3 stayed in the same position in all tests
, whereas sensor 2 was placed on the B2 level (north side wall) in test 1, on the B1 level (north side wall) in test 2, and on the B2 level (east side wall) in test 3. Each test lasted for 45 minutes during a workday afternoon when the building was relatively busy.

\begin{figure}
    \centering
    \includegraphics[width=1\linewidth]{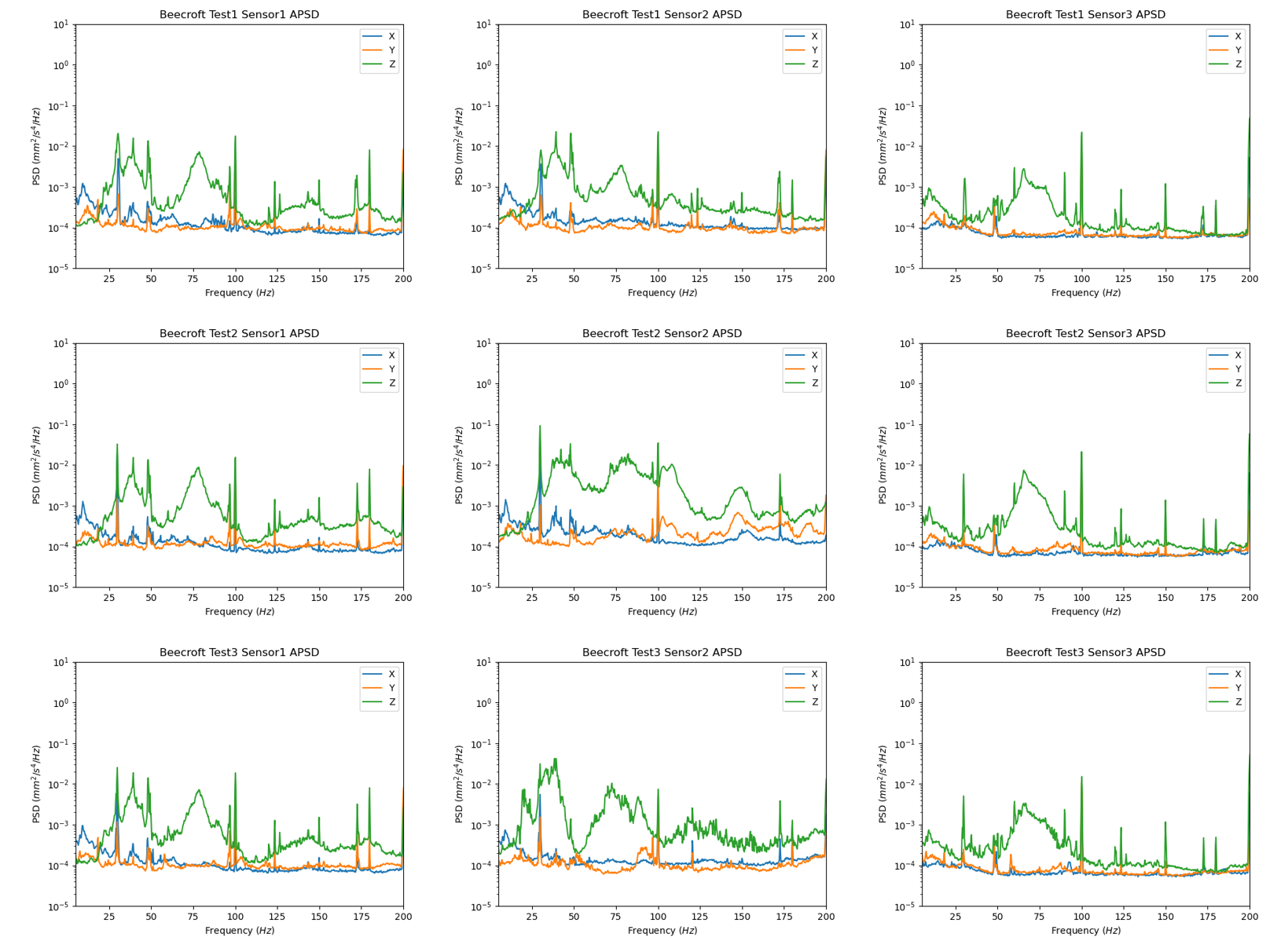}\\
    \caption{Acceleration power spectrum densities (APSDs) of the measured vibrations.}
    \label{fig:VibrationTest3}
\end{figure}

The acceleration power spectrum densities (APSDs) of the measured vibrations are illustrated in Figure~\ref{fig:VibrationTest3}. Peaks were found around frequencies including $30\,\textrm{Hz}$, $40\,\textrm{Hz}$, $50\,\textrm{Hz}$ and $100\,\textrm{Hz}$. The $30\,\textrm{Hz}$ and $40\,\textrm{Hz}$ peaks may be related to service machinery in the building, whereas the $50\,\textrm{Hz}$ and $100\,\textrm{Hz}$ peaks are thought to be related to electrical interference. The results agree well with the previous measurements detailed in Fig.~\ref{fig:Acceleration+Displacement}, with similar peaks in the 25-50Hz region. These new results broaden the range of frequencies up to 200Hz. Further data processing will be carried out at a later stage to prepare the data for use in vibration analysis.
     
\clearpage

\section{Magnetic Analysis}
\label{sec:magnetic}
\subsection{Magnetic Shielding}
\label{sec:magshielding}
A stable and controlled magnetic field environment is necessary for atom interferometry to perform optimally. This puts requirements on both the field strength in the interferometry region and on its stability and homogeneity.
The requirements for the magnetic field have been set by the collaboration as:
\begin{itemize}
\item
A tuneable horizontal field from $10\,\textrm{mG}$ to $10\,\textrm{G}$ with a $<5\,\textrm{mG}$ inhomogeneity and homogenous within $<5\,\textrm{mrad}$. The magnetic field must be able to be applied to the two horizontal field axes ($x$ and $y$).
\item
The magnetic field noise must be $< 1\mu$G/$\sqrt{\rm Hz}$.
\end{itemize}

To work towards meeting the requirements, various geometries of mu-metal magnetic shielding have been modelled that are practical for the purpose of shielding a long tube of relatively small diameter.
The magnetic modelling has been performed using the COMSOL electromagnetic modelling software package~\cite{COMSOL}, based on a FEA algorithm to model the magnetic fields. The modelling has been done for a single 5m beam pipe section, using continuous shields (neglecting any holes for mounting and fixings). 


To summarise the modelling below, we conclude that a double layer of shielding, with octagonal and staggered geometry, is required to provide the shielding to mG levels in the shields. The magnetic field noise density amplitude is measured on-location (see below) and will meet the required levels for frequencies $> 10^{-4}\,\textrm{Hz}$ when shielded by a factor of 1000. Furthermore, we conclude that a simple race-track coil configuration system will provide the required field strength in the interferometry region, but the homogeneity and divergence requirements will not be met. For this, saddle-type coil configurations are required, see Section \ref{sec:magguide}.

\subsubsection*{Design Choices}

\begin{figure}
    \centering
    \includegraphics[width=0.8\linewidth]{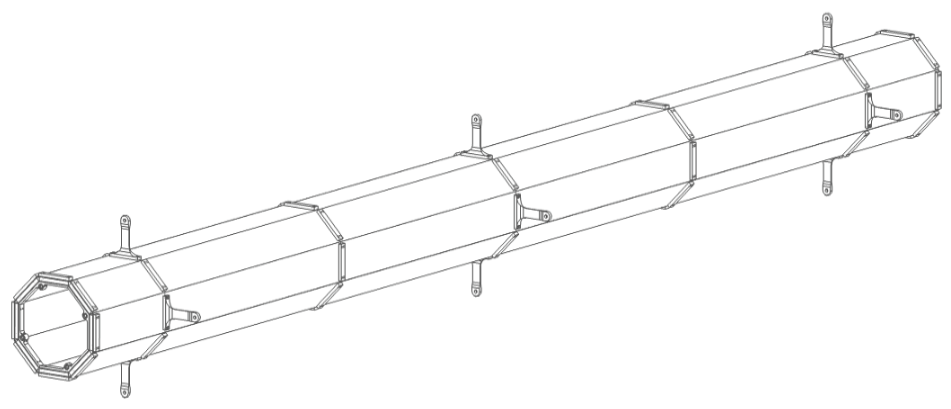}\\
    \includegraphics[width=0.8\linewidth]{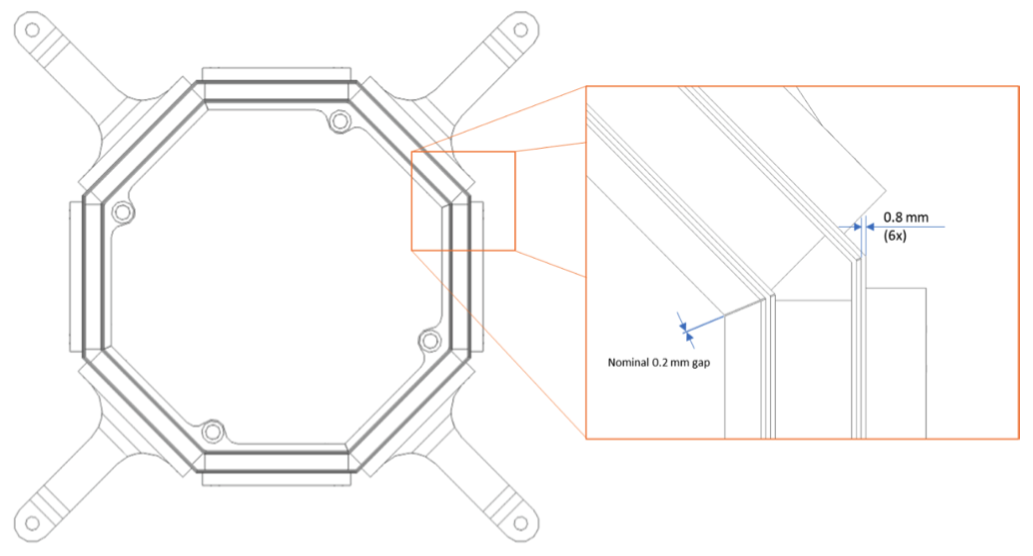}\\
     \includegraphics[width=0.8\linewidth]{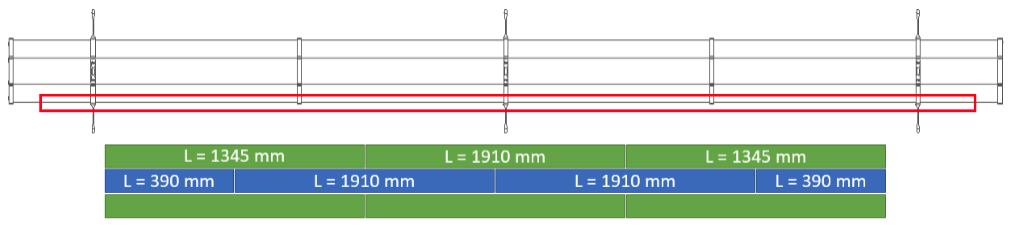}
   \caption{Top panel: CAD image of the magnetic shielding.
   Middle panel: Zoomed-in view of the transverse cross-section of the shielding. Note the staggered V approach to avoid gaps through the full shielding thickness.
   Bottom panel: Representative sketch showing the longitudinal staggering between shielding layers 1-3 (inner shield). An identical scheme is present in shielding layers 4-6 (outer shield).}
    \label{fig:Shielding}
\end{figure}  

The design of the magnetic shielding is shown in Fig.~\ref{fig:Shielding}. It has been driven not only by the shielding requirements, but also by practical requirements such as ease of installation and assembly, annealing needs and de-Gaussing capabilities. With the interferometer region being located in long cylindrical tubes, the magnetic field modelling has been carried out for cylindrical and octagonal shields of mu-metal. These have a typical thickness of $2-3\,\textrm{mm}$, in both a single-layer configuration and a double-layer configuration.

The design has converged to octagonal shields, which is also the design choice of our MAGIS collaborators in the US~\cite{MAGIS-100:2021etm} and of the VLBAI interferometer in Hannover~\cite{schlippert2020matter, wodey2020scalable}. This allows mounting the mu-metal in a staggered geometry (removing seams at the joints) and can use flat sheets of mu-metal rather than bent sheets.

A single magnetic shield consists of three layers of $0.8\,\textrm{mm}$ thick mu-metal, arranged in an octagonal prism, for a total thickness of $2.4\,\textrm{mm}$. To compress the layers, ensure minimal gaps and provide regular interfaces to both the tower and inner vacuum chamber, the layers are fixed regularly using a modular mild-steel collar and clamping system. The second magnetic shield has the same general design, with increased guiding dimensions to allow a nominal $15\,\textrm{mm}$ gap between the inner and outer shields. The assembly method is discussed below.

As mu-metal cannot typically be sourced in lengths longer than $3\,\textrm{m}$, and to promote good assembly practice, the assembly must contain both longitudinal and transverse splits. Longitudinally, the octagon is formed by 4 vee-shaped cross-sections for each layer. The middle layer is then offset by 45 degrees to both the outer and inner layers to minimise the impact of gaps between mu-metal plates. In the transverse direction, the middle layer is similarly staggered, with the intent of minimising all shielding leaks.

The shield will be fixed to the UHV chamber on the upper flange of the tied bellow assembly on the lower end of each beam pipe assembly. At the middle and upper interfaces, it will be fixed in 5 degrees of freedom using low-friction linear bearings to bespoke UHV flanges. This provides mechanical stability, and provides a pathway to support the UHV chamber at the centre of the exposed $4.6\,\textrm{m}$ span.
It may be possible to implement a further radial mount in the centre of each $2\,\textrm{m}$ chamber, though initial stability simulation indicates that this will make a negligible overall contribution to the stiffness of the structure or component subassemblies.


 
  
\subsubsection*{Shielding Analysis}
\begin{figure}
    \centering
    \includegraphics[width=0.45\linewidth]{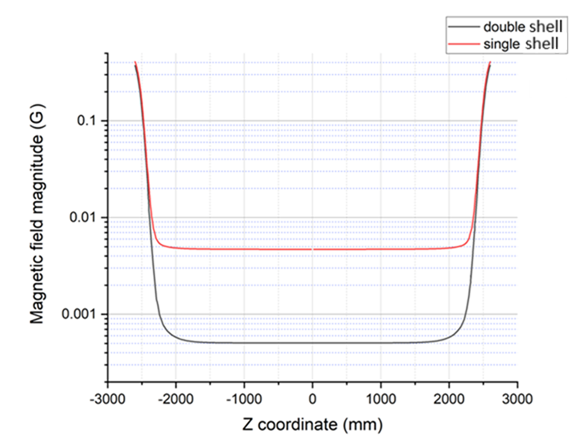}
    \includegraphics[width=0.45\linewidth]{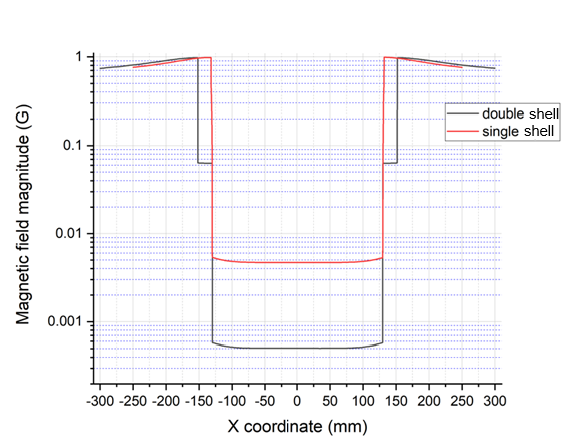}
   \caption{{\it Left panel}: The magnetic field magnitude in the centre of a single layer (red line) and double layer (black line) octagonal $2\,\textrm{mm}$ shielding as function of vertical position ($z$). The Earth magnetic field ($0.6\,\textrm{G}$ outside the shields) is seen to penetrate a distance of $\sim 500\,\textrm{mm}$ from the ends. The magnetic field is $4.4\,\textrm{mG}$ at the centre of the shield for the single-layer shielding and $0.5\,\textrm{mG}$ for the double-layer shielding configuration.
   {\it Right panel}: The magnetic field magnitude as function of distance to the centre of the tube for a single-layer (red line) and double-layer (black line) shield configuration. An Earth magnetic field of $0.6\,\textrm{G}$ is applied outside the shields.}
    \label{fig:SingleDouble}
\end{figure}

{\bf Single- and double-layer octagonal shields:}
Results of modelling single- and double-layer octagonal geometries are shown in Fig.~\ref{fig:SingleDouble} ($B$ in the centre of the tube as function of height $z$ in the left panel and $B$ as function of the distance to the centre of the tube in the right panel). In this representation, the $x$-axis is purely horizontal, with the model assuming a field of $0.6\,\textrm{G}$ in this direction. The field inside the single layer shields is shown to be at the level of $4.4\,\textrm{mG}$, so we have a static shielding factor of 130 in this configuration. Considering the requirement of setting magnetic fields to mG level homogeneities and anticipating compromises to the shielding when allowing for access ports, etc., we conclude that this single-layer shielding configuration will not be sufficient. The results for double-layer octagonal shield geometry modelling show an increased shielding down to $0.5\,\textrm{mG}$ inside the shields and field gradients $< 0.01\,\textrm{G/m}$. We can conclude that a double-layer octagonal shield would be suitable for the AION set-up.
 
\begin{figure}
    \centering
    \includegraphics[width=0.6\linewidth]{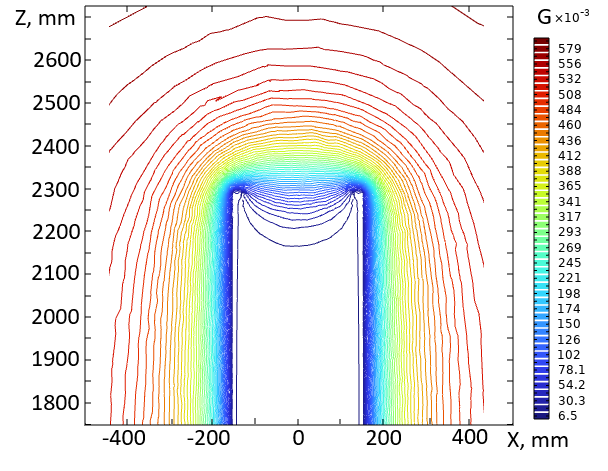}
   \caption{The magnetic field lines near the top section of the shields illustrating how the magnetic field will penetrate into the top section of the shields.}
    \label{fig:Bleak}
\end{figure}

{\bf Magnetic field near the end-sections of the shielding:}
The magnetic field near the tops of the shields will leak inside the shields, as can be seen in Fig.~\ref{fig:Bleak}, showing that the field penetrates $\sim150\,\textrm{mm}$ into the shields to a level of $6.5\,\textrm{mG}$ for the double-layer shielding configuration. The effects of this can be mitigated by end-cap screening and compensation coils but we anticipate that there will be a distance in the top and bottom range of the shields that cannot be used for interferometry purposes, and the shields will have to be slightly longer than the designed interferometry length.

\subsubsection*{Magnetic field measurements and noise}
\begin{figure}
    \centering
    \includegraphics[width=0.3\linewidth]{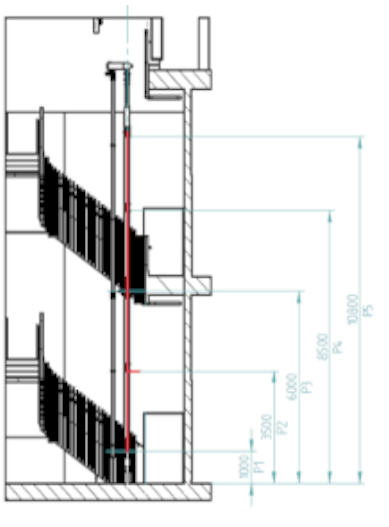}
    \includegraphics[width=0.6\linewidth]{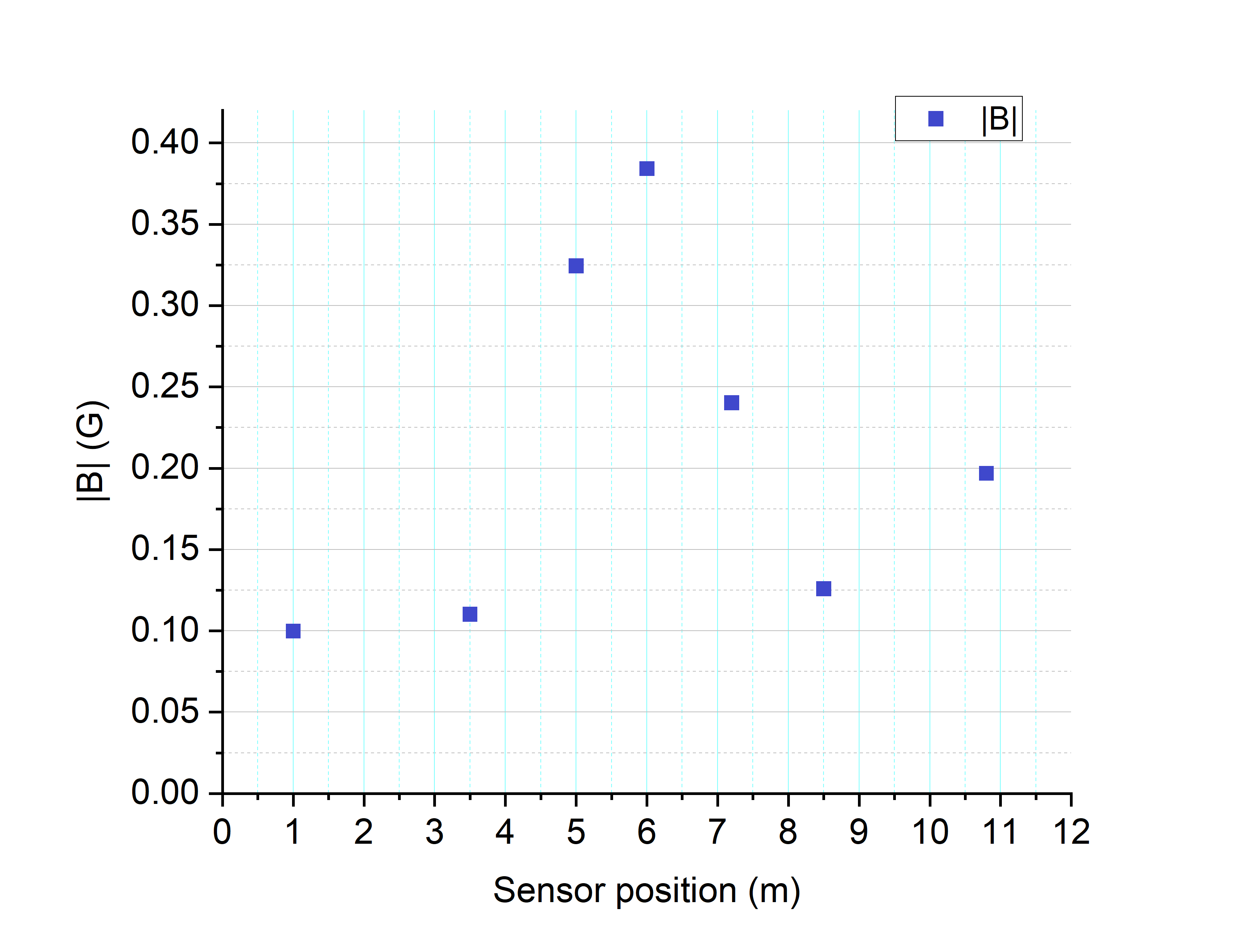}
   \caption{Ambient magnetic field measurements in the Beecroft building. The magnetic field (right panel) has been measured at various heights in the building (left panel).}
    \label{fig:BBeecroft}
\end{figure}

Magnetic field measurements have been conducted in the Beecroft building using fluxgate magnetometry. These measurements have been taken at the location where the interferometer assembly will be positioned, at different heights along the vertical axis. The measurements have been taken for periods up to one week, allowing us to construct the ambient magnetic noise spectral density in a frequency range of $1\,\textrm{$\mu$Hz}$ to $50\,\textrm{Hz}$. Fig.~\ref{fig:BBeecroft} shows a set of magnetic field measurements at various heights in the Beecroft building. There is a noticeable fluctuation in the ambient magnetic field as a function of height, which can be explained by the difference in surroundings over the full length of the experimental location, ranging from a basement environment, through the vicinity of the (ferromagnetic) stairwell structures and closeness to large windows at the top. The measurements have been repeated and were found to be stable in time. The static shielding factor of 1000 implies that these magnetic field variations should not represent issues for interferometry purposes.

\begin{figure}
    \centering
    \includegraphics[width=0.6\linewidth]{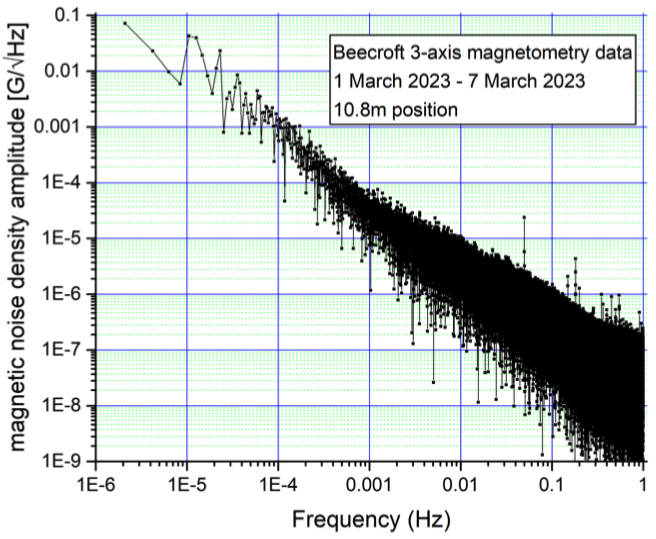}
   \caption{The ambient magnetic noise density amplitude based on measurements in the Beecroft building. Including a factor of 1000 shielding, the requirement of noise density level $< 1\mu$G/$\sqrt{\rm Hz}$ is met for frequencies above $0.1\,\textrm{mHz}$.}
    \label{fig:Bnoise}
\end{figure}

Figure~\ref{fig:Bnoise} shows the ambient magnetic noise spectral density at a height of $10.8\,\textrm{m}$ from the floor. The dynamic shielding of the double-layer shielding is expected to be of the order of 1000, which would imply a frequency region above $0.1\,\textrm{mHz}$ where we fulfil the requirement that the noise density amplitude be $< 1\mu$G/$\sqrt{\rm Hz}$, covering the frequency region where AION would take data.

\subsection{Magnetic Guide Field}
\label{sec:magguide}
\subsubsection*{General guide field coils arrangement}
\begin{figure}
    \centering
    \includegraphics[width=0.75\linewidth]{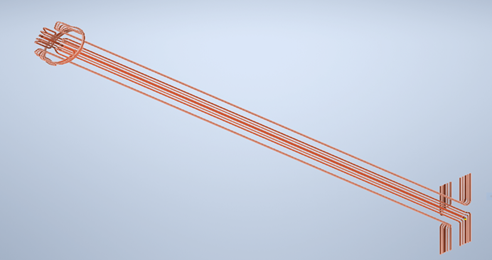}
   \caption{CAD image of the guide field coil assembly.}
    \label{fig:FieldCoil}
\end{figure}

To provide a stable and tuneable horizontal magnetic field the system contains two field coil assemblies (one for each interferometry region). Fig.~\ref{fig:FieldCoil} shows a CAD image of the field coil assembly. The wires are assembled onto a semi-cylindrical shell measuring $4.6\,\textrm{m}$ in length, and a maximum envelope diameter of $272\,\textrm{mm}$, including the wires. Due to the size of the shell, it is expected that this will be fabricated in parts before being mechanically clamped together. At such breaks, the connection between the wires will be specified carefully to maintain magnetic field homogeneity. 
The two semi-cylindrical field coil halves are joined together and 
connected to the 
same expansion joint as the magnetic shielding, to take up thermal expansion.
For further radial constraint, linear bearings are employed in a similar fashion to the magnetic shielding. This is to allow for differential thermal expansion and heating.

\subsubsection*{Field homogeneity and field divergence modelling}

\begin{figure}
    \centering
    \includegraphics[width=0.6\linewidth]{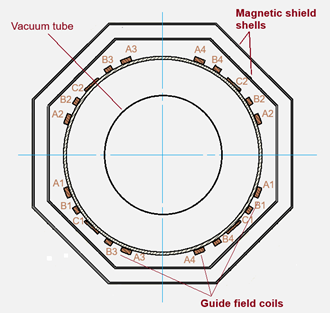}
   \caption{Field coil geometry showing the two sets of coils.}
    \label{fig:CoilGeom}
\end{figure}

\begin{figure}
    \centering
    \includegraphics[width=0.75\linewidth]{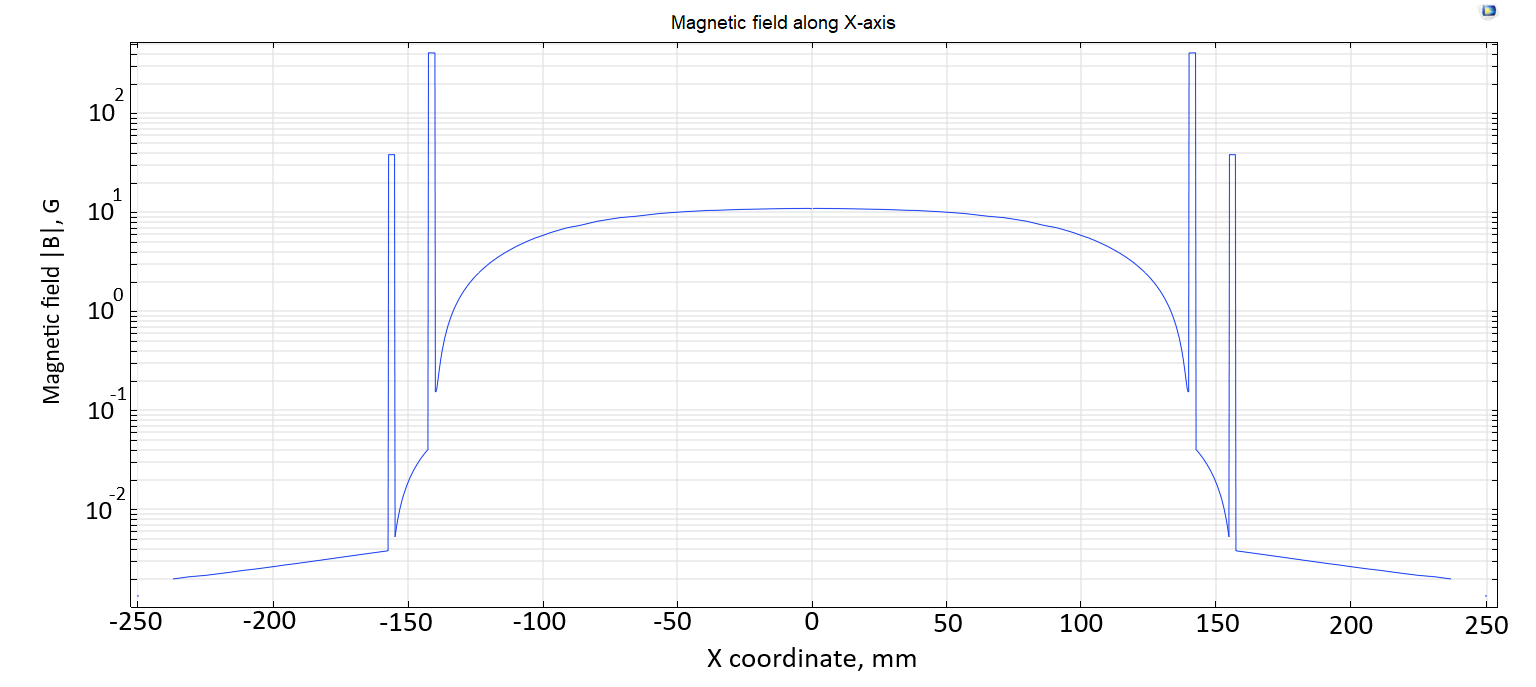}
    \caption{Modelling of the field generated by the magnetic guide field coils inside the double layer magnetic shield.}
    \label{fig:fieldmodelled}
\end{figure}

The magnetic guide field has been modelled based on a set of five pairs of coils $4.6\,\textrm{m}$ long, as sketched in Fig.~\ref{fig:CoilGeom}. By selecting specific coil pairs the horizontal orientation of the resulting magnetic field can be set. Fig.~\ref{fig:fieldmodelled} shows the magnetic field generated in the tube, inside the double magnetic shields. The shielding for a cylindrical shield geometry is known analytically (see, e.g.,~\cite{wodey2020scalable}), we have used COMSOL multiphysics v6.3 in all the shielding and magnetic guide field calculations~\cite{COMSOL}. The current in coils A1 and A2 are set to $4.6\,\textrm{A}$, in coils B1 and B2 to $3.5\,\textrm{A}$ and in coils C1 and C2 to $3.3\,\textrm{A}$. A field strength of $10.8\,\textrm{G}$ is generated. The field in the interferometry region is below $5\,\textrm{mG}$. 
 
\subsubsection*{Magnetic Field Orientation}

\begin{figure}
    \centering
    \includegraphics[width=0.75\linewidth]{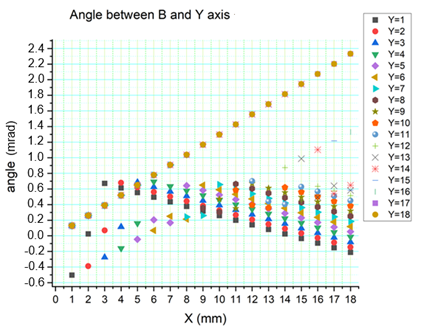}
    \caption{The magnetic field orientation as a function of the $x$ coordinate for a set of $y$ coordinates. The guide field is set along the $y$ axis.}
    \label{fig:fielddivergence}
\end{figure}

The orientation of the field generated by this set of coils has also been modelled. Fig.~\ref{fig:fielddivergence} shows the fluctuations in field orientation  [mrad] as a function of the $x$ position calculated for a given $y$ position. The maximum deviation for the ideal orientation is well below $5\,\textrm{mrad}$ in the $19\,\textrm{mm}$ range from the centre of the tube.

\subsubsection*{Tuning the Field Orientation}

\begin{figure}
    \centering
    \includegraphics[width=0.75\linewidth]{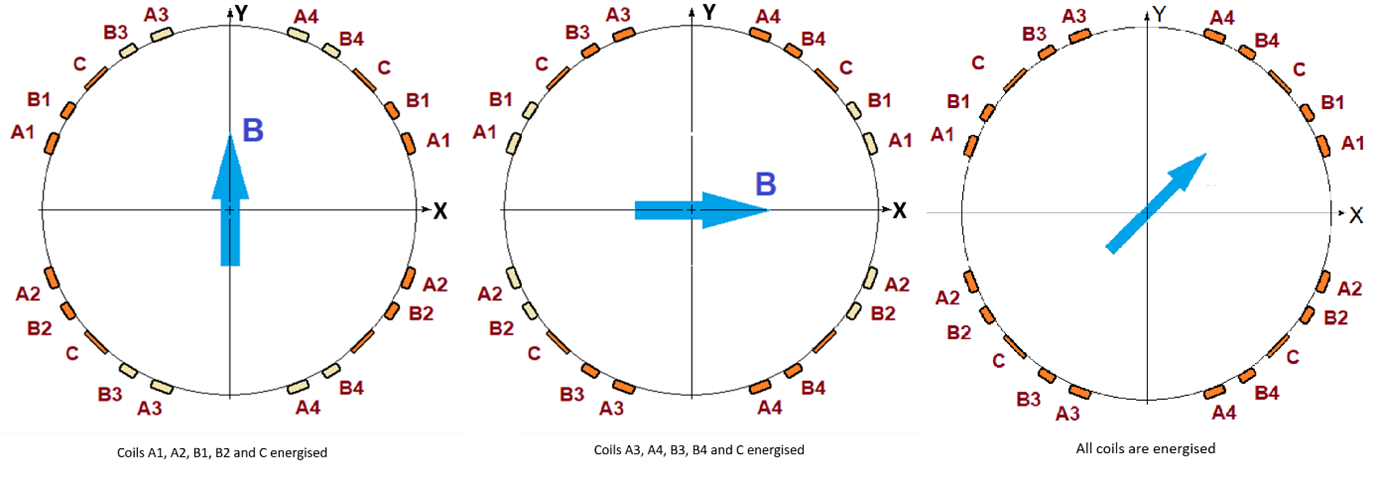}
    \caption{Illustration of field orientations generated by the different coils.}
    \label{fig:fieldorient}
\end{figure}

The field direction can be set by selecting the various coils as illustrated in Fig.~\ref{fig:fieldorient}. The field is applied along the $y$-axis with one set of coils activated and applied along the $x$-axis when the second set of coils is activated. When activating both set of coils the field can be directed at any angle in the $(x, y)$ plane. 

\begin{figure}
    \centering
    \includegraphics[width=0.75\linewidth]{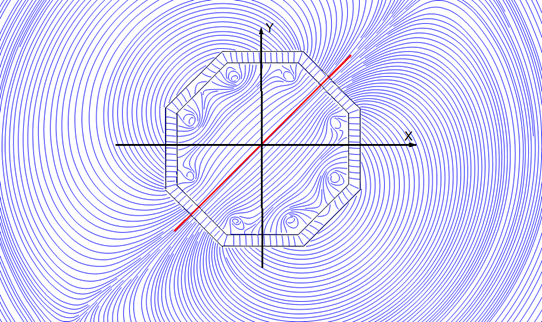}
    \caption{Magnetic field lines generated with equal currents through the coil sets. The magnetic field orientation is along the $x = y$ line.}
    \label{fig:fieldlines}
\end{figure}

\begin{figure}
    \centering
    \includegraphics[width=1\linewidth]{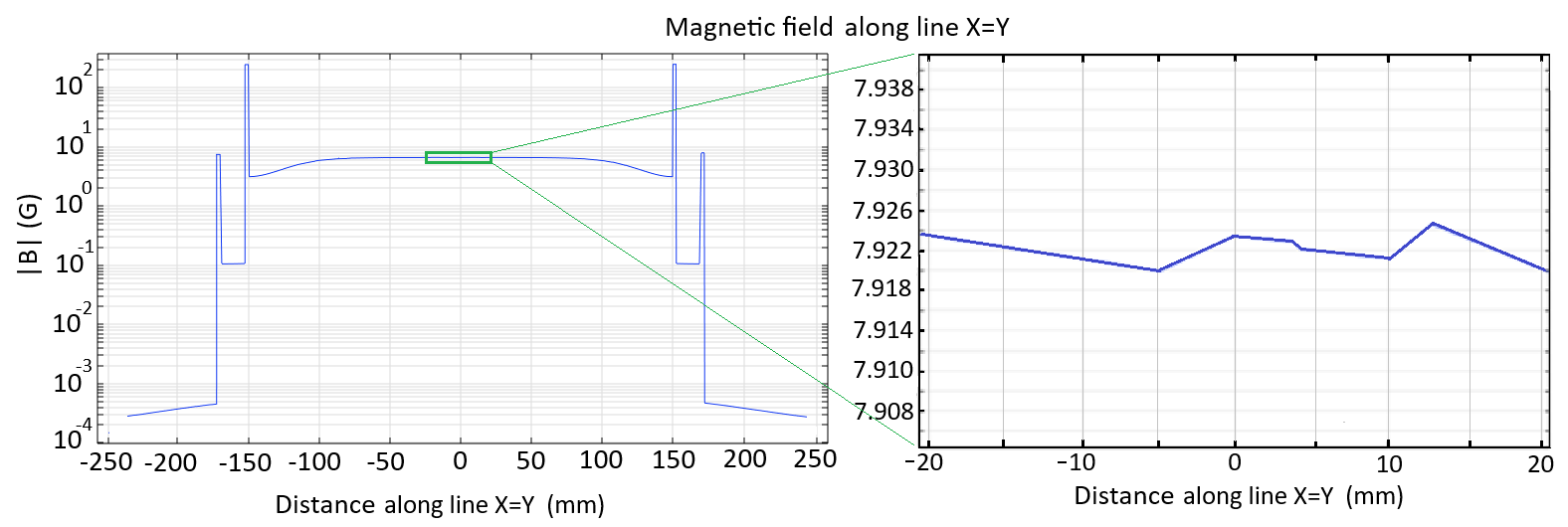}
    \caption{The magnetic field magnitude generated along the $x = y$ line. The inset shows the field in the central region with a homogeneity below $5\,\textrm{mG}$.}
    \label{fig:fieldmag}
\end{figure}

Figure~\ref{fig:fieldlines} shows the field lines as modelled with the coils arranged for a $45^\circ$ angle in the $(x, y)$ plane. This field orientation is obtained by applying the same current settings to each coil set. The corresponding field magnitude along the $x = y$ axis is shown in Fig.~\ref{fig:fieldmag} showing that, as one would expect in this symmetrical layout, the field homogeneity stays below $5\,\textrm{mG}$.

\subsubsection*{Correction Coils}
\begin{figure}
    \centering
    \includegraphics[width=0.75\linewidth]{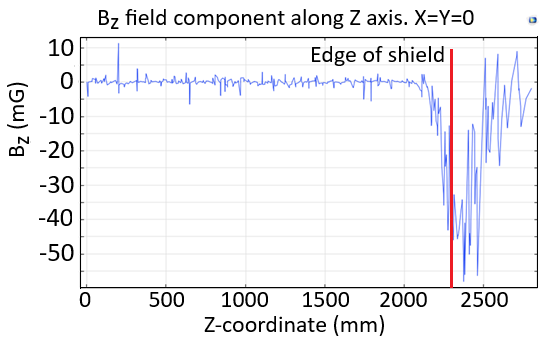}
    \caption{The magnetic field $B_z$ inside the shields: near the edge of the shield the external field will penetrate into the shields. The spike at z=$\sim200\,\textrm{mm}$ is due to mesh refinement.}
    \label{fig:fieldpen}
\end{figure}

The magnetic field near the shield edges will naturally be affected by the external magnetic field (the Earth's field and the additional magnetic environment surrounding the experiment). As this effectively reduces the available length over which interferometry can be performed, it is of importance to minimise the region where the field is affected. Fig.~\ref{fig:fieldpen} shows the magnetic field as modelled near the shield edges: the field is disturbed by the external environment up to a distance of $100\,\textrm{mm}$ from the edge. This would thus reduce the available interferometry length by $200\,\textrm{mm}$.

\begin{figure}
    \centering
    \includegraphics[width=0.4\linewidth]{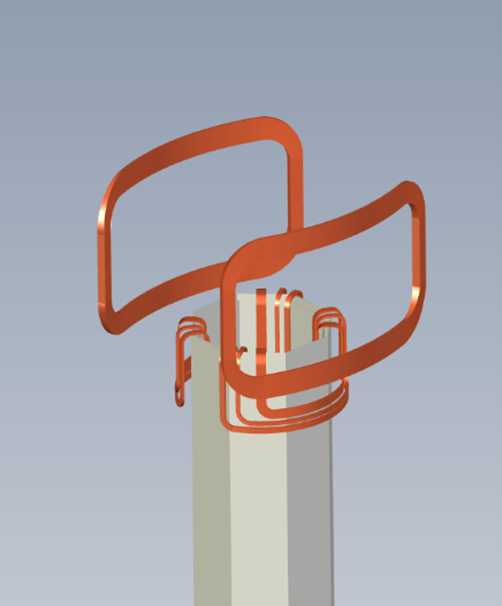}
    \includegraphics[width=0.5\linewidth]{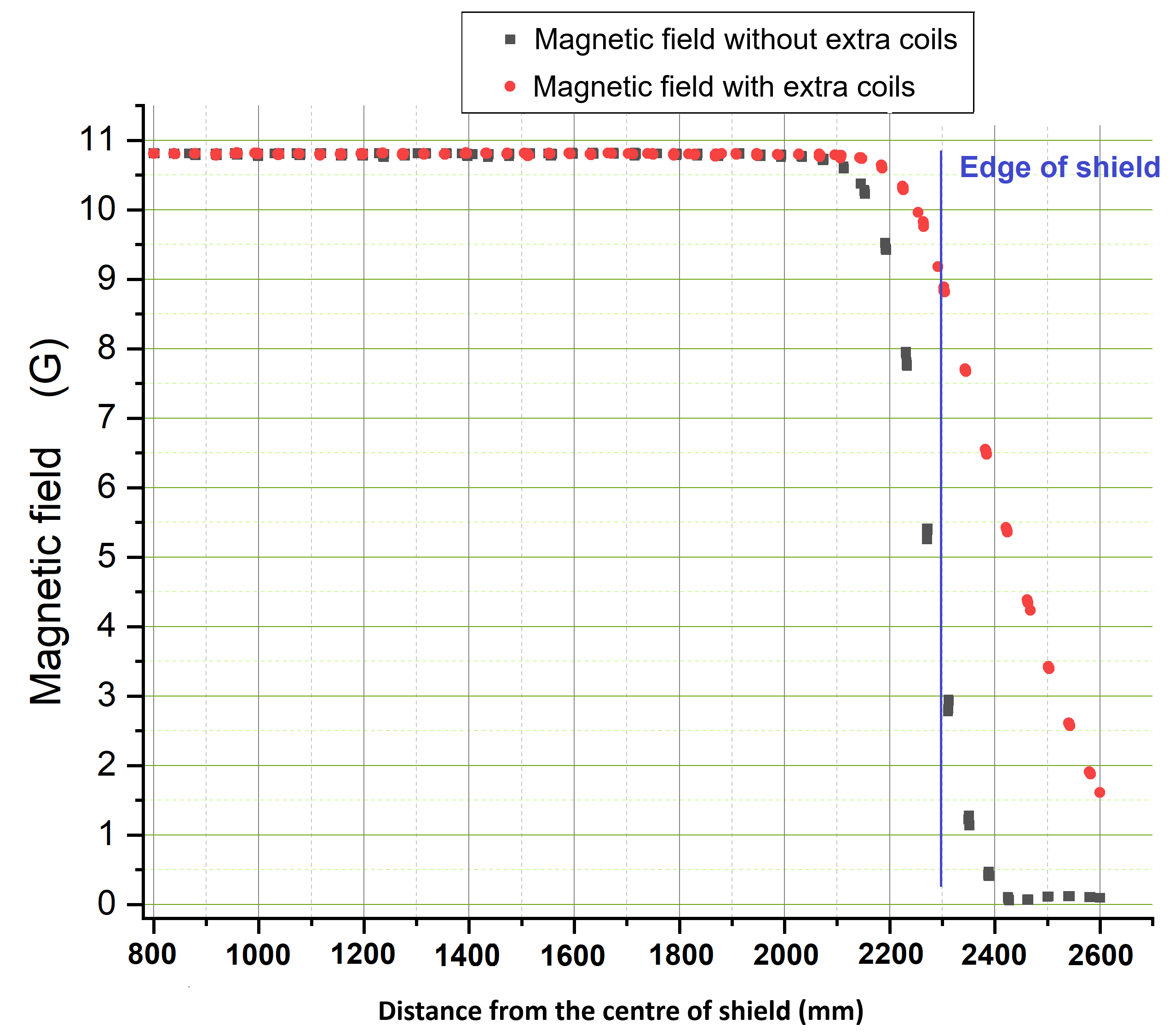}
    \caption{Left panel: A pair of correction coils placed near the field edge. Right panel: The magnetic field in the shields modelled with (red) and without (black) the compensation coil energised.}
    \label{fig:compcoils}
\end{figure}

To maximise the available interferometry region in the tube, correction coils will need to be added near the edges of the shields. A pair of correction coils has been modelled to provide the correction to the external magnetic field near the shield edges (see Fig.~\ref{fig:compcoils}). Modelling the guide field with the correction coils activated show that the guide field is extended closer to the shield edge and provides us with a longer interferometry region.

\begin{figure}
    \centering
    \includegraphics[width=1\linewidth]{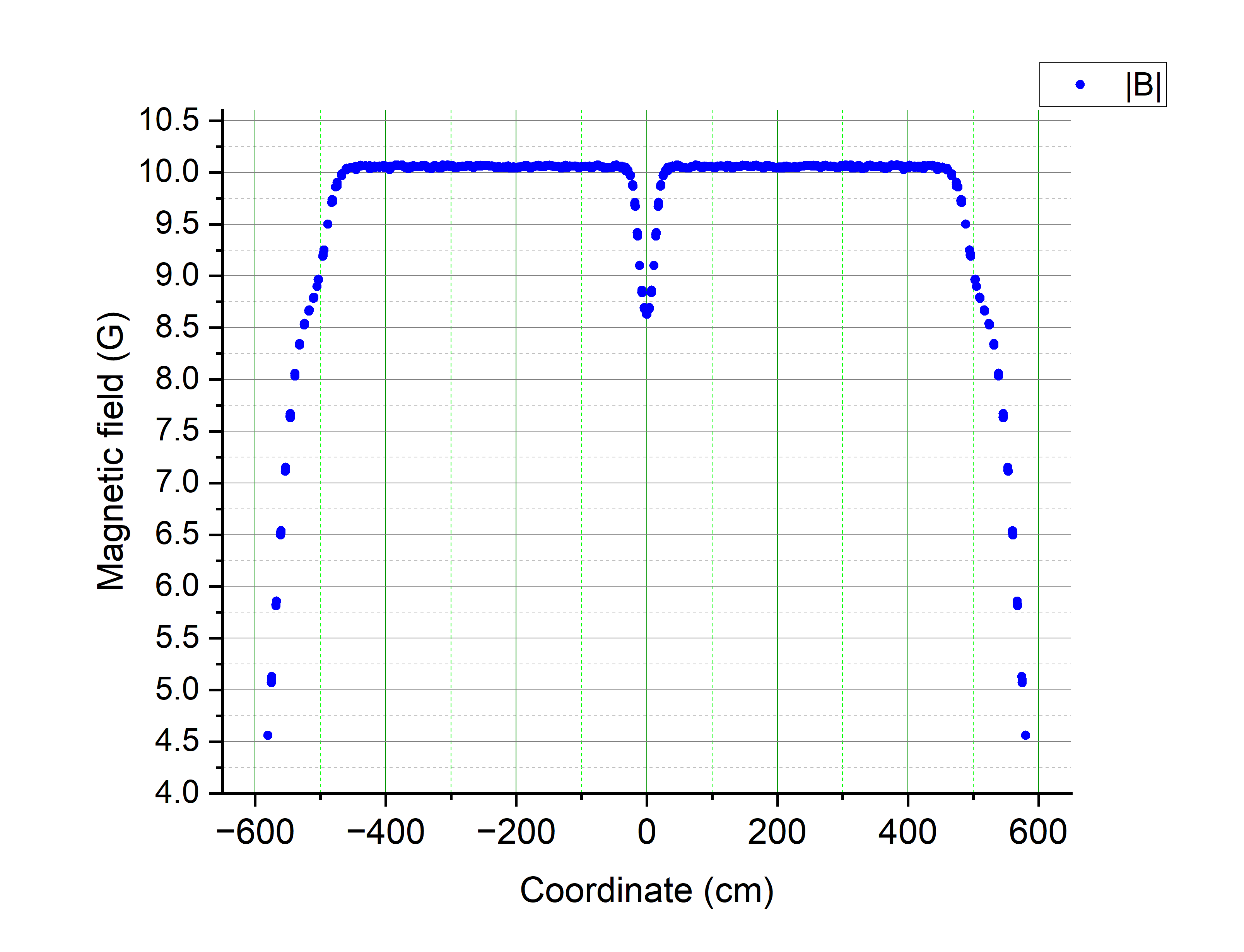}
    \caption{Magnetic field over the full length of the interferometer applying guide fields and compensation coils.}
    \label{fig:fullfield}
\end{figure}

The magnetic field has been modelled over the full length of the interferometer (i.e., over the two separate interferometer and shields sections). The magnetic field, with the guide field coils and the compensation coils applied, is shown in Fig.~\ref{fig:fullfield}. 

\section{Vacuum Analysis}
\label{sec:vacuum}
A key objective of the vacuum design is to understand the approximate pumping requirements to achieve the specified vacuum level.  This can be done by using the surface area of the vessel and the known outgassing properties of the material it is made from.  The treatment of the vessel material is critical to reach the required vacuum, and we first address the requirement to bake the system out.
\subsection{Initial Estimates}
We have assumed that the vessel is a tube of diameter $153\,\textrm{mm}$ and length $11.1\,\textrm{m}$, an underestimate of the surface area as we have not included any of the chambers or internal optics.
If we wish to avoid a bake-out of the vacuum system in-situ we must prepare the vessel (vacuum clean and bake-out) before moving to the site.  The sections must then be vented to connect them together into the full tower before final pump down of the system.
Assuming the use of pumps with a pumping rate of $1000\,\textrm{l/s}$ and specifying a $5 \times 10^{-10}$ mbar vacuum level in a stainless-steel vessel, we would require 22 pumps on the interferometer.  This increases to 1076 pumps if the ideal XHV conditions of $1 \times 10^{-11}$ mbar are to be met. We can reduce this to 38 pumps by using a titanium vessel, which has a lower material outgassing rate.  All these options are clearly unfeasible for the instrument and hence an in-situ bake out will be required and design considerations for this must be considered.

To reach the challenging ideal specification of XHV $1 \times 10^{-11}$~mbar we will require an in-situ bake of the stainless-steel vessel, and if we can also increase the pumping rate to $2000\,\textrm{l/s}$ this will allow the use of only 3 pumps on the interferometer.  A possible pump would be a combined NEG - ion pump such as NEXTorr D2000-10~\cite{NEXTorr}.  This has a pumping speed of $2000\,\textrm{l/s}$, a weight of $6.7\,\textrm{kg}$ and mounts on a CF100 flange.  This is a large flange and consideration should be given to the possibility of using two smaller pumps.  Other options can also be considered, depending on the material choice such as a titanium or NEG coated stainless steel vessel.
Once all details on the final in-vacuum components are finalized, more detailed modelling will be undertaken to determine the pressure profile along the length of the tower.  This will inform optimised pump placement to reach the required vacuum specification.

\subsection{Molflow+ Vacuum Simulation}
We now outline the initial vacuum calculations performed for the AION-10 project, which were made using the computer program code {\tt Molflow+}~\cite{Molflow} based on the Monte-Carlo method.  Monte-Carlo simulations can be performed only for a closed volume; therefore the pumping ports are closed by virtual surfaces with the following properties: 
\begin{itemize}
\item Gas desorption or gas injection is described by a desorption coefficient; 
\item Pumping is described in terms of the surface area $A$, and a sticking probability (or capture factor) $\alpha$, where $\alpha = 0$ corresponds to no pumping speed and $\alpha = 1$ corresponds to an ideal pumping speed (a `black hole').
\end{itemize}
 
\begin{figure}
    \centering
    \includegraphics[width=0.45\linewidth]{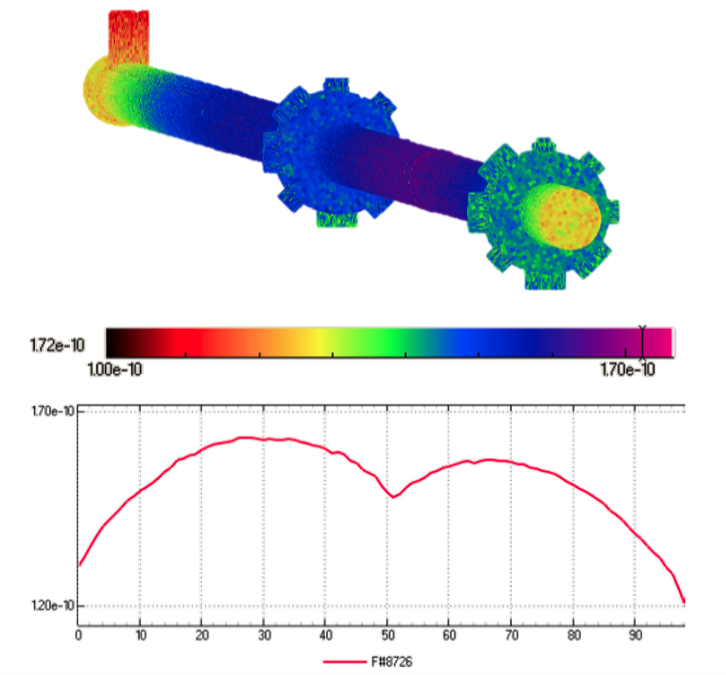}
    \includegraphics[width=0.45\linewidth]{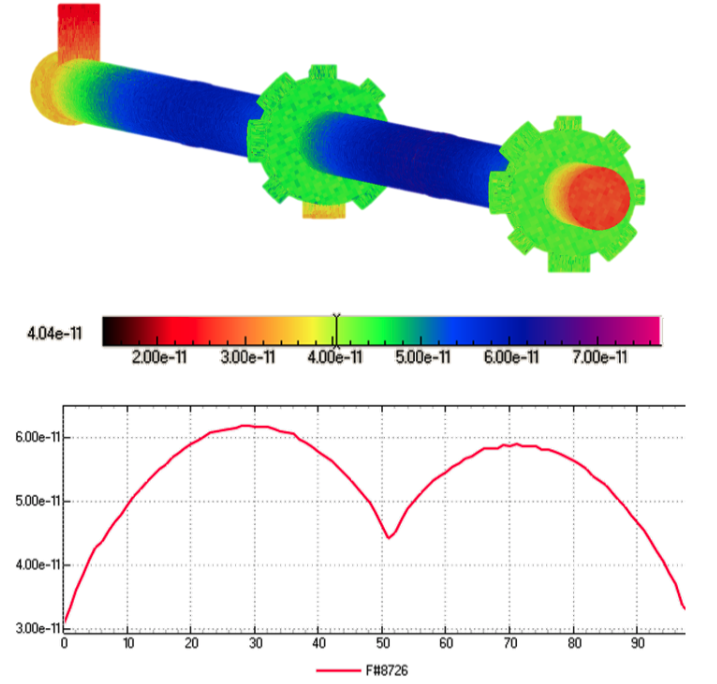}\\
    \caption{AION-10 vacuum simulations with (left panel) three $75\,\textrm{l/s}$ pumps and (right panel) three $300\,\textrm{l/s}$ pumps.}
    \label{fig:VacSims}
\end{figure}

For the purposes of the vacuum calculations an outgassing rate for stainless steel was assigned of $5 \times 10^{-13}$~mbar~l/s/cm$^2$, which is recognised as a very good outgassing rate that can be achieved assuming the stainless steel has been treated in accordance with good UHV practices~\cite{calder1967reduction}.
Two vacuum simulations were performed:
\begin{itemize}
\item Outgassing rate of all internal vacuum surfaces were set to $5 \times 10^{-13}$~mbar~l/s/cm$^2$, and three pumping ports were assumed with a pumping speed of 75 l/s as shown in the left panel of Fig.~\ref{fig:VacSims};
\item Outgassing rate of all internal vacuum surfaces were set to $5 \times 10^{-13}$~mbar~l/s/cm$^2$, three pumping ports were assumed with a pumping speed of 300 l/s as shown in the right panel of Fig.~\ref{fig:VacSims}.
\end{itemize}

Fig.~\ref{fig:VacSims} shows the AION-10 structure and the vacuum calculations that have been performed. Each figure shows the AION-10 structure in multiple colours, which can be matched with the colour bar beneath the plot that shows how the pressure varies along the interferometry beam pipe. The pressure profile plot underneath shows the average pressure through the centre of the AION-10 vacuum vessel from one end to the other, whereas the 3D plot shows the average pressure at the vacuum surfaces. In the majority of simulations both the plots show identical results, however, there may be special situations where the two differ.

The results shown in Fig.~\ref{fig:VacSims} that in both cases good UHV pressures can be achieved. However, the $300\,\textrm{l/s}$ pumping speed produces better results, with the average pressure in the $10^{-11}\,\textrm{mbar}$ pressure range, which is the minimum requirement for the AION-10 project.  For the purposes of the AION-10 project more details are needed to make the vacuum simulations more representative of the real situation, e.g., no details have been input for the strontium side-arm source or the laser transport in these initial calculations.

Although modelling has been done with $300\,\textrm{l/s}$ up to this point, we will need to increase this to meet the agreed specification.  The pressure profile indicates a factor of 3 increase near the centre point between pumping ports, which should be assessed to understand the impact on the interferometer.  This can be reduced by more distributed pumping using additional pumping ports or coating, and further work is required to find the optimum solution.

It should be noted that improved performance can be achieved with careful consideration of the material choices for AION-10 and the manufacturing processes and specifications that must be met.  Detailed vacuum specifications that are adhered to can produce outgassing rates down to the $10^{-14}$ to $10^{-15}$~mbar~l/s/cm$^2$ range~\cite{Young}. Therefore, in comparison to the calculations summarised here a reduction in pressure by an order of magnitude below the results in Fig.~\ref{fig:VacSims} is possible.

\section{Conclusions}
\label{sec:conclusions}

\subsection*{Summary}
Considerable progress has been made in the AION-10 tower design and analysis.
These include a mature design of the tower structure, the development of the magnetic shielding, the incorporation of active vibration isolation features, completion of the telescope stability analysis, the development of a multi-input vibration model, and a new vibration test in the Beecroft building.

The design of the tower structure has proceeded to a mature stage with proper consideration of the assembly method and sequence for the modules. The refinement of the design was conducted alongside analysis to minimize the impact of mechanical vibration on the working of the atom interferometer with particular attention paid to the lenses within the telescope module. The analysis suggests that the current design achieves satisfactory structural stability.

As a tool for further improving the structure design, a new multi-input vibration model has been developed that tackles the limitation of the current analysis, which accommodates only one vibration input. The proposed model allows the contributions of different vibration sources to be considered and combined. This enables a more realistic simulation of the AION-10 structure in which vibrations can be input through multiple supports. As the new model requires vibration data obtained simultaneously from different locations, a new vibration test was carried out in the Beecroft building. The test involved placing sensors near the frame supports and sampled synchronized vibration data. Further data collection will be performed to prepare for analysis in the next stage.

The design of the magnetic shield has also undergone significant development. The shield has a modular design that facilitates installation and can adapt to future dimension changes without the need for reworking. The completed magnetic shield design has been validated by the manufacturer and is deemed to be feasible for manufacturing.

\subsection*{Plans for Future Work}
The following work is planned for the next stage of the project.

Immediate priority tasks include finalising the active vibration isolation system and completing the detailed mechanical design of launch lattices and interconnects.

To refine the design of the support tower, detailed design of the connections between the instrument, the magnetic shield and the support tower frame needs to be developed. This includes any temporary structures that may be required to support the instrument and magnetic shield during the horizontal assembly of the long main modules. The support tower frame geometry may also need to be refined around the interconnect chamber so as to accommodate components that go around the chamber, servicing requirements for the optics, integration of the bias coils, and installation of the launch lattice that extends out from the interconnect chamber into the main vacuum pipe below and above it. Further refinement and integration testing of the split launch lattice and interconnect mechanical assemblies will be pursued during the final construction and commissioning phases. The bracing patterns on the support tower can also be refined to optimise the structural stiffness of the frame. In particular, the beam transfer pipe module at the top of the tower requires further support from extra bracings that are yet to be developed. This can be done through modal analyses on iterations of bracing pattern designs.

A thermal model will be developed to examine the effect of thermal expansion on the instrument’s stress and stability during bakeout and operation. Further vibration analyses need to be conducted on the updated support tower and instrument model to ensure the stability requirement is still met. Additional vibration analysis will be needed to check the stability of other optics in the instrument, e.g., the beam conditioning pipe and top chamber.

Further development is required to apply the proposed multi-input vibration model to the stability analysis. It will also be necessary to extend the vibration tests in the Beecroft building. The tests reviewed in this report only collected data for $45\,\textrm{min}$, which is too short to capture sufficient information to reflect the overall vibration behaviour of the building. In future work, it is recommended to perform tests that last more than 2 days. The hardware and installation are currently constraints on the test. The locations of the vibration sensors are limited by cable length and accessibility in the stairwell, so it was not possible to place the sensors at the most desired locations. To relax these limitations in the future, innovative approaches may be considered, e.g., using wireless sensors and data-acquisition systems to collect data, and the use of robots/drones to deploy the sensors.

These developments will prepare the Collaboration for the Final Design Report and subsequent construction phase.

\subsection*{Perspectives}
Successfully delivering AION-10 will establish a key capability for using quantum sensors to detect gravitational waves and dark matter, laying the foundation for future large-scale atom interferometry experiments such as AION-100, AION-km~\cite{Badurina:2019hst} and AEDGE~\cite{AEDGE:2019nxb}.
By pioneering quantum sensor technologies at $10\,\textrm{m}$ scale, the AION Collaboration aims for the forefront of fundamental physics exploration in the coming decades.

\section*{Acknowledgements}

This work was supported by UKRI through its Quantum Technology for Fundamental Physics programme, via the following grants from EPSRC and STFC in the framework of the AION Consortium: ST/T006536/1 and ST/W006448/1 to the University of Birmingham; ST/T006579/1, ST/W006200/1 and ST/X004864/1 to the University of Cambridge; ST/T007001/1 to the University of Liverpool, ST/T006994/1 and ST/W006332/1 to Imperial College London; ST/T00679X/1 to King’s College London; ST/T006633/1 to the University of Oxford; ST/T006358/1 and ST/W006510/1 to the STFC Rutherford-Appleton Laboratory. L.B. acknowledges support from the STFC Grant No. ST/T506199/1. Di.B. acknowledges support from the R\&D\&i project PID2023-146686NB-C31 funded by MICIU/AEI/10.13039/501100011033/ and by ERDF/EU and from the ERC grant ERC-2024-SYG 101167211. IFAE is partially funded by the CERCA program of the Generalitat de Catalunya. J.M. acknowledges support from the University of Cambridge Isaac Newton Trust. Je.S. acknowledges support from the Rhodes Trust. For the purpose of open access, the authors have applied for a Creative Commons Attribution (CC-BY) licence to any Author Accepted Manuscript version arising from this submission.
Close collaboration between Liverpool, Stanford and Northwestern Universities and Fermilab was crucial for this work. We thank Lucy Nobrega, James Santucci, Linda Valerio and Ronald J. Kellet (Fermilab), Tim Kovachy (Northwestern), Ben Garber and Jason Hogan (Stanford). Finally, we thank Liverpool’s  Detector Development Manufacturing Facility staff for manufacturing all custom parts.

\bibliographystyle{JHEP}
\bibliography{CDR}

\end{document}